\documentclass[a4paper,10pt]{article}

\usepackage{jheppub} 

\usepackage[T1]{fontenc} 
\usepackage{bbold}
\usepackage{amsmath}
\usepackage{empheq}
\usepackage{booktabs}

\newcommand{\as}{\alpha_s}

\newcommand{\css}{\text{\scriptsize CSS}}
\newcommand{\zi}{z_i^{(\ell_i)}}

\newcommand{\Vsc}{V}

\newcommand{\Vfull}{V}

\newcommand{\dZ}{d{\cal Z}[\{R', k_i\}]}

\title{\boldmath Momentum-space resummation for transverse observables
  and the Higgs $p_\perp$ at N$^3$LL+NNLO}

\author[a]{Wojciech Bizo\'{n},}
\author[b]{Pier Francesco Monni,}
\author[b,c]{Emanuele Re,}
\author[a]{Luca Rottoli,}
\author[d]{Paolo Torrielli.}

\affiliation[a]{Rudolf
  Peierls Centre for Theoretical Physics,University of Oxford, Keble
  Road, Oxford OX1 3NP, UK}
\affiliation[b]{CERN, Theoretical Physics Department, CH-1211 Geneva 23,
Switzerland}
\affiliation[c]{LAPTh, CNRS, Universit\'e Savoie Mont Blanc, BP110, F-74941 Annecy-le-Vieux Cedex, France}
\affiliation[d]{Dipartimento di Fisica, Universit\`a di Torino and INFN, Sezione di Torino,
Via P. Giuria 1, I-10125, Turin, Italy}

\emailAdd{wojciech.bizon@physics.ox.ac.uk}
\emailAdd{pier.monni@cern.ch}
\emailAdd{emanuele.re@lapth.cnrs.fr}
\emailAdd{luca.rottoli@physics.ox.ac.uk}
\emailAdd{torriell@to.infn.it}

\abstract{We present an approach to the momentum-space resummation of
  global, recursively infrared and collinear safe observables that can
  vanish away from the Sudakov region.
  We focus on the hadro-production of a generic colour singlet, and we consider
  the class of observables that depend only upon the total transverse
  momentum of the radiation, prime examples being the transverse
  momentum of the singlet, and $\phi^*$ in Drell-Yan pair production.
  We derive a resummation formula valid up to
  next-to-next-to-next-to-leading-logarithmic accuracy for the
  considered class of observables. We use this result to compute
  state-of-the-art predictions for the Higgs-boson
  transverse-momentum spectrum at the LHC at
  next-to-next-to-next-to-leading-logarithmic accuracy matched to
  fixed next-to-next-to-leading order.
  Our resummation formula reduces exactly to the
  customary resummation performed in impact-parameter space in the
  known cases, and it also predicts the correct power-behaved scaling
  of the cross section in the limit of small value of the observable.
  We show how this formalism is efficiently implemented by means of
  Monte Carlo techniques in a fully exclusive generator that allows
  one to apply arbitrary cuts on the Born variables for any colour
  singlet, as well as to automatically match the resummed results to
  fixed-order calculations.
}

\preprint{OUTP-17-05P, CERN-TH-2017-111, LAPTH-018-17}

\allowdisplaybreaks

\begin{document} 
\maketitle
\flushbottom

\section{Introduction}
After the discovery of the Higgs
boson~\cite{Aad:2012tfa,Chatrchyan:2012xdj}, the precise measurements
from Run 2 of the LHC programme have so far confirmed the Standard
Model with remarkable precision. Given that signals of new physics
will most likely be elusive, it is important to define and study
observables that can be both experimentally measured and theoretically
predicted with a few-percent uncertainty.  In this scenario, a
prominent role is played by processes featuring the production of a
colour singlet of high invariant mass, for instance gluon-fusion Higgs
and Drell-Yan, where quantities like the transverse momentum of the
singlet or angular observables defined on its decay products have been
studied with increasing accuracy in the last decades.

The differential study of these processes not only is important from a
purely phenomenological perspective, but also because it represents
the ideal baseline for a more fundamental understanding of the
underlying theory. Their structural
simplicity indeed allows one to provide predictions that include 
several orders of
perturbative corrections, hence probing in depth many
non-trivial features of QCD.

In this paper, we consider the hadro-production of a heavy colour
singlet, and we study the class of observables, henceforth denoted by
the symbol $v$, which are both {\it transverse} (i.e. which do not
depend on the rapidity of the radiation) and {\it inclusive}
(i.e. that depend only upon the total momentum of the radiation). As
such, they only depend on the total transverse momentum of the
radiation. Specifically, we concentrate on the transverse-momentum
distribution of a Higgs boson in gluon fusion, but we stress that the
same formulae hold for the whole class of transverse and inclusive
observables, for instance the $\phi^*$ angle in Drell-Yan pair
production. Moreover, although we limit ourselves to inclusive
observables, the formalism presented in this work can be
systematically extended to all transverse observables in
colour-singlet hadro-production.

Inclusive and differential distributions for gluon-fusion Higgs
production are nowadays known with very high precision.  The inclusive
cross section is now known at next-to-next-to-next-to-leading-order
(N$^3$LO) accuracy in QCD~\cite{Anastasiou:2015ema,Anastasiou:2016cez}
in the heavy top-quark limit. The N$^3$LO correction amounts to a few
percent of the total cross section, indicating that the perturbative
series has started to manifest convergence and that missing
higher-order corrections are now getting under theoretical
control. Current estimates show that they are very moderate in
size~\cite{Bonvini:2016frm}.  The state-of-the-art results for the
Higgs transverse-momentum spectrum in fixed-order perturbation theory
are the next-to-next-to-leading-order (NNLO) computations of
refs.~\cite{Boughezal:2015dra,Boughezal:2015aha,Caola:2015wna,Chen:2016zka},
which have been obtained in the heavy top-quark limit.  The impact of
quark masses on differential distributions in the
large-transverse-momentum limit is still poorly known beyond leading
order, while in the moderate-$p_t$ region, next-to-leading-order (NLO)
QCD corrections to the top-bottom interference contribution were
recently
computed~\cite{Melnikov:2016qoc,Melnikov:2017pgf,Lindert:2017pky}.

Although fixed-order results are crucial to obtain reliable
theoretical predictions away from the soft and collinear regions of
the phase space ($v\sim 1$), it is well known that 
regions dominated by soft and collinear QCD radiation --- which give rise to the
bulk of the total cross section --- are affected by
large logarithmic terms of the form $\alpha_s^n \ln^k(1/v)/v$, with
$k\leq 2n-1$, which spoil the convergence of the perturbative series
at small $v$.
In order to have a finite calculation in this limit, the subtraction
of the infrared and collinear divergences requires an all-order
resummation of the logarithmically divergent terms. The logarithmic
accuracy is commonly defined in terms of the perturbative series of the
\emph{logarithm} of the cumulative cross section $\Sigma$ as
\begin{align}
\label{eq:cumulant-initial}
  \ln \Sigma(v) &\equiv \ln\int_0^v dv' \frac{d \sigma(v')}{d v'} \notag\\
&= \sum_n \left\{{\cal O}\left(\alpha_s^n\ln^{n+1}(1/v)\right) + {\cal O}\left(\alpha_s^n\ln^{n}(1/v)\right) + {\cal O}\left(\alpha_s^n\ln^{n-1}(1/v)\right)+\dots\right\}.
\end{align}
One refers to the dominant terms $\alpha_s^n \ln^{n+1}(1/v)$ as
leading logarithmic (LL), to terms $\alpha_s^n
\ln^{n}(1/v)$ as next-to-leading logarithmic (NLL), to
$\alpha_s^n \ln^{n-1}(1/v)$ as next-to-next-to-leading logarithmic
(NNLL), and so on.

The resummation of the $p_t$ spectrum of a heavy colour singlet was
first analysed in the seminal work by Parisi and
Petronzio~\cite{Parisi:1979se}, where it was shown that in the
low-$p_t$ region the spectrum vanishes as $ d\sigma/dp_t \sim p_t$,
instead of vanishing exponentially as suggested by Sudakov
suppression. This power-law behaviour is due to configurations in
which $p_t$ vanishes due to cancellations among the non-vanishing
transverse momenta of all emissions.  Around and below the peak of the
distribution, this mechanism dominates with respect to kinematical
configurations where $p_t$ becomes small due to all the emissions
having a small transverse momentum, i.e.~the configurations which
would yield an exponential suppression.  In order to properly deal
with these two competing mechanisms, in ref.~\cite{Collins:1984kg} it
was proposed to perform the resummation in the impact-parameter ($b$)
space, where both effects leading to a vanishing $p_t$ are handled
through a Fourier transform.

Using the $b$-space formulation, the Higgs $p_t$ spectrum was resummed
at NNLL accuracy in~\cite{Bozzi:2003jy,Bozzi:2005wk} using the
formalism developed in~\cite{Collins:1984kg,Catani:2000vq}, as well as
in~\cite{Becher:2012yn} by means of a soft-collinear-effective-theory
(SCET) approach~\cite{Becher:2010tm,GarciaEchevarria:2011rb}. A study
of the related theory uncertainties in the SCET formulation was
presented in ref.~\cite{Neill:2015roa}.  More recently, all the
necessary ingredients for the N$^3$LL resummation were
computed~\cite{Catani:2011kr,Catani:2012qa,Gehrmann:2014yya,Li:2016ctv,Vladimirov:2016dll},
with the exception of the four-loop cusp anomalous dimension which is
currently unknown.  This paves the way to more precise predictions for
transverse observables in the infrared region. The impact of both
threshold and high-energy resummation on the small-transverse-momentum
region was also studied in detail in
refs.~\cite{Li:1998is,Laenen:2000ij,Kulesza:2003wn,Marzani:2015oyb,Forte:2015gve,Caola:2016upw,Lustermans:2016nvk,Marzani:2016smx,Muselli:2017bad}.

The problem of the resummation of the transverse momentum distribution
in direct ($p_t$) space received substantial attention throughout the
years~\cite{Ellis:1997ii,Frixione:1998dw,Kulesza:1999sg}, but remained
unsolved until recently. Due to the vectorial nature of these
observables, it is indeed not possible to define a resummed cross
section at a given logarithmic accuracy in direct space that is
simultaneously free of any subleading logarithmic contributions and of
spurious singularities at finite values of $p_t > 0$.  Last year some
of us proposed a solution to this problem by formulating a resummation
formalism in direct space up to NNLL order~\cite{Monni:2016ktx}, and
used it to match the NNLL resummation to the NNLO Higgs $p_t$
spectrum. The problem of direct-space resummation for the
transverse-momentum distribution was also considered more recently in
ref.~\cite{Ebert:2016gcn} following a SCET approach, where the
renormalisation-group evolution is addressed directly in momentum
space. In this article we explain in detail the formalism introduced
in~\cite{Monni:2016ktx}. Furthermore, we extend it to N$^3$LL, and
formulate it in general terms, so that a direct application at this
logarithmic accuracy to all transverse, inclusive observables is
possible. We point out that our final result lacks the contribution of
the unknown four-loop cusp anomalous dimension, which is set to zero
in the following.

The paper is structured as follows: in Section~\ref{sec:nll} we sketch
the main features of our formalism, based on and extending the one
developed in ref.~\cite{Banfi:2004yd}, through the derivation of a
simplified NLL formula relevant to the case of scale-independent
parton densities. Section~\ref{sec:resolution} discusses the choice of
the resolution variable and kinematic ordering in the evolution of the
radiation. In Section~\ref{sec:higher-order} we discuss the structure
of higher-order corrections, and in particular in
Section~\ref{sec:recoil} we treat the inclusion of parton densities
and of hard-collinear radiation, thereby making our formalism fully
capable of dealing with initial-state radiation. In
Section~\ref{sec:compare-to-bspace} we prove that our method is
formally equivalent to the more common $b$-space formulation of
transverse-momentum resummation. Section~\ref{sec:n3ll} shows how to
evaluate our formula to N$^3$LL order and in
Section~\ref{sec:small-pt-limit} we present a study of the scaling
property of the differential distribution in the $p_t\to 0$ limit, and
compare our findings to the classic result by Parisi and
Petronzio~\cite{Parisi:1979se}. Finally, in Section~\ref{sec:results}
we discuss the matching to NNLO, and in Section~\ref{sec:pheno} we
present N$^3$LL accurate predictions for the Higgs-boson transverse
momentum spectrum at the LHC, matched to NNLO.

In Appendix~\ref{app:parton-shower} we show that, at NLL, the approach
used here is equivalent to a backward-evolution algorithm for this
class of observables, while Appendix~\ref{app:radiator} collects some
of the relevant equations used in the article.

\section{Derivation of the master formula}
We consider the resummation of a continuously global, recursive
infrared and collinear (rIRC) safe~\cite{Banfi:2004yd} observable $V$
in the reaction $pp\to B$, $B$ being a generic colourless system with
high invariant mass $M$.  It is instructive to work out in detail the
case of NLL resummation first. This will be done in
Section~\ref{sec:nll}, where we assume that the parton densities are
independent of the scale. We then discuss the inclusion of
higher-order corrections in Section~\ref{sec:higher-order}, and the
correct treatment of the parton luminosity will be dealt with in
Section~\ref{sec:recoil}. Finally, in
Section~\ref{sec:compare-to-bspace}, we discuss the connection to the
impact-parameter space formulation for transverse-momentum
resummation.

\subsection{Cancellation of IRC divergences and NLL resummation}
\label{sec:nll}
In the present subsection we assume that the parton densities are
independent of the scale and set to one for the sake of simplicity. To
set up the notation we work in the rest frame of the produced colour
singlet, and we introduce two reference light-like momenta that will
serve to parametrise the radiation
\begin{equation}
\label{eq:com-frame}
\tilde{p}_1 = \frac{M}{2}(1,0,0,1)\,,\qquad \tilde{p}_2 = \frac{M}{2}(1,0,0,-1) \,,
\end{equation}
where $M$ is the invariant mass of the colour singlet with momentum
$p_B$ that in this frame reads
\begin{equation}
p_B=\tilde p_{1}+\tilde p_{2}.
\end{equation}
The directions of the two momenta in Eq.~\eqref{eq:com-frame} coincide
with the beam axis at the Born level.
Beyond the Born level, radiation of gluons and quarks takes place, so
that the final state consists in general of $n$ partons with {\it
  outgoing} momenta $k_1,\dots,k_n$, and of the colour singlet. Due to
this radiation, the singlet acquires a transverse momentum with
respect to the beam direction. We express the final-state momenta by
means of the Sudakov parametrisation
\begin{equation}
  \label{eq:Sudakov}
  k_i = (1-y_i^{(1)})\tilde{p}_1 +  (1-y_i^{(2)})\tilde{p}_2 + \tilde{\kappa}_{ti}\,,
\end{equation}
where $\tilde{\kappa}_{ti}$ are space-like four-vectors, orthogonal to
both $\tilde{p}_1$ and $\tilde{p}_2$. In the reference
frame~\eqref{eq:com-frame} each $\tilde{\kappa}_{ti}$ has no time
component, and can be written as
$\tilde{\kappa}_{ti}=(0,\vec{\tilde{k}}_{ti})$, such that
$\tilde{\kappa}_{ti}^2 = - \tilde{k}_{ti}^2$.  Notice that since $k_i$
is massless
\[
\tilde{k}_{ti}^2 = (1-y_i^{(1)})(1-y_i^{(2)})M^2 =\frac{2(\tilde{p}_1 k_i) 2(\tilde{p}_2 k_i)}{2(\tilde{p}_1 \tilde{p}_2)}\,.
\]
In the chosen parametrisation, the emission's (pseudo-)rapidity
$\eta_i$ in this frame is
\begin{equation}
  \label{eq:rapidity}
  \eta_i = \frac{1}{2}\ln \frac{1-y_i^{(1)}}{1-y_i^{(2)}}.
\end{equation}

The observable $V$ is in general a function of all momenta, and we
denote it by $\Vfull(\{\tilde p\},k_1,\dots,k_n)$; without loss of
generality we assume that it vanishes in Born-like kinematic
configurations. The {\it transverse} observables considered in this
paper are those which obey the following general parametrisation for a
single soft emission $k$ collinear to leg $\ell$:
\begin{equation}
\label{eq:v-scaling}
 \Vsc(\{\tilde{p}\},k) \equiv \Vsc(k) = d_\ell \,g_\ell(\phi) \left(\frac{k_t}{M}\right)^a\,,
\end{equation}
where $k_t$ is the transverse momentum with respect to the beam axis,
$g_{\ell}(\phi)$ is a generic function of the angle $\phi$
that $\vec k_t$ forms with a fixed reference
vector $\vec n$ orthogonal to the beam axis, $d_\ell$ is a
normalisation factor, and $a > 0$ due to collinear and infrared
safety. In particular, in this work we focus on the family of
inclusive observables that will be defined in the next
section. Examples of such observables are the transverse momentum of
the colour-singlet system (corresponding to
$d_\ell=g_\ell(\phi)=a=1$)\footnote{Without loss of generality we have
  introduced a dimensionless version of the transverse momentum by
  dividing by the singlet's mass.}, and $\phi^*$~\cite{Banfi:2010cf}
(corresponding to $d_\ell=a=1\,,\,g_\ell(\phi)=|\sin(\phi)|$). In the
latter case, the reference vector $\vec{n}$ is chosen along the
direction of the dilepton system in the rest frame of the $Z$ boson.

The transverse momentum of the parametrisation~\eqref{eq:Sudakov} is
related to the one relative to the beam axis, which enters the
definition of the observable, by recoil effects due to hard-collinear
emissions off the same leg $\ell$. To find the relationship, we
consider the radiation collinear to $\tilde{p}_1$. The momentum of the
initial-state parton before any radiation $p_1$ is related to the latter
as follows
\begin{equation}
p_1 = \tilde{p}_1 + \sum_{j\in 1} k_j,
\end{equation}
where the notation $j\in 1$ indicates all emissions $k_i$ radiated off
leg $1$. The above equation can be recast as
\begin{equation}
p_1 = (1+\sum\limits_{j \in 1}(1-y_j^{(1)}))\tilde{p}_1 +
\sum\limits_{j \in 1}(1-y_j^{(2)})\tilde{p}_2 + \sum_{j\in 1} \tilde{\kappa}_{tj}.
\end{equation}
We can use the above equation to express $\tilde{p}_1$ as a function
of $p_1$. By plugging the resulting equation into
Eq.~\eqref{eq:Sudakov}, we find that the transverse momentum of
emission  $k_i$ with respect to $p_1$ is 
\begin{align}
\vec{k}_{ti} = \vec{\tilde k}_{ti} - \frac{1-y_i^{(1)}}{1+\sum\limits_{j \in 1}(1-y_j^{(1)})}\left(\sum_{j\in 1}\vec{\tilde k}_{tj}\right).
\end{align}
Generalising the above equation for $k_i$ emitted off any leg $\ell=1,2$  we obtain
\begin{align}
\label{eq:kt-rel}
\vec{k}_{ti} = \vec{\tilde k}_{ti} - \frac{1-y_i^{(\ell)}}{1+\sum\limits_{j \in \ell}(1-y_j^{(\ell)})}\left(\sum_{j\in \ell}\vec{\tilde k}_{tj}\right),
\end{align}
where with the notation $j\in \ell$ we refer to partons that are
emitted off the same leg $\tilde{p}_\ell$ as $k_i$. When only one
emission is present, the above relation reduces to
\begin{equation}
\vec{k}_{ti} = \frac{\vec{\tilde k}_{ti}}{2-y_i^{(\ell)}}.
\end{equation}
%
%
In the soft approximation the two quantities coincide as
$y_i^{(\ell)}\simeq 1$. In the present section we work under the
assumption of soft kinematics in order to introduce the notation and
derive the NLL result. The treatment of hard-collinear emissions will
be discussed in detail in Section~\ref{sec:recoil}, where we extend the
results derived here to the general case of initial-state radiation.
\\

The central quantity under study is the resummed cumulative cross
section for $V$ smaller than some value $v$, $\Sigma(v)$, defined as
\begin{equation}
  \label{eq:Sigma}
\Sigma(v) = \int_0^v dv' \frac{d \sigma(v')}{d v'}.
\end{equation}

In the infrared and collinear (IRC) limit, $\Sigma(v)$ receives
contributions from both virtual corrections and soft and/or collinear
real emissions. The IRC divergences of the form factor exponentiate at
all orders (see, for instance, refs.~\cite{Dixon:2008gr,Magnea:1990zb}
and references therein), and we denote them by $ {\cal V}(\Phi_B)$ in
the following discussion, where $\Phi_B$ is the phase space of the
underlying Born. Therefore we can recast Eq.~\eqref{eq:Sigma} as
follows
\begin{equation}
  \label{eq:Sigma-2}
  \Sigma(v) = \int d\Phi_B {\cal V}(\Phi_B) \sum_{n=0}^{\infty}
  \int\prod_{i=1}^n [dk_i]
  |M(\tilde{p}_1,\tilde{p}_2,k_1,\dots ,k_n)|^2\,\Theta\left(v-V(\{\tilde p\},k_1,\dots,k_n)\right)\,,
\end{equation}
where $M$ is the matrix element for $n$ real emissions (the case with
$n=0$ reduces to the Born matrix element $M_B$), and $[dk_i]$ denotes
the phase space for the emission $k_i$. The $\Theta$ function
represents the measurement function for the observable under
consideration. Finally, to keep the notation concise, we have defined
$d\Phi_B\equiv d x_1 d x_2 d\Phi_{n}
(2\pi)^d\delta(\tilde{p}^\mu_1+\tilde{p}^\mu_2-p^\mu_B)$,
where $d\Phi_{n}$ is the $n$-body phase space of the singlet system,
and we have absorbed the partonic flux factor
$1/(4 \,\tilde{p}_1\!\cdot\!\tilde{p}_2)$ into the squared amplitude
$|M|^2$ (and analogously in $|M_B|^2$ below).

The renormalised squared amplitude for $n$ real emissions ($p p\to B +
n$ gluons) can be conveniently decomposed as~\footnote{The
  decomposition above can be extended to the case in which some of the
  $n$ emissions are quarks by properly changing the multiplicity
  factors in front of each term.}
\begin{align}
\label{eq:nPC}
&|M(\tilde{p}_1,\tilde{p}_2,k_1,\dots ,k_n)|^2 = |M_B(\tilde{p}_1,\tilde{p}_2)|^2\left\{\left(\frac{1}{n!}\prod_{{\substack{i=1\\\phantom{x}}}}^{n} \left|M(k_i)\right|^2\right) +\right.\notag\\
&\left.  \left[\sum_{a > b}\frac{1}{(n-2)!}\left(\prod_{\substack{i=1\\ i\neq a,b}}^{n} \left|M(k_i)\right|^2 \right)\left|\tilde{M}(k_a, k_b)\right|^2+\right.\right. \notag\\
&\left.\left.\sum_{a > b}\sum_{\substack{ c > d\\ c,d\neq a,b}}\frac{1}{(n-4)!2!}\left(\prod_{\substack{i=1\\ i\neq a,b,c,d}}^{n} \left|M(k_i)\right|^2 \right)\left|\tilde{M}(k_a, k_b)\right|^2 \left|\tilde{M}(k_c, k_d)\right|^2\right. + \dots \right]\notag\\
&\left. + \left[\sum_{a > b > c}\frac{1}{(n-3)!}\left(\prod_{\substack{i=1\\ i\neq a,b,c}}^{n} \left|M(k_i)\right|^2 \right)\left|\tilde{M}(k_a, k_b,k_c)\right|^2 +\dots\right]+\dots\right\},
\end{align}
where we have introduced the $n$-particle correlated matrix elements squared $|\tilde{M}(k_a, \dots,k_n)|^2$, which are defined recursively as follows
\begin{align}
|\tilde{M}(k_a)|^2 =&\frac{|M(\tilde{p}_1,\tilde{p}_2,k_a)|^2}{|M_B(\tilde{p}_1,\tilde{p}_2)|^2}=~|M(k_a)|^2,\notag\\
|\tilde{M}(k_a,k_b)|^2 =&~\frac{|M(\tilde{p}_1,\tilde{p}_2,k_a,k_b)|^2}{|M_B(\tilde{p}_1,\tilde{p}_2)|^2}-\frac1{2!}|M(k_a)|^2|M(k_b)|^2,\notag\\
|\tilde{M}(k_a,k_b,k_c)|^2 =&~\frac{|M(\tilde{p}_1,\tilde{p}_2,k_a,k_b,k_c)|^2}{|M_B(\tilde{p}_1,\tilde{p}_2)|^2}-\frac1{3!}|M(k_a)|^2|M(k_b)|^2|M(k_c)|^2\notag\\
&~-|\tilde{M}(k_a,k_b)|^2|M(k_c)|^2-|\tilde{M}(k_a,k_c)|^2|M(k_b)|^2-|\tilde{M}(k_b,k_c)|^2|M(k_a)|^2,
\end{align}
and so on. These represent the contributions to the $n$-particle
squared matrix element that vanish in strongly-ordered kinematic
configurations, that can not be factorised in terms of
lower-multiplicity squared amplitudes.
Each of the correlated squared amplitudes admits a perturbative
expansion
\begin{equation}
\label{eq:nPC-def}
|\tilde{M}(k_a, \dots,k_n)|^2~\equiv~ \sum_{j=0}^{\infty}\left(\frac{\alpha_s(\mu)}{2\pi}\right)^{n+j}n\mbox{PC}^{(j)}(k_a, \dots,k_n),
\end{equation}
where $\mu$ is a common renormalisation scale, and $\alpha_s$ is the
strong coupling constant in the $\overline{\rm MS }$ scheme. The
notation $n\mbox{PC}$ in Eq.~\eqref{eq:nPC} stands for ``$n$-particle
correlated'' and it will be used throughout the article.

The rIRC safety of the observables considered here guarantees a hierarchy
between the different blocks in the decomposition~\eqref{eq:nPC}, in
the sense that, generally, correlated blocks with $n$ particles start
contributing at one logarithmic order higher than correlated blocks
with $n-1$ particles~\cite{Banfi:2004yd,Banfi:2014sua}.
In the present article, we focus on the family of inclusive
observables $V$ for which
\begin{equation}
\label{eq:inclusive}
V(\{\tilde{p}\},k_1,\dots, k_n) = V(\{\tilde{p}\},k_1+\dots +k_n)\,.
\end{equation}
In this case, we can integrate the {\it n}PC blocks for $n>1$
inclusively prior to evaluating the observable. Hence, starting from
Eq.~\eqref{eq:nPC} for the pure gluonic case, we can replace it with
the following squared amplitude
\begin{align}
\label{eq:nPC-inclusive}
\sum_{n=0}^{\infty}|M(\tilde{p}_1,&\tilde{p}_2,k_1,\dots ,k_n)|^2 \longrightarrow |M_{B}(\tilde{p}_1,\tilde{p}_2)|^2\notag\\ & \times
\sum_{n=0}^{\infty}\frac{1}{n!}\left\{\prod_{i=1}^{n} \left(|M(k_i)|^2+\int [d k_a][d k_b]|\tilde{M}(k_a,k_b)|^2\delta^{(2)}(\vec{k}_{ta}+\vec{k}_{tb}-\vec{k}_{ti})\delta(Y_{ab}-Y_i)\right.\right.\notag\\ &\left.\left. +
     \int [d k_a][d k_b][d k_c]|\tilde{M}(k_a,k_b,k_c)|^2\delta^{(2)}(\vec{k}_{ta}+\vec{k}_{tb}+\vec{k}_{tc}-\vec{k}_{ti})\delta(Y_{abc}-Y_i)
     + \dots\right.\bigg)\right.\Bigg\}\notag\\
& \equiv |M_{B}(\tilde{p}_1,\tilde{p}_2)|^2\sum_{n=0}^{\infty}\frac{1}{n!}\prod_{i=1}^{n} |M(k_i)|_{\rm inc}^2,
\end{align}
where $Y_{abc...}$ is the rapidity of the $k_a+k_b+k_c+\dots$ system
in the centre-of-mass frame of the collision. We refer to this
treatment of the squared amplitude as to the {\it inclusive
  approximation}.\footnote{For non-inclusive observables, namely the ones that
do not fulfil Eq.~\eqref{eq:inclusive}, this reorganisation is not
correct starting at NNLL. Therefore in that case one must correct for
the non-inclusive nature of the observables. The full set of NNLL
corrections for a generic global, rIRC safe observable is defined in
refs.~\cite{Banfi:2014sua,Banfi:2016zlc}. In the rest of this article
we refer to observables of the type~\eqref{eq:inclusive}.}
With the above notation, we can rewrite Eq.~\eqref{eq:Sigma-2} as
\begin{equation}
  \label{eq:Sigma-start-1}
  \Sigma(v) = \int d\Phi_B |M_{B}(\tilde{p}_1,\tilde{p}_2)|^2{\cal V}(\Phi_B) \sum_{n=0}^{\infty}
  \frac{1}{n!}  \int\prod_{i=1}^n [dk_i]
  |M(k_i)|_{\rm inc}^2\,\Theta\left(v-V(\{\tilde p\},k_1,\dots,k_n)\right)\,,
\end{equation}
where $|M(k_i)|_{\rm inc}^2$ is defined in
Eq.~\eqref{eq:nPC-inclusive}.\\

Once the logarithmic counting for the squared amplitude has been set
up, as a next step we need to discuss the cancellation of the
exponentiated divergences of virtual origin against the real ones. At
all perturbative orders at a given logarithmic accuracy, we need to
single out the IRC singularities of the real matrix elements, which
can again be achieved by
exploiting~\cite{Banfi:2003je,Banfi:2004yd,Banfi:2014sua} the rIRC
safety of the observable $V(\{\tilde p\},k_1,\dots,k_n)$ that we are
computing.

We then order the inclusive blocks described by $|M(k_i)|_{\rm inc}^2$
according to their contribution to the observable $\Vsc(k_i)$,
i.e. $\Vsc(k_1) > \Vsc(k_2) > \dots > \Vsc(k_n)$. We consider
configurations where the radiation corresponding to the first
(hardest) block $|M(k_1)|_{\rm inc}^2$ has occurred, where we use the
fact that the contribution with $n=0$ in Eq.~\eqref{eq:Sigma-start-1}
(which does not have any real emissions) vanishes since it is
infinitely suppressed by the pure virtual corrections
${\cal V}(\Phi_B)$ The rIRC safety of the observable allows us to
introduce a resolution parameter $\epsilon \ll 1$ independent of the
observable such that all inclusive blocks with
$\Vsc(k_i) < \epsilon \Vsc(k_1)$ can be neglected in the computation
of the observable up to power-suppressed corrections
${\cal O}(\epsilon^p\Vsc(k_1))$, that eventually will vanish once we
take the limit $\epsilon\to 0$.  Therefore, we classify inclusive
blocks $k$ as {\it resolved} if $\Vsc(k)> \epsilon \Vsc(k_1)$, and as
{\it unresolved} if $\Vsc(k)< \epsilon \Vsc(k_1)$. This definition is
collinear safe at all perturbative orders.  With this separation
Eq.~\eqref{eq:Sigma-start-1} becomes
\begin{align}
\label{eq:Sigma-start-2}
\Sigma(v) &= \int d\Phi_B |M_{B}(\tilde{p}_1,\tilde{p}_2)|^2{\cal V}(\Phi_B)\notag\\
&\times\int [dk_1]
|M(k_1)|_{\rm inc}^2 \left(\sum_{l=0}^{\infty}
\frac{1}{l!}  \int\prod_{j=2}^{l+1} [dk_j]
|M(k_j)|_{\rm inc}^2\,\Theta(\epsilon \Vsc(k_1)- \Vsc(k_j))\right)\notag\\
&\times\left(\sum_{m=0}^{\infty}
\frac{1}{m!}  \int\prod_{i=2}^{m+1} [dk_i]
|M(k_i)|_{\rm inc}^2\,\Theta(\Vsc(k_i)-\epsilon \Vsc(k_1))\Theta\left(v-V(\{\tilde p\},k_1,\dots,k_{m+1})\right)\right)\,.
\end{align}
The phase space of the unresolved real ensemble is now solely
constrained by the upper resolution scale, since it does not
contribute to the evaluation of the observable. As a consequence, it
can be exponentiated directly in Eq.~\eqref{eq:Sigma-start-2} and
employed to cancel the divergences of the virtual corrections
${\cal V}(\Phi_B)$.
\\
 
\noindent We can now proceed with an explicit evaluation of
Eq.~\eqref{eq:Sigma-start-2} at NLL order. As we mentioned earlier, at
different logarithmic orders the cross section will receive
contributions from different classes of correlated blocks. This, for
instance, means that double-logarithmic terms of the form $\alpha_s^n
\ln^{2n}(1/v)$ entirely arise from 1PC$^{(0)}$ blocks, in particular
from their soft-collinear part. \\ If one wants to control all the
leading-logarithmic terms of order $\alpha_s^n \ln^{n+1}(1/v)$ in
$\ln\left(\Sigma(v)\right)$ (Eq.~\eqref{eq:cumulant-initial}) then the
leading (soft-collinear) term of the 1PC$^{(1)}$ and 2PC$^{(0)}$
blocks must be included as well. In particular, within the inclusive
approximation defined in Eq.~\eqref{eq:nPC-inclusive} we find that
\begin{align}
\label{eq:2PC-inclusive}
|M(k)|_{\rm inc}^2&\simeq|M(k)|^2 + \int [dk_a][dk_b]|\tilde{M}(k_a,k_b)|^2\delta^{(2)}(\vec{k}_{ta}+\vec{k}_{tb}-\vec{k}_{t})\delta(Y_{ab}-Y)\notag\\
&= \frac{\alpha_s(\mu)}{2\pi}1{\rm PC}^{(0)}(k)\left(1+\alpha_s(\mu)\left(\beta_0\ln\frac{k_{t}^2}{\mu^2} + \frac{K}{2\pi}\right)+\dots\right),
\end{align}
where $\beta_0$ is the leading term of the QCD beta function (see
Appendix~\ref{app:radiator}). Moreover, the QCD coupling is
renormalised in the $\overline{\rm MS}$ scheme. The contribution of
the one-loop cusp anomalous dimension $K$, defined as
\begin{equation}
\label{eq:K}
K = \left(\frac{67}{18}-\frac{\pi^2}{6}\right)C_A - \frac{5}{9} n_f\,,
\end{equation}
enters at NLL order, and it will be considered later in this section.
Up to, and including, the NLL term proportional to $K$ in
Eq.~\eqref{eq:2PC-inclusive}, one can integrate inclusively over the
invariant mass of the 2PC$^{(0)}$ block, while keeping the bounds on
the rapidity $Y$ as computed from the massless kinematics. This
approximation neglects terms which are at most NNLL, and are denoted
by the ellipsis in the second line of Eq.~\eqref{eq:2PC-inclusive}.

We notice that the leading soft-collinear terms proportional to
$\beta_0$ in Eq.~\eqref{eq:2PC-inclusive} can be entirely encoded in
the running of the coupling of the single-emission squared amplitude
$1{\rm PC}^{(0)}(k)$ through a proper choice of the scale $\mu$ at which the
latter is evaluated. It is indeed easy to see from
Eq.~\eqref{eq:2PC-inclusive} that this is achieved by setting $\mu$ to
the $k_t$ (equal to $\tilde{k}_t$ for soft radiation) of each emission
$k$ in the
parametrisation~\eqref{eq:Sudakov}~\cite{Catani:1992ua,Catani:1989ne}. The
inclusive matrix element squared and phase space controlling all
$\alpha_s^n \ln^{n+1}(1/v)$ terms are thus
\begin{equation}
  \label{eq:sc-amplitude}
[dk] | M(k)|_{\rm inc}^2\simeq[dk]M^2_{\rm sc}(k)= \sum_{\ell=1,2} 2 C_\ell \frac{\alpha_s(k_{t})}{\pi}\frac{dk_{t}}{k_{t}} 
 \frac{d z^{(\ell)}}{1-z^{(\ell)}}\, \Theta\left((1-z^{(\ell)}) - k_{t}/M\right) \Theta(z^{(\ell)})
\frac{d\phi}{2\pi}\,,
\end{equation}
where we use $M_{\rm sc}(k)$ to denote the amplitude in the soft
approximation. We denoted by $C_\ell$ the Casimir relative to the
emitting leg ($C_\ell=C_F$ for quarks, and $C_\ell=C_A$ for gluons).

For initial-state radiation, $1-z^{(\ell)}$ is the fraction of the
incoming momentum (entering the emission vertex) that is carried by
the emitted parton. This will in general differ from the $y^{(\ell)}$
fractions of the Sudakov parametrisation~\eqref{eq:Sudakov} when some
emissions are not soft. In particular, while $(1-z^{(\ell)})\leq 1$,
this is not true in general for the $(1-y^{(\ell)})$ appearing in our
initial parametrisation. However, in the soft limit, the energy of the
emission is much smaller than the singlet's mass $M$, which restricts
$y_i^{(\ell)}$ to positive values in this limit. \\ For a single
emission, the two variables are related by
\begin{equation}
\label{eq:z-to-y}
1-y^{(\ell)} = \frac{1-z^{(\ell)}}{z^{(\ell)}},
\end{equation}
from which is clear that in the soft limit $z^{(\ell)}\simeq 1$ one
has $z^{(\ell)}\simeq y^{(\ell)}$. The upper bound for $z^{(\ell)}$ in
the single-emission case can be worked out by imposing that
$y^{(\ell)} < 1-\tilde{k}_t/M$, and subsequently relating
$\tilde{k}_t$ to $k_t$ relative to the beam axis. This yields
\begin{equation}
\label{eq:z-limit}
z^{(\ell)} < 1-k_t/M + {\cal O}(k_t^2).
\end{equation}

To extend the above discussion to all NLL terms of order $\alpha_s^n
\ln^n(1/v)$ in the {\it logarithm} of $\Sigma(v)$, we must include the
less singular part of the 1PC$^{(1)}$ and 2PC$^{(0)}$ blocks in the
soft limit, that is the term proportional to $K$ in
Eq.~\eqref{eq:2PC-inclusive} that was previously ignored. This simply
amounts to replacing the inclusive (soft) matrix element in the
r.h.s. of ~\eqref{eq:sc-amplitude} with
\begin{equation}
  \label{eq:sc-amplitude-CMW}
[dk]M^2_{\rm CMW}(k)= \sum_{\ell=1,2} 2 C_\ell
\frac{\alpha_s(k_{t})}{\pi}\left (1 +  \frac{\alpha_s(k_t)}{2\pi}
  K\right)\frac{dk_{t}}{k_{t}} 
\frac{d z^{(\ell)}}{1-z^{(\ell)}}\, \Theta\left((1-z^{(\ell)}) - k_{t}/M\right) \Theta(z^{(\ell)})
 \frac{d\phi}{2\pi}\,.
\end{equation}
The above operation is also known as the Catani-Marchesini-Webber
(CMW) scheme~\cite{Catani:1990rr} for the running
coupling.\footnote{Although in the present article we are considering
  only inclusive observables, it can be
  shown~\cite{Banfi:2003je,Banfi:2004yd,Banfi:2014sua} that for all
  rIRC safe observables (also non-inclusive ones) the inclusive
  approximation is accurate at NLL order.}\\

At this logarithmic order the cross section also receives
contributions from the hard-collinear part of the 1PC$^{(0)}$ block,
that we ignored so far. Thus, one has to modify
Eq.~\eqref{eq:sc-amplitude-CMW} as
\begin{align}
  \label{eq:single-emsn}
  [dk] | M(k)|_{\rm inc}^2&= [dk]M^2_{\rm CMW}(k)
     \notag \\ &+
    \sum_{\ell=1,2}
    \frac{dk_t^2}{k_t^2}\frac{dz^{(\ell)}}{1-z^{(\ell)}}\frac{d\phi}{2\pi}\frac{\alpha_s(k_t)}{2\pi}\left(
    (1-z^{(\ell)})P^{(0)}(z^{(\ell)})-\lim_{z^{(\ell)}\to 1}\left[(1-z^{(\ell)})P^{(0)}(z^{(\ell)})\right]
    \right)\,,
\end{align}
where $P^{(0)}(z^{(\ell)})$ is the leading-order unregularised
splitting function, reported in Appendix~\ref{app:radiator}.\footnote
{For emissions off gluonic legs, $P^{(0)}$ receives contributions from
  both $P^{(0)}_{gg}$ and $P^{(0)}_{gq}$, as it will be discussed in
  Sec.~\ref{sec:master-initial-state}. In this case, we implicitly exploit the
  symmetry of $P^{(0)}_{gg}$ under $z\leftrightarrow 1-z$ to recast it
  such that it has only a $z\to 1$ singularity.}  At NLL order, the
above hard-collinear contribution can be treated by neglecting the
effect of recoil both in the phase-space boundaries of other emissions
and in the observable, both of which enter at NNLL order. Therefore,
also for this contribution we can use the soft kinematics derived in
the first part of this section. Moreover, in colour-singlet
production, we can use the azimuthally averaged splitting functions
(see Appendix~\ref{app:radiator}) up to NNLL accuracy. At N$^3$LL,
corrections from azimuthal correlations arise~\cite{Catani:2010pd},
and they will be introduced in Section~\ref{sec:master-initial-state}.
\\

We insert Eq.~\eqref{eq:single-emsn} back into
Eq.~\eqref{eq:Sigma-start-2}. At NLL accuracy, we can neglect the
constant terms of the virtual corrections. The remaining singular
structure of the virtual corrections only depends upon the invariant
mass of the singlet $M^2$
\begin{equation}
{\cal V}(\Phi_B) \simeq {\cal V}(M^2) = \exp\left\{-\int [dk]|M(k)|_{\rm inc}^2\right\} \,\,\,{\rm at \,\,\,NLL}.
\end{equation}
The combination of unresolved real and virtual contributions is thus
finite and gives rise to a Sudakov suppression factor
\begin{align}
{\cal V}(M^2)&\exp\left\{\int [dk]|M(k)|_{\rm inc}^2\,\Theta(\epsilon
  \Vsc(k_1)- \Vsc(k))\right\}\notag\\
\simeq &\exp\left\{-\int [dk]|M(k)|_{\rm
    inc}^2\,\Theta(\Vsc(k) -\epsilon
  \Vsc(k_1))\right\} = e^{-R(\epsilon \Vsc(k_1))},
\end{align}
where $R$ is the radiator which at this order
reads~\cite{Banfi:2004yd,Banfi:2014sua}
\begin{equation}
  \label{eq:radiator}
  \begin{split}
    R(v) \simeq R_{\rm NLL}(v) & \equiv \int [dk] M_{\rm
      CMW}^2(k)\Theta\left(\ln
      \left(\frac{k_t}{M}\right)^{a} -\ln
      v\right) +
    \int [dk] M_{\rm CMW}^2(k) \ln\bar d_{\ell } \,\delta\left(\ln \left(\frac{k_t}{M}\right)^{a}-\ln v\right) \\
    & + \sum_{\ell=1,2}C_\ell B_\ell\int
    \frac{dk_t^2}{k_t^2}\frac{\alpha_s(k_t)}{2\pi}\Theta\left(\ln\left(\frac{k_t}{M}\right)^{a}-\ln v
    \right)\,,
  \end{split}
\end{equation}
where
\begin{equation}
\label{eq:lndbar}
\ln\bar d_\ell = \int_0^{2\pi} \frac{d\phi}{2\pi}\ln d_\ell g_\ell(\phi)\,,
\end{equation}
and
\begin{equation}
  C_\ell B_\ell = \int_0^1 \frac{d z^{(\ell)}}{1-z^{(\ell)}}\left((1-z^{(\ell)}) P^{(0)}(z^{(\ell)}) - \lim_{z^{(\ell)}\to 1}\left[(1-z^{(\ell)}) P^{(0)} (z^{(\ell)})\right]\right)\,.
\end{equation}

The next and final step is to treat the resolved real blocks $k_i$ for
which $V(k_i)>\epsilon \Vsc(k_1)$. It is therefore necessary to work
out the kinematics and phase space in the presence of additional
radiation, which modifies the relations~\eqref{eq:z-to-y}
and~\eqref{eq:z-limit} obtained in the single-emission case.  For this
we use the fact that the radiation is ordered in $\Vsc(k_i)$. For a
given inclusive block of total momentum $k_i$, one then
has\footnote{See also discussion in the appendix E of
  ref.~\cite{Banfi:2004yd}.}
\begin{equation}
\label{eq:z-to-y-all}
1-y_i^{(\ell)}=\frac{1-z_i^{(\ell)}}{z_1^{(\ell)}z_2^{(\ell)}\dots z_i^{(\ell)}},
\end{equation}
where emissions $k_1,k_2,\dots,k_{i-1}$ have been radiated off the
same hard leg before $k_i$. In general, this implies that the phase
space available for each emissions is changed by the previous resolved
radiation. At the NLL order considered in this section, as already
stressed, the real-radiation kinematics can be approximated with its
soft limit~\cite{Banfi:2004yd,Banfi:2014sua}. This allows us to
approximate $y_i^{(\ell)}\simeq z^{(\ell)}_i$ and
$k_t\simeq {\tilde k}_t$ for all real emissions and therefore the
phase space of each emission becomes in fact independent of the
remaining radiation in the event.

The squared matrix element~\eqref{eq:single-emsn} and phase space for a
resolved real emission can be parametrised by introducing the
functions
\begin{equation}
  \label{eq:R'1,2}
  \begin{split}
    R'_1\left(\frac{v}{d_1 g_1(\bar \phi)}\right)&= \int [dk]
    |M(k)|^2_{\rm inc}
    \,(2\pi) \delta(\phi-\bar \phi)\, v\delta\left(v-\Vsc(k)\right)\Theta(y^{(2)}
-y^{(1)})  \,, \\
    R'_2\left(\frac{v}{d_2 g_2(\bar \phi)}\right)& = \int [dk]
    |M(k)|^2_{\rm inc}\,(2\pi) \delta(\phi-\bar \phi)\,
    v\delta\left(v-\Vsc(k)\right)\Theta(y^{(1)}
-y^{(2)})\,,
  \end{split}
\end{equation}

and
\begin{equation}
  \label{eq:R'}
  R'(v,\phi) =  R'_1\left(\frac{v}{d_1
      g_1(\phi)}\right)+R'_2\left(\frac{v}{d_2 g_2(\phi)}\right)\,.
\end{equation}
From the generic form of the rIRC safe observable
$\Vsc(k)$~\eqref{eq:v-scaling}, it is easy to verify that the $R'$
functions only depend upon the ratio $v/(d_\ell g_\ell(\bar \phi))$ up
to regular terms, which are
neglected~\cite{Banfi:2004yd,Banfi:2014sua}. Indeed, the only
non-trivial integration in Eqs.~\eqref{eq:R'1,2} is the one over the
rapidity of $k$, which can be performed inclusively since the
observable $V(k)$ does not depend on it (see
Eq.~\eqref{eq:v-scaling}). Then the final integral only depends on the
ratio of the two remaining scales, i.e. the invariant mass of the
singlet $M$, and its transverse momentum that is set to $(v/(d_\ell
g_\ell(\bar \phi)))^{1/a} M$ by the constraint
$\delta\left(v-\Vsc(k)\right)$.
Upon inclusive integration over the rapidity of momentum $k$, by using
Eq.~\eqref{eq:single-emsn}, we can
parametrise the inclusive squared amplitude and its phase space as
\begin{equation}
  \label{eq:single-emsn-rp}
  \begin{split}
    [dk_i] |M(k_i)|^2_{\rm inc} & = \frac{dv_i}{v_i}
    \frac{d\phi_i}{2\pi}\sum_{\ell_i=1,2}
    R'_{\ell_i}\left(\frac{v_i}{d_{\ell_i} g_{\ell_i}(\phi_i)}\right)
    = \frac{d\zeta_i}{\zeta_i} \frac{d\phi_i}{2\pi} \sum_{\ell_i=1,2}
    R'_{\ell_i}\left(\frac{\zeta_i v_1}{d_{\ell_i}
        g_{\ell_i}(\phi_i)}\right) \,,
  \end{split}
\end{equation}
where we defined $v_i=\Vsc(k_i)$ and $\zeta_i = \Vsc(k_i)/\Vsc(k_1)$.

With the above considerations, Eq.~\eqref{eq:Sigma-start-2} finally becomes
\begin{align}
\label{eq:master-NLL}
  \Sigma(v) &= \sigma^{(0)}\int \frac{d v_1}{v_1} \int_0^{2\pi} \frac{d\phi_1}{2\pi}
              e^{-R(\epsilon v_1)}\sum_{\ell_1=1,2} R_{\ell_1}'\left(\frac{v_1}{d_{\ell_1}
              g_{\ell_1}(\phi_1)}\right) \times \notag\\ &\times \sum_{n=0}^{\infty}\frac{1}{n!}
                                                           \prod_{i=2}^{n+1}
                                                           \int_{\epsilon}^{1}\frac{d\zeta_i}{\zeta_i}\int_0^{2\pi}
                                                           \frac{d\phi_i}{2\pi} \sum_{\ell_i=1,2}
                                                           R_{\ell_i}'\left(\frac{\zeta_i v_1}{d_{\ell_i}
                                                           g_{\ell_i}(\phi_i)}\right)\, \Theta\left(v-V(\{\tilde{p}\},k_1,\dots, k_{n+1})\right)\,,
\end{align}
where we introduced the total Born cross section
\begin{equation}
\sigma^{(0)}=\int d\Phi_B |M_{B}(\tilde{p}_1,\tilde{p}_2)|^2.
\end{equation}

Eq.~\eqref{eq:master-NLL} resembles equation (2.34) of
ref.~\cite{Banfi:2004yd} which after a number of approximations leads
to the general NLL formula of the {\tt CAESAR} method for global rIRC
observables in processes with two hard legs.  We remind the reader
that additional corrections coming from the parton luminosities start
at NLL order, and they will be discussed in Section~\ref{sec:recoil}.

Eq.~\eqref{eq:master-NLL} can be directly evaluated using Monte-Carlo
(MC) techniques since it is finite in four dimensions. However, as it
is formulated now it contains effects that are logarithmically
subleading with respect to the formal NLL accuracy we are considering
in this section. For observables that vanish only in the Sudakov
limit, these subleading effects can be systematically disposed of by
means of a few approximations, as described in
ref.~\cite{Banfi:2004yd}. We now briefly review such approximations on
Eq.~\eqref{eq:master-NLL}, and show that in the case of observables
that vanish away from the Sudakov region they lead to a divergent
result, hence they cannot be trivially performed.

In order to neglect subleading corrections from
Eq.~\eqref{eq:master-NLL}, we need to consistently treat the resolved
squared amplitude and the corresponding Sudakov radiator. In
particular, with NLL accuracy, ref.~\cite{Banfi:2004yd} suggests to
perform the following Taylor expansions in Eq.~\eqref{eq:master-NLL}
\begin{align}
R(\epsilon v_1) =&  R(v) + \frac{dR(v)}{d\ln(1/v)}\ln\frac{v}{\epsilon v_1} + {\cal O}\left(\ln^2\frac{v}{\epsilon v_1}\right),\notag\\
R'_{\ell_i}\left(\frac{v_i}{d_{\ell_i}
              g_{\ell_i}(\phi_i)}\right) =&  R'_{\ell_i}(v) + {\cal O}\left(\ln\frac{v d_{\ell_i}
              g_{\ell_i}(\phi_i) }{v_i}\right).
\end{align}
This is motivated by the fact that at NLL the resolved real emissions
are such that $v_i\sim v_1\sim v$, and hence the terms neglected in the above
expansions are at most NNLL. Only by expanding consistently (i.e. to
the same logarithmic order) the $\epsilon$ dependence in the Sudakov
and in the resolved real emissions we are sure that the result is
completely $\epsilon$-independent.

We observe that, since we expanded out the $\phi_i$ dependence in
$R'$, we have $dR(v)/d\ln(1/v)=\sum_{\ell}R'_\ell(v)$ and
Eq.~\eqref{eq:master-NLL} becomes
\begin{align}
\label{eq:master-NLL-CAESAR}
  \Sigma(v) &\simeq \sigma^{(0)}\int \frac{d v_1}{v_1} \int_0^{2\pi} \frac{d\phi_1}{2\pi}
              e^{-R(v)}e^{-\sum_\ell R_\ell'(v)\ln\frac{v}{\epsilon v_1}}\sum_{\ell_1=1,2} R_{\ell_1}'\left(v\right) \times \notag\\ &\times \sum_{n=0}^{\infty}\frac{1}{n!}
                                                           \prod_{i=2}^{n+1}
                                                           \int_{\epsilon}^{1}\frac{d\zeta_i}{\zeta_i}\int_0^{2\pi}
                                                           \frac{d\phi_i}{2\pi} \sum_{\ell_i=1,2}
                                                           R_{\ell_i}'\left(v\right)\, \Theta\left(v-V(\{\tilde{p}\},k_1,\dots, k_{n+1})\right)\,.
\end{align}
At this stage, the integration over $v_1$ can be performed
analytically, and Eq.~\eqref{eq:master-NLL-CAESAR} reproduces exactly
the known {\tt CAESAR} formula.\footnote{Some extra simplifications
  can be made at NLL: in the resolved real squared matrix elements
  $R'_\ell$ one can keep only the term proportional to $M^2_{\rm sc}$
  as remaining terms are subleading. In order to guarantee the
  cancellation of the divergences in the $\epsilon$ regulator, the
  same approximation has to be made in the term $\sum_\ell
  R_\ell'(v)\ln\frac{v}{\epsilon v_1}$ coming from the expansion of
  the Sudakov radiator. Finally, the observable can be treated in its
  soft-collinear approximation given that, at NLL, the real emissions
  constitute an ensemble of soft-collinear gluons.}

However, in order to perform the latter expansions about the
observable's value $v$, one has to make sure that the ratio $v_i/v$
remains of order one in the real-emission phase space. rIRC safety
ensures that emissions with $v_i \ll v$ do not contribute to the
observable, and are fully exponentiated and accounted for in the
Sudakov radiator. Therefore, the condition $v_i/v\sim 1$ is fulfilled
only if configurations in which $v_i \gg v$ never occur.

While the latter condition holds true for most rIRC observables, it is
clearly violated for observables that vanish away from the Sudakov
limit. An example is given by the transverse momentum of a colour
singlet, which can vanish even in the presence of several emissions
with a finite (non-zero) transverse momentum. In that case, as shown
in ref.~\cite{Monni:2016ktx}, Eq.~\eqref{eq:master-NLL-CAESAR} has a
divergence at $\sum_\ell R'_\ell(v) \simeq 2$. For a different
observable vanishing away from the Sudakov limit, the divergence
will occur at a different, non-zero value of $v$.

For such observables, Eq.~\eqref{eq:master-NLL} cannot be expanded
around $v$. As we will discuss in detail in
Section~\ref{sec:mom-space}, we suggest to perform the following
alternative expansion about the observable's value of the hardest
block $v_1$
\begin{align}
\label{eq:expansion-right}
R(\epsilon v_1) =&  R(v_1) + \frac{dR(v_1)}{d\ln(1/v_1)}\ln\frac{1}{\epsilon} + {\cal O}\left(\ln^2\frac{1}{\epsilon}\right),\notag\\
R'_{\ell_i}\left(\frac{v_i}{d_{\ell_i}
              g_{\ell_i}(\phi_i)}\right) =&  R'_{\ell_i}(v_1) + {\cal O}\left(\ln\frac{v_1 d_{\ell_i}
              g_{\ell_i}(\phi_i) }{v_i}\right).
\end{align}

In this way, the rIRC safety of the observable guarantees that
$v_i\sim v_1$ ($\zeta_i\sim 1$) and therefore the terms neglected in
Eqs.~\eqref{eq:expansion-right} are at most NNLL. However, a class of
higher-order terms still remains in Eq.~\eqref{eq:expansion-right}
through the dependence of the considered terms on $v_1$. These
higher-order terms cannot be disposed of entirely, as they regularise
the divergence discussed above.  Therefore, while the resulting
equation is finite and accurate at NLL order also for rIRC-safe observables that
vanish away from the Sudakov limit, subleading corrections beyond NLL
cannot be entirely removed.

The above approximations make the evaluation of
Eq.~\eqref{eq:master-NLL} considerably simpler than its original form,
as it will be shown in Section~\ref{sec:n3ll}. Its implementation can
be carried out efficiently with MC methods as described in detail in
Section~\ref{sec:event-generation}.

\subsection{Choice of the resolution and ordering variable}
\label{sec:resolution}
The derivation that we carried out for the resummation formalism
relies to a large extent on the introduction of a resolution variable
that separates resolved real blocks from unresolved ones as discussed
in the previous section. This resolution variable acts on the total
momentum of each of the correlated blocks.

One has some freedom in choosing the resolution variable. In
principle, the only necessary property for a good resolution variable
is that it must guarantee, at all orders, the cancellation of the IRC
divergences of the exponentiated virtual corrections, and hence has to
be rIRC safe. A particular choice is motivated by convenience in
formulating the calculation. For instance, choosing a variable that
shares the same leading logarithms with the resummed observable allows
for a much easier implementation of the all-order result, as it will
be discussed in Section~\ref{sec:n3ll}. A natural choice, which
fulfils the above requirements, is the value of observable in its
soft-collinear approximation, as discussed in
refs.~\cite{Banfi:2001bz,Banfi:2004yd,Banfi:2014sua,Banfi:2016zlc}.

However, we note that for the {\it whole} class of transverse
observables (that scale like Eq.~\eqref{eq:v-scaling} for a single
emission), a more convenient choice for the resolution variable is
$V(k)=(k_t/M)^a$, $k$ being the sum of the four-momenta in each
correlated block. While this exactly coincides with the above
prescription for observables with $d_\ell=g_\ell(\phi)=1$, it is a
legitimate choice also for observables with $d_\ell\neq 1$,
$g_\ell(\phi)\neq 1$ since the dependence on $d_\ell g_\ell(\phi)$
first enters at NLL order, hence the leading logarithms of the
resolution variable are the same as for the resummed observable.

The advantage of the latter choice, besides the simplifications in the
implementation to be discussed in Section~\ref{sec:n3ll}, is that it
leads to a universal Sudakov radiator for all observables with the
same $a$ in the parametrisation~\eqref{eq:v-scaling}, while the
resolved real radiation will correctly encode the full observable
dependence through the measurement function
$\Theta\left(v-V(\{\tilde{p}\},k_1,\dots, k_{n+1})\right)$. In the
present article, we adopt this choice, and we present explicitly the
case for $a=1$. The generalisation to any $a>0$ is straightforward
following our derivation. With this choice, Eq.~\eqref{eq:master-NLL}
reads
\begin{align}
\label{eq:master-NLL-kt}
  \Sigma(v) &= \sigma^{(0)}\int \frac{d k_{t1}}{k_{t1}} \int_0^{2\pi} \frac{d\phi_1}{2\pi}
              e^{-R(\epsilon k_{t1})}\sum_{\ell_1=1,2} R_{\ell_1}'\left(k_{t1}\right) \times \notag\\ &\times \sum_{n=0}^{\infty}\frac{1}{n!}
                                                           \prod_{i=2}^{n+1}
                                                           \int_{\epsilon}^{1}\frac{d\zeta_i}{\zeta_i}\int_0^{2\pi}
                                                           \frac{d\phi_i}{2\pi} \sum_{\ell_i=1,2}
                                                           R_{\ell_i}'\left(\zeta_i k_{t1}\right)\, \Theta\left(v-V(\{\tilde{p}\},k_1,\dots, k_{n+1})\right)\,,
\end{align}
where, with a little abuse of notation, we redefined
$\zeta_i=k_{ti}/k_{t1}$. As it will be described in
Section~\ref{sec:event-generation}, the above equation can be
efficiently evaluated as a simplified shower of primary emissions off
the initial-state legs, ordered in transverse momentum. This choice of
the ordering variable is dictated by the choice of the resolution
scale, that in turn leads to the Sudakov radiator for a $k_t$ ordered
evolution in Eq.~\eqref{eq:master-NLL-kt}.

\subsection{Structure of higher-order corrections}
\label{sec:higher-order}
In deriving the main result of the previous section,
Eq.~\eqref{eq:master-NLL}, we made two approximations. Firstly, we
ignored $n$PC correlated blocks with $n>2$ in the squared
amplitudes~\eqref{eq:nPC-inclusive}. Secondly, we did not specify a
complete treatment of hard-collinear radiation. Indeed, the only
hard-collinear contribution entering at NLL (in
Eq.~\eqref{eq:single-emsn}) has been treated with soft kinematics. We
discuss how to relax both approximations in the next two subsections.

\subsubsection{Correlated blocks at higher-logarithmic order} 
Higher-order corrections require the inclusion of higher-multiplicity
and higher-order blocks with respect to those relevant to
Eq.~\eqref{eq:master-NLL}. The relevant blocks necessary to a given
order are summarised in Table~\ref{tab:log-order}.
\begin{table}[h]
\begin{center}
\begin{tabular}{ c  c }
  {Logarithmic order} & { Blocks required} \\
\midrule
  LL       & \{1PC$^{(0)}$ (sc)\} \\
  \cmidrule(){1-2}
  NLL      & \{1PC$^{(0)}$, 1PC$^{(1)}$ (sc)\}; \{2PC$^{(0)}$ (sc)\} \\
    \cmidrule(){1-2}
  NNLL     & \{1PC$^{(m\leq 1)}$, 1PC$^{(2)}$ (sc)\}; \{2PC$^{(0)}$, 2PC$^{(1)}$ (sc)\};\\
          & \{3PC$^{(0)}$ (sc)\} \\
            \cmidrule(){1-2}
  N$^3$LL  & \{1PC$^{(m\leq 2)}$, 1PC$^{(3)}$ (sc)\}; \{2PC$^{(m\leq 1)}$, 2PC$^{(2)}$ (sc)\}; \\
           & \{3PC$^{(0)}$, 3PC$^{(1)}$ (sc)\}; \{4PC$^{(0)}$ (sc)\} \\ 
 \cmidrule(){1-2}
$\vdots$ & $\vdots$ \\
  \cmidrule(){1-2}
  N$^k$LL  & \{1PC$^{(m\leq k-1)}$, 1PC$^{(k)}$ (sc)\}; $\cdots$ ; \{$(k+1)$PC$^{(0)}$ (sc)\}\\
\end{tabular}
\caption{Blocks to be included in the squared-amplitude decomposition
  at a given logarithmic order. At each order, the higher-rank blocks
  are to be included in the soft-collinear limit (``sc'' in the table).}
\label{tab:log-order}
\end{center}
\end{table}
For instance, at NNLL, for the observables~\eqref{eq:inclusive}, one
has to include 2PC$^{(0)}$ (i.e.~the fully correlated double
emission), and 1PC$^{(1)}$ both in the soft and in the hard-collinear
limit, and 3PC$^{(0)}$, 2PC$^{(1)}$, and 1PC$^{(2)}$ blocks in the
soft-collinear limit. Given the inclusive nature of the
observables~\eqref{eq:inclusive} that we are treating in this article,
the inclusion of higher-order blocks can be done in a simple
systematic way by adding more terms to the r.h.s. of
Eq.~\eqref{eq:nPC-inclusive}.

We remind the reader of the fact that, while at NLL the bounds for
rapidity $Y_i$ of the inclusive block $|M(k_i)|^2_{\rm inc}$ can be
approximated with their massless limit (see
Eq.~\eqref{eq:2PC-inclusive} and comments below it), starting at NNLL
the integration over the rapidity $Y_i$ must be performed exactly.

\subsubsection{Hard-collinear emissions and treatment of recoil}
\label{sec:recoil}
In order to repeat the procedure that led to Eq.~\eqref{eq:master-NLL}
at higher logarithmic accuracy, we need to handle the phase space in
the multiple-emission kinematics. In the NLL case derived in the
previous section, indeed, all resolved real emissions are soft and
collinear and therefore they do not modify each other's phase
space. However, starting at NNLL one or more real emissions can be
hard and collinear to the emitting leg and this changes the available
phase space for subsequent real emissions. More precisely, at NNLL we
need to work out the corrections due to a single hard-collinear resolved
emission within an ensemble of soft-collinear radiation. Similarly, at
N$^3$LL, one has to consider up to two resolved hard-collinear
emissions embedded in an ensemble of soft-collinear radiation. The
kinematics and the proper treatment of hard-collinear emissions, still
missing in our formulation, will be discussed in this section.

To correctly include the evolution of the hard-collinear radiation in
our formulation, we first consider how initial-state radiation
modifies the real-emission kernels, illustrating this in the
single-emission case for the sake of clarity. Throughout this section
and in the rest of this article we use the tree-level splitting
functions as reported in Appendix~\ref{app:radiator}.

We start by formulating the single-emission probability for a
gluon-initiated process. For the sake of concreteness, all prefactors
in this subsection are given under the assumption that the colour
singlet is a single particle, e.g. a Higgs boson. We express the
probability of emitting either a gluon or a quark off leg $1$ (an
analogous term can be written for an emission off leg $2$), for an
observable $v$, as
\begin{align}
\label{eq:first-emission}
&\Sigma(v) = 2\pi\,|M_{B}|_{gg}^2\int d x_1 d x_2\delta(x_1 x_2 s - M^2)\int \frac{dk_t}{k_t}\frac{\as}{\pi}
\frac{d\phi}{2\pi}\notag\\
&\times\,\bigg(\int_{x_1}^{1-k_t/M} dz \left[{2}\frac{P^{(0)}_{gg}(z)}{z}
                                                                 f_{g}(\mu_F,\frac{x_1}{z})+
  \frac{P^{(0)}_{gq}(z)}{z} \left(f_{q}(\mu_F,\frac{x_1}{z}){+
  f_{\bar q}(\mu_F,\frac{x_1}{z})}\right)\right]f_{g}(\mu_F,x_2)
  \Theta(v-v(k)) \notag\\
&-\int_{0}^{1-k_t/M} dz \left[P^{(0)}_{gg}(z)
                                                                 +
 n_f P^{(0)}_{qg}(z) \right]f_{g}(\mu_F,x_1) f_{g}(\mu_F,x_2)\notag\\
& -\left(\hat{P}^{(0)}_{gg}\otimes f_{g}\right)(x_1) f_{g}(\mu_F,x_2)
  - \left(P^{(0)}_{gq}\otimes f_{q}\right)(\mu_F,x_1) f_{g}(\mu_F,x_2) -\left(P^{(0)}_{gq}\otimes f_{\bar q}\right)(\mu_F,x_1) f_{g}(\mu_F,x_2)
\bigg)\,\notag\\
&+ \textrm{constant terms}\, ,
\end{align}
where $f_g(\mu_F,x)$ is the gluon density renormalised in the ${\rm
  \overline{MS}}$ scheme, evaluated at a factorisation scale $\mu_F$,
and $\hat{P}$ denotes the regularised splitting function. Since
$\hat{P}^{(0)}_{gq}(z)=P^{(0)}_{gq}(z)$ (see
Appendix~\ref{app:radiator}), the regularised label
$``\,\,\hat{}\,\,"$ applies only to $P_{gg}^{(0)}$. The second, third,
and fourth line of Eq.~\eqref{eq:first-emission} denote the real
emission, the virtual corrections, and collinear counterterm,
respectively. For the virtual correction, we simply use the
first-order expansion of the resummed form factor
${\cal V} (\Phi_B)$~\cite{Dixon:2008gr} expressed in terms of
leading-order splitting functions, of which we take the limit in four
dimensions. The unregulated soft and collinear divergences of the
four-dimensional virtual corrections manifestly cancel against the
ones in the real emissions at the integrand level. We stress once
again that in colour-singlet production we can use the azimuthally
averaged splitting functions (see Appendix~\ref{app:radiator}) up to
NNLL accuracy. At N$^3$LL, corrections from azimuthal correlations
arise~\cite{Catani:2010pd}, and they will be introduced in
Section~\ref{sec:master-initial-state}.

In general, the upper bound of the $z$ integration in the virtual
corrections is different from the one in the real correction when more
than one hard-collinear emission is present, since the available phase
space for the real emissions is changed by the presence of the
hard-collinear radiation. However, for the single-emission case
treated in Eq.~\eqref{eq:first-emission}, the upper bound, derived in
Eq.~\eqref{eq:z-limit}, is identical for the real and virtual
contributions.

Eq.~\eqref{eq:first-emission} also contains constant contributions arising
from both the finite terms of the virtual form factor in ${\rm
  \overline{MS}}$, and the ${\cal O}(\as)$ collinear coefficient
functions.  For the sake of simplicity, in the following discussion we
neglect these NNLL constant terms, which we will however include in
our final formula.

We now add and subtract the term 
\begin{align}
&2\pi\,|M_{B}|_{gg}^2\int d x_1 d x_2\delta(x_1 x_2 s - M^2)\int \frac{dk_t}{k_t}\frac{\as}{\pi}
\frac{d\phi}{2\pi}\notag\\
&\times\int_{0}^{1-k_t/M} dz \left[P^{(0)}_{gg}(z)
                                                                 +
  n_f P^{(0)}_{qg}(z) \right]f_{g}(\mu_F,x_1) f_{g}(\mu_F,x_2) \Theta(v-v(k))\,,
\end{align}
and recast Eq.~\eqref{eq:first-emission} as
\begin{align}
\Sigma(v) &= 2\pi\,|M_{B}|_{gg}^2\int d x_1 d x_2\delta(x_1 x_2 s - M^2)\int \frac{dk_t}{k_t}\frac{\as}{\pi}
\frac{d\phi}{2\pi}\notag\\
&\times\bigg(\int_{x_1}^{1-k_t/M} dz \,{2}\frac{P^{(0)}_{gg}(z)}{z}
                                                                 f_{g}(\mu_F,\frac{x_1}{z})f_{g}(\mu_F,x_2)
  \Theta(v-v(k)) - \int_{x_1}^1 dz \frac{\hat{P}^{(0)}_{gg}(z)}{z}
                                                                 f_{g}(\mu_F,\frac{x_1}{z}) f_{g}(\mu_F,x_2)\notag\\
&-\int_{0}^{1-k_t/M} dz \left[P^{(0)}_{gg}(z)
                                                                 +
  n_f P^{(0)}_{qg}(z) \right]f_{g}(\mu_F,x_1) f_{g}(\mu_F,x_2) \Theta(v-v(k))\notag\\
&+\int_{0}^{1-k_t/M} dz \left[P^{(0)}_{gg}(z)
                                                                 +
  n_f P^{(0)}_{qg}(z) \right]f_{g}(\mu_F,x_1) f_{g}(\mu_F,x_2)\left( \Theta(v-v(k))-1\right)\notag\\
& 
  + \int_{x_1}^1 dz \frac{P^{(0)}_{gq}(z)}{z}
  \left(f_{q}(\mu_F,\frac{x_1}{z}){+f_{\bar q}(\mu_F,\frac{x_1}{z})}\right)f_{g}(\mu_F,x_2) \left( \Theta(v-v(k))-1\right) \notag\\&- \int_{1-k_t/M}^1 dz \frac{P^{(0)}_{gq}(z)}{z} \left(f_{q}(\mu_F,\frac{x_1}{z}){+f_{\bar q}(\mu_F,\frac{x_1}{z})}\right)f_{g}(\mu_F,x_2)
  \Theta(v-v(k))\bigg).
\end{align}
By using the symmetry of the $P_{gg}$ splitting function under $z\leftrightarrow
1-z$, one finds that
\begin{equation}
\label{eq:Pgg-regularised}
\int_{x_1}^1 dz \,{2}\frac{P^{(0)}_{gg}(z)}{z} f_{g}(\mu_F,\frac{x_1}{z}) - \int_{0}^1 dz \left(P^{(0)}_{gg}(z)
                                                                 +
 n_f  P^{(0)}_{qg}(z) \right)f_{g}(\mu_F,x_1) = \int_{x_1}^1 dz \frac{\hat{P}^{(0)}_{gg}(z)}{z} f_{g}(\mu_F,\frac{x_1}{z})\,,
\end{equation}
which allows us to recast the previous equation as
\begin{align}
\label{eq:first-emission-G}
\Sigma(v) &= 2\pi\,|M_{B}|_{gg}^2\int d x_1 d x_2\delta(x_1 x_2 s - M^2)\int \frac{dk_t}{k_t}\frac{\as}{\pi}
\frac{d\phi}{2\pi}\notag\\
&\times\bigg\{\int_{x_1}^{1} dz \frac{\hat{P}^{(0)}_{gg}(z)}{z}
                                                                 f_{g}(\mu_F,\frac{x_1}{z})f_{g}(\mu_F,x_2)
  \left(\Theta(v-v(k))-1\right)\notag\\
&+\int_{0}^{1-k_t/M} dz \left[P^{(0)}_{gg}(z)
                                                                 +
  n_f P^{(0)}_{qg}(z) \right]f_{g}(\mu_F,x_1) f_{g}(\mu_F,x_2)\left( \Theta(v-v(k))-1\right)\notag\\
& 
  + \int_{x_1}^1 dz \frac{P^{(0)}_{gq}(z)}{z}
  \left(f_{q}(\mu_F,\frac{x_1}{z}){+f_{\bar q}(\mu_F,\frac{x_1}{z})}\right)f_{g}(\mu_F,x_2) \left(
  \Theta(v-v(k))-1\right)\notag\\
& - \int_{1-k_t/M}^1 dz \left( {2}\frac{P^{(0)}_{gg}(z)}{z}
                                                                 f_{g}(\mu_F,\frac{x_1}{z})f_{g}(\mu_F,x_2)
  - \left[P^{(0)}_{gg}(z)
                                                                 +
  n_f P^{(0)}_{qg}(z) \right]f_{g}(\mu_F,x_1) f_{g}(\mu_F,x_2)\right.\notag\\
&\left. + \frac{P^{(0)}_{gq}(z)}{z}
  \left(f_{q}(\mu_F,\frac{x_1}{z}){+f_{\bar q}(\mu_F,\frac{x_1}{z})}\right)f_{g}(\mu_F,x_2) \right)
  \Theta(v-v(k))\bigg\}\,.
\end{align}
Analogously, it is straightforward to show that the logarithmic part
for a quark-initiated process with an emission off the leg $1$ reads
\begin{align}
\label{eq:first-emission-Q}
\Sigma(v) &= 2\pi\,|M_{B}|_{q\bar{q}}^2\int d x_1 d x_2\delta(x_1 x_2 s - M^2)\int \frac{dk_t}{k_t}\frac{\as}{\pi}
\frac{d\phi}{2\pi}\notag\\
&\times\,\bigg\{\int_{x_1}^{1} dz \frac{P^{(0)}_{qg}(z)}{z}
                                                                 f_{g}(\mu_F,\frac{x_1}{z})f_{\bar
  q}(\mu_F,x_2)
  \left(\Theta(v-v(k))-1\right)\notag\\
&+\int_{0}^{1-k_t/M} dz P^{(0)}_{qq}(z)f_{q}(\mu_F,x_1) f_{\bar q}(\mu_F,x_2)\left( \Theta(v-v(k))-1\right)\notag\\
& 
  + \int_{x_1}^1 dz \frac{\hat{P}^{(0)}_{qq}(z)}{z}
  f_{q}(\mu_F,\frac{x_1}{z})f_{\bar q}(\mu_F,x_2) \left(
  \Theta(v-v(k))-1\right)\notag\\
& - \int_{1-k_t/M}^1 dz \left( \frac{P^{(0)}_{qq}(z)}{z}
                                                                 f_{q}(\mu_F,\frac{x_1}{z})f_{\bar
  q}(\mu_F,x_2)
  - P^{(0)}_{qq}(z)f_{q}(\mu_F,x_1) f_{\bar q}(\mu_F,x_2) \right.\notag\\
&\left.+ \frac{P^{(0)}_{qg}(z)}{z}
  f_{g}(\mu_F,\frac{x_1}{z})f_{\bar q}(\mu_F,x_2) \right)
  \Theta(v-v(k))\bigg\}\,,
\end{align}
where we have set ${\hat P}^{(0)}_{qg}(z) = P^{(0)}_{qg}(z)$.

In Eqs.~\eqref{eq:first-emission-G} and~\eqref{eq:first-emission-Q},
the last integral from $1-k_t/M$ to $1$ gives rise to regular terms
and can therefore be neglected. As far as the remaining terms are
concerned, we notice that the squared matrix element for an
initial-state emission, which corresponds to the terms containing a
$\Theta$ function in Eqs.~\eqref{eq:first-emission-G}
and~\eqref{eq:first-emission-Q}, can be separated into two pieces:
\begin{itemize}
\item The first one, encoded in the third line of
  Eqs.~\eqref{eq:first-emission-G} and~\eqref{eq:first-emission-Q},
  modifies neither the flavour nor the momentum fraction of the
  incoming partons, and the bounds of the relative $z$ integration are
  those of the corresponding virtual phase space. This contribution is
  fully analogous to the case treated in Sec.~\ref{sec:higher-order},
  that gives rise to $R'$ in Eq.~\eqref{eq:master-NLL}. When
  evaluating this term explicitly, we can further split it, as done in
  Eq.~\eqref{eq:single-emsn}, into a soft term and a hard-collinear
  contribution. The exact upper bound of the $z$ integral is only
  relevant in the soft contribution, while it can be extended up to
  $1$ in the hard-collinear term up to regular (non logarithmic)
  terms. In the following, we will refer to this term as the $R'$
  contribution.
\item The second one (second and fourth lines of
  Eqs.~\eqref{eq:first-emission-G} and~\eqref{eq:first-emission-Q})
  does modify both flavour and momentum fraction. This contribution
  corresponds to an exclusive step of DGLAP evolution. The
  corresponding $z$ integration can be extended up to the soft limit
  ($z=1$) as this limit is regularised by the plus distribution in the
  corresponding splitting function. We stress once again that the
  latter extension of the upper bound of the $z$ integration in the
  hard-collinear radiation's phase space is correct up to regular
  terms that are ignored in our treatment. We will refer to this term
  as the {\it exclusive DGLAP} evolution step.
\end{itemize}
This decomposition is only a convenient way of expressing the squared
amplitude and phase space for an initial-state emission, and only the
sum of all logarithmic terms in Eqs.~\eqref{eq:first-emission-G}
and~\eqref{eq:first-emission-Q} is physically well defined. The
considerations above will be useful in the rest of this section when
the all-order kinematics is discussed.
\\

As anticipated in the beginning of this subsection, in order to
achieve N$^3$LL accuracy, one has to consider configurations with up
to two resolved hard-collinear emissions together with any number of
soft-collinear partons in the final state.
We therefore study how the presence of hard-collinear emissions
affects the phase space of the remaining radiation in the all-order
picture.\footnote{We thank A.~Banfi for fruitful discussions on this
  point.} We consider again the emissions ordered according to their
transverse momentum. In this picture, the relation between the
$z^{(\ell)}$ variable and the Sudakov variable $y^{(\ell)}$ for a
given emission $k$ will be modified by the radiation that occurred
before $k$ as described in Eq.~\eqref{eq:z-to-y-all}.

We consider the case of an ensemble of resolved emissions off a leg
$\ell$ of which a single one is hard and collinear, while all the
remaining radiation is soft. We can group the emissions into the
following three sets: the soft emissions that occur {\it before} the
hard-collinear parton is emitted (i.e. at larger transverse momenta),
the hard-collinear emission itself, and the soft emissions that occur
{\it after} the hard-collinear one (at smaller transverse
momenta). The soft radiation emitted before the hard-collinear
emission has $z_i^{(\ell)}\simeq y_i^{(\ell)}\simeq 1$ and therefore
$k_{ti}\simeq \tilde{k}_{ti}$, so its phase space boundaries are as
described in Section~\ref{sec:nll}. For the hard-collinear emission
$k^{hc}$ the relation between $z_{hc}^{(\ell)}$ and $y_{hc}^{(\ell)}$
is reported in Eq.~\eqref{eq:z-to-y} and the corresponding
$z_{hc}^{(\ell)}$ integration bound is in
Eq.~\eqref{eq:z-limit}. Finally, soft emissions that occur after the
hard-collinear one will again have $k_{ti}\simeq \tilde{k}_{ti}$ but
now $1-y_i^{(\ell)} \simeq (1-z_i^{(\ell)})/z^{(\ell)}_{hc}$. The
upper bound of their $z_i^{(\ell)}$ integral is therefore
\begin{equation}
\label{eq:virt-bound}
z_i^{(\ell)} < 1-z_{hc}^{(\ell)} k_{ti}/M.
\end{equation}
From the above equation we see that the phase space of the soft
radiation emitted after the hard-collinear emission is modified by the
presence of the latter. However, the squared amplitude and phase space
for emissions in the soft
limit only depend on $z_i^{(\ell)}$ through $d
z_i^{(\ell)}/(1-z_i^{(\ell)})$. Therefore, using the relation
\begin{equation}
\label{eq:int-simpl-soft}
\frac{d z_i^{(\ell)}}{1-z_i^{(\ell)}} = \frac{d y_i^{(\ell)}}{1-y_i^{(\ell)}},
\end{equation} 
and using the fact that $k_{ti}\simeq \tilde{k}_{ti}$ for these
emissions, we can replace the integral over $z_i^{(\ell)}$ with an
integral over $y_i^{(\ell)}$ whose upper bound is given by 
\begin{equation}
y_i^{(\ell)} < 1- k_{ti}/M.
\end{equation}
This allows one to disentangle the phase space of all emissions in the
considered kinematic configuration and, hence, to iterate the
procedure at all orders.

The remaining kinematic configuration to be considered in a N$^3$LL
resummation is given by an ensemble of soft-collinear emissions
accompanied by two hard-collinear ones. We label the two hard
collinear emissions by $k^{hc}_{1}$ and $k^{hc}_{2}$ and we assume,
without any loss of generality, that $k^{hc}_{1}$ is emitted before
$k^{hc}_{2}$ (hence it has a larger transverse momentum in our
picture). The upper bounds of the corresponding $z^{(\ell)}$ integrals
for the real contribution will now be complicated functions of the
transverse momenta $k^{hc}_{t1}$ and $k^{hc}_{t2}$ that can be
obtained starting from
Eqs.~\eqref{eq:kt-rel},~\eqref{eq:z-to-y-all}. However, things are
much simplified if we use the decomposition described in the first
part of this section, as follows. We recall that the real matrix
element can be decomposed as a sum of the $R'$ contribution (that does
not modify the momentum fraction of the emitter, and whose kinematics
is soft by construction), and an exclusive DGLAP step that modifies
the momentum fraction of the emitting leg, as shown in
Eqs.~\eqref{eq:first-emission-G},~\eqref{eq:first-emission-Q}. In the
latter term, the upper bound of the $z^{(\ell)}$ integration can be
extended to $1$ (hence it becomes independent of the kinematics of the
rest of the event) since the soft limit is regularised by the plus
prescription in the corresponding splitting functions. As for the $R'$
contributions relative to $k^{hc}_{1}$ and $k^{hc}_{2}$, they can be
further decomposed into a soft-collinear term and a term that contains
the hard-collinear part of the matrix element (which however does not
modify the momentum fraction of the emitting leg). Once again, in the
latter contribution the $z^{(\ell)}$ integration can be extended to
$1$, while in the soft-collinear contribution one can simply replace
the $z^{(\ell)}$ integral with an integral over $y^{(\ell)}$ by means
of Eq.~\eqref{eq:int-simpl-soft}. Moreover, using the fact that for a
soft emission $\tilde{k}_t\simeq k_t$, the corresponding upper bound
of the $y^{(\ell)}$ integral can be replaced by $1-k_t/M$.

This procedure allows one to disentangle completely the phase space of
the $R'$ contributions (whose kinematics is soft by construction) from
that of the exclusive DGLAP evolution step which are by construction
hard and collinear. The lower bounds in the $z^{(\ell)}$ integrals of
multiple resolved DGLAP evolution steps are entangled as each of them
modifies significantly the momentum available for the subsequent
hard-collinear ones, resulting in a convolution between the splitting
kernels and the corresponding parton density.

The above treatment of the double-hard-collinear case is valid up to
regular terms. In this section we neglected the constant terms that
arise from the finite part of the renormalised form factor, and from
the collinear coefficient functions, which are relevant already for a
NNLL resummation. For inclusive observables considered in this
article, the collinear coefficient functions factorise in front of the
Sudakov factor and, for the processes considered here, they were
computed to ${\cal O}(\as^2)$ in
refs.~\cite{Catani:2011kr,Catani:2012qa,Gehrmann:2014yya}. These will
be introduced in the following section when we iterate the arguments
discussed here at all perturbative orders in $\as$.

\subsubsection{Resummed formula for initial-state radiation}
\label{sec:master-initial-state}

The arguments derived in the previous section can be used to formulate
the structure of the cross section at all orders by iterating the
single-emission picture defined above. Given the inclusive nature of the observables
studied here, the inclusion of higher-order logarithmic corrections
can be achieved in a simple way by just adding the relevant correlated
blocks (as reported in Table~\ref{tab:log-order}) in the inclusive
approximation~\eqref{eq:nPC-inclusive}. The contribution to the cross
section from each inclusive block, in turn, can be split into an
$R'$-type contribution (which does not modify either the momentum
fraction or the flavour of the emitting leg), and a DGLAP step
(inclusive in the content of each correlated block, but differential
in its transverse momentum), and hence it can be treated in a fully analogous way to what done for single emissions in the previous subsection. This simple prescription allows us to discuss the inclusion of
the parton densities by referring to emissions (for the sake of
simplicity), while keeping in mind that they are to be thought of as
inclusive sums of correlated blocks as defined in
Eq.~\eqref{eq:nPC-inclusive}.

\begin{figure}[h!]
  \centering
  \includegraphics[width=0.9\columnwidth]{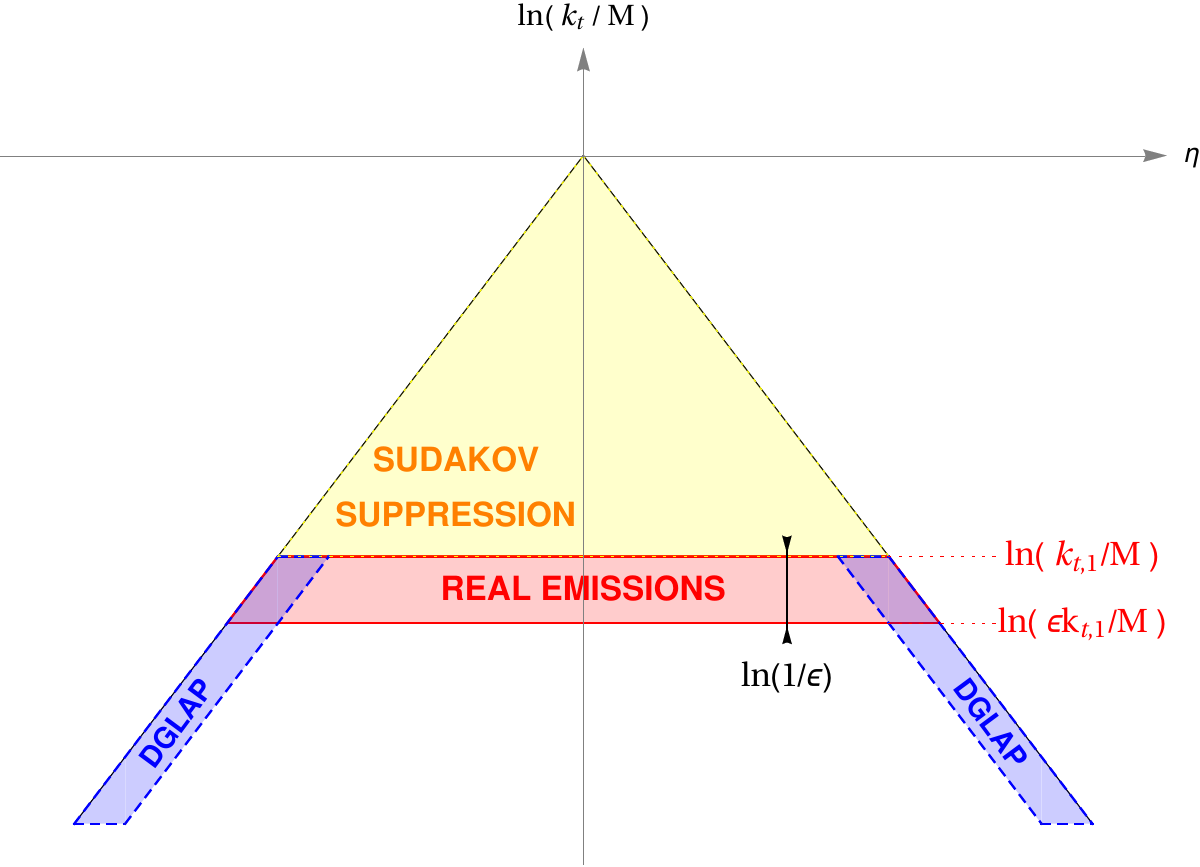} 
  \caption{Phase space for a secondary real emission.}
  \label{fig:lund}
\end{figure}

To show how the parton densities are accounted for, we start by
evaluating them at a scale $\mu_0$ that is assumed to be smaller than
all transverse momenta in the event. We consider the situation in
which the emissions are ordered in transverse momentum, and the
hardest (resolved) emission $k_1$ occurred. The phase-space diagram
for any secondary emission $k_{i}$ with $i>1$ is depicted in
Fig.~\ref{fig:lund} in the $\ln(k_t/M)-\eta$ (Lund) plane, where now
$\eta$ denotes the rapidity in the centre-of-mass frame of the
incoming partons which are extracted from the proton at a
factorisation scale $\mu_0$, and the transverse momentum $k_t$ is
taken with respect to the beam direction. As stated in
Section~\ref{sec:nll}, due to rIRC safety, only emissions that take
place in the strip between $\epsilon k_{t1}$ and $k_{t1}$ (labelled
with ``REAL EMISSIONS'' in Fig.~\ref{fig:lund}) modify the observable
significantly and are resolved. The remaining unresolved real
emissions ($k_{ti}<\epsilon k_{t1}$) are combined with the virtual
corrections, which populate the whole region below the two diagonal
lines that denote the upper rapidity limits. The result of this
combination is indeed the Sudakov form factor associated with the
first emission that vetoes secondary emissions in the yellow region
(labelled with ``SUDAKOV SUPPRESSION'' in Fig.~\ref{fig:lund}) of the
Lund plane. In addition, the combination of virtual and unresolved
emissions gives also rise to a constant term that multiplies the
Sudakov and encodes both the finite part of the virtual corrections
and the constant contribution due to soft and/or collinear emissions
exactly at the edges of their phase space, encoded in the collinear
coefficient functions.

In the initial-state-radiation case at hand, hard-collinear emissions
define the evolution of the parton densities. These emissions occur on
a strip (labelled with ``DGLAP'' in Fig.~\ref{fig:lund}) along the
upper rapidity bounds, and their evolution is encoded in the DGLAP
equations.
In the unresolved region ($k_{ti}<\epsilon k_{t1}$), the DGLAP
evolution can be performed {\it inclusively} since emissions in this
phase-space region do not affect the value of the observable. On the
other hand, when $k_{t1}>k_{ti}>\epsilon k_{t1}$ the corresponding
hard-collinear emissions modify significantly the observable's value
and therefore must be treated {\it exclusively}, namely unintegrated
in $k_t$.

In addition to the parton densities, starting at ${\cal O}(\as)$, one
needs to include the coefficient functions that emerge from their
renormalisation, and originate from emissions that occur at the edges
of the phase space in Fig.~\ref{fig:lund}. The coefficient functions
contribute to the logarithmic structure only through the scale of
their running coupling, which is the transverse momentum of the
emission(s) they are associated with. As done for the parton
densities, one can evaluate them initially at a scale $\mu_0$ smaller
than any transverse momentum in the event, and subsequently evolve
them inclusively up to the resolution scale $\epsilon k_{t1}$. Their
evolution must be instead treated exclusively in the resolved strip
$k_{t1}>k_{ti}>\epsilon k_{t1}$.

In order to introduce the all-order result, it is convenient to
simplify the flavour structure of the evolution for the time being. We
neglect real-emission kernels that modify the flavour of the emitting
leg, namely those that do not have a soft singularity $P_{qg}$ and
$P_{gq}$. This ensures that the flavour of the initial parton
densities is only modified by the coefficient functions and is
conserved by the resolved real radiation. This approximation is made
without any loss of generality, and for the only sake of
simplicity. The extension to the full flavour case will be trivial
once the final formula is obtained.

For the remaining part of the section, it is useful to introduce a
matrix notation to simplify the structure of our expressions in
flavour space. We define ${\bf f}$ as the array containing the
$2n_f+1$ partonic densities, where $n_f$ denotes the number of active
flavours.  To handle different Born configurations with different
incoming flavours ${c_\ell}$, we then define the coefficient-function
matrix ${\bf C}^{c_\ell}$ as a $(2 n_f +1 ) \times (2 n_f +1 ) $
diagonal matrix in flavour space whose entries are
\begin{align}
\label{eq:coeff-funct-mellin}
[{\bf C}^{c_\ell}]_{a b} = C_{c_\ell f(a)} \delta_{a b}, 
\end{align}
where $C_{ij}$ are the collinear coefficient functions, $c_\ell$ is
the flavour of the leg $\ell$ entering the Born process, and $f(a)$ is
the flavour corresponding to the $a$-th entry of the parton-density
array.  For instance, we explicitly show the above convention in the
case of Higgs production, considering only a single quark flavour
$q$. By defining the array $\mathbf f = (f_g,f_q,f_{\bar q})^T$, the matrix
${\bf C}^g$ reads
\begin{eqnarray}
{\bf C}^g &=& \left( \begin{array}{ccc}
C_{gg} & 0 & 0 \\
0 & C_{gq} & 0 \\
0 & 0 & C_{g\bar{q}} \end{array} \right).
\end{eqnarray}

The evolution of~\eqref{eq:coeff-funct-mellin} between two scales is
entirely encoded in the evolution of the running coupling. By
introducing the corresponding anomalous-dimension matrix
${\bf \Gamma}^{(C)}$
\begin{equation}
\label{eq:gammaC}
{\bf \Gamma}^{(C)}(\alpha_s(k_{t})) = 2
  \beta(\alpha_s(k_{t}))\frac{d \ln{\bf
  C}^{c_\ell}(\alpha_s(k_{t}))}{d
  \alpha_s(k_{t})},
\end{equation}
we can write the Renormalisation-Group evolution (RGE) of the coefficient function
matrix as
\begin{align}
\label{eq:RGE}
{\bf C}^{c_\ell}(\alpha_s(\mu)) = \exp\left\{-\int_\mu^{\mu_0}\frac{d k_t}{k_t} {\bf \Gamma}^{(C)}(\alpha_s(k_{t}))\right\} {\bf C}^{c_\ell}(\alpha_s(\mu_0)).
\end{align}
In principle, the matrix ${\bf \Gamma}^{(C)}$ should also explicitly
carry a label $c_\ell$ to specify that it evolves the coefficient
function ${\bf C}^{c_\ell}$ associated with the Born flavour
$c_\ell$. We omit this label as the notation in what
follows is unambiguous. We stress however that the flavour of the
coefficient function is not modified by its RG evolution, indeed it is
manifestly flavour diagonal.

The iterative structure of the squared amplitudes appears more
transparent if we work in Mellin space, where convolutions become
products. We therefore introduce the Mellin transform of a function
$g(x)$ as
\begin{equation}
g_{N_\ell}\equiv\int_0^1d x\,x^{N_\ell-1} g(x).
\end{equation}

The DGLAP~\cite{Gribov:1972ri,Altarelli:1977zs,Dokshitzer:1977sg}
evolution of the parton-density vector ${\bf f}$ can be conveniently
written in Mellin space as
\begin{equation}
\label{eq:DGLAP}
{\bf  f}_{N_\ell}(\mu) =
{\cal P} \exp\left\{-\int_{\mu}^{\mu_0}\frac{dk_t}{k_t}\frac{\alpha_s(k_t)}{\pi}{\bf
    \Gamma}_{N_\ell}(\alpha_s(k_t))\right\}
{\bf  f}_{N_\ell}(\mu_0).
\end{equation}
In the previous equation $\cal P$ is the path-ordering symbol, and the
matrix ${\bf \Gamma}$ is defined as 
\begin{align}
\left[{\bf \Gamma}_{N_\ell}(\alpha_s(\mu))\right]_{ab} = \int_0^1 d z\,z^{N_\ell-1} \hat{P}_{f(a) f(b)}(z,\as(\mu)) \equiv  \gamma_{N_\ell;f(a)f(b)} =
\sum_{n=0}^{\infty} \left(\frac{\alpha_s(\mu)}{2\pi}\right)^{n} \gamma_{N_\ell;f(a)f(b)}^{(n)},
\end{align}
where $\hat{P}_{f(a)f(b)}$ are the regularised splitting functions
(see Appendix~\ref{app:radiator}). We stress that, within the
simplifying assumption made above on flavour-conserving real-emission
kernels, no splitting functions involving a real quark emission are
included, therefore the matrix ${\bf \Gamma}$ is diagonal. Within this
assumption, the path ordering in Eq.~\eqref{eq:DGLAP} can be lifted.

With this notation, the hadronic cumulative cross section,
differential with respect to the Born phase space $\Phi_B$, can be
written as
\begin{align}
&\frac{d\Sigma(v)}{d\Phi_B} =\int_{{\cal C}_1} \frac{d N_1}{2\pi i} \int_{{\cal C}_2}\frac{dN_2}{2\pi i}x_1^{-N_1} x_2^{-N_2}\,\sum_{c_1, c_2}\frac{d|M_B|_{c_1c_2}^2}{d\Phi_B}
{\bf f}^{T}_{N_1}(\mu_0) \hat{\bf \Sigma}^{c_1,c_2}_{N_1,N_2}(v) {\bf f}_{N_2}(\mu_0),
\label{eq:hadxs}
\end{align}
where the sum runs over all possible Born configurations and we employed a double inverse Mellin transform.
The contours ${\cal C}_1$ and ${\cal C}_2$ are understood to lie along
the imaginary axis to the right of all singularities of the
integrand. 
In Eq.~\eqref{eq:hadxs}, and from now on, we define the notation
\begin{equation*}
\frac{d|M_B|_{c_1c_2}^2}{d\Phi_B} \equiv \int d\Phi^\prime_B
|M_B|_{c_1c_2}^2 \delta(x_1-x_1^\prime)
\delta(x_2-x_2^\prime)\delta(\Omega_B-\Omega_B^\prime)
\,,
\end{equation*}
where $\Omega_B$ denotes any set of internal phase-space variables
used to parametrise the colour-singlet system. The right-hand side
differs from the squared amplitude $|M_B|_{c_1c_2}^2$ simply by a
jacobian factor.\\

The matrix $\hat{\bf \Sigma}$ encodes the effect of the all-order
radiation that evolves the partonic cross section and the
corresponding parton densities. To write down an all-order expression
for $\hat{\bf \Sigma}$ for the observables~\eqref{eq:inclusive}, we
need to iterate the single-emission probability derived in the
previous section. Given that the phase space of the $R'$ contributions
and the exclusive DGLAP evolution steps are completely disentangled in
the resolved real radiation, this operation can be performed
straightforwardly in Mellin space, yielding
\begin{align}
\label{eq:partxs-mellin}
\hat{\bf \Sigma}^{c_1,c_2}_{N_1,N_2}(v) &= \left[{\bf
             C}^{c_1; T}_{N_1}(\alpha_s(\mu_0)) H(\mu_R)
    {\bf C}^{c_2}_{N_2}(\alpha_s(\mu_0)) \right]\,\int_0^{M}\frac{d k_{t1}}{k_{t1}} \int_0^{2\pi}
                     \frac{d\phi_1}{2\pi}\notag\\& \times e^{-{\bf R}(\epsilon k_{t1})}\exp\left\{-\sum_{\ell=1}^{2}\left(
\int_{\epsilon k_{t1}}^{\mu_0}\frac{dk_t}{k_t}\frac{\alpha_s(k_t)}{\pi}{\bf
  \Gamma}_{N_\ell}(\alpha_s(k_t)) + \int_{\epsilon
                                                   k_{t1}}^{\mu_0}\frac{dk_t}{k_t} {\bf \Gamma}_{N_\ell}^{(C)}(\alpha_s(k_{t}))\right)\right\}\notag\\
&\sum_{\ell_1=1}^2\left(
  {\bf R}_{\ell_1}'\left(k_{t1}\right) + \frac{\alpha_s(k_{t1})}{\pi}{\bf
  \Gamma}_{N_{\ell_1}}(\alpha_s(k_{t1})) + {\bf \Gamma}_{N_{\ell_1}}^{(C)}(\alpha_s(k_{t1}))\right) \notag\\
&\times\sum_{n=0}^{\infty}\frac{1}{n!}
                                                     \prod_{i=2}^{n+1}
                                                     \int_{\epsilon}^{1}\frac{d\zeta_i}{\zeta_i}\int_0^{2\pi}
                                                     \frac{d\phi_i}{2\pi}
  \sum_{\ell_i=1}^2\left({\bf R}_{\ell_i}'\left(k_{ti}\right) +\frac{\alpha_s(k_{ti})}{\pi}{\bf
  \Gamma}_{N_{\ell_i}}(\alpha_s(k_{ti})) + {\bf \Gamma}_{N_{\ell_i}}^{(C)}(\alpha_s(k_{ti}))\right)\notag\\
& \times\Theta\left(v-V(\{\tilde{p}\},k_1,\dots, k_{n+1})\right),
\end{align}
where now $\zeta_i=k_{ti}/k_{t1}$ since we are using the transverse
momentum as a resolution and ordering variable. ${\bf R}_{\ell}'$ is a
diagonal matrix in flavour space: given the flavour $c_\ell$ of the
Born leg $\ell$, it describes the flavour-conserving resolved
radiation off leg $\ell$. It is defined as
\begin{equation}
\label{eq:rp-array}
[{\bf R}_{\ell}']_{ab} = R_\ell' \delta_{ab},
\end{equation}
and $R_\ell'$ is defined in Eq.~\eqref{eq:R'1,2}. The Sudakov operator
${\bf R}$ is then defined as
\begin{equation}
\label{eq:rad-mat}
{\bf R}(\epsilon k_{t1}) = \sum_{\ell=1}^2\int_{\epsilon
  k_{t1}}^{M}\frac{dk_{t}}{k_{t}}{\bf R}_{\ell}'(k_t).
\end{equation}
The terms proportional to ${\bf R}'$ in Eq.~\eqref{eq:partxs-mellin}
encode the contribution of the radiation which is flavour-diagonal,
and does not modify the momentum fraction of the incoming
partons. This is the analogue of what has been derived in
Sec.~\ref{sec:nll} in the case of scale-independent parton densities. In
addition, the real emission probability now involves the exclusive
evolution for the parton densities and coefficient functions.

The matrices $\hat{\bf \Sigma}^{c_1,c_2}$ are diagonal in flavour
space within the flavour assumption that we are making here.
The first line of Eq.~\eqref{eq:partxs-mellin} contains the factor
$\left[{\bf C}^{c_1; T}_{N_1}(\alpha_s(\mu_0)) H(\mu_R) {\bf
    C}^{c_2}_{N_2}(\alpha_s(\mu_0)) \right]$
that encodes the hard-virtual corrections to the form factor and the
collinear coefficient functions. Explicit expressions for these
quantities will be given later (see Sec.~\ref{sec:mom-space} and
references therein). As discussed above, the coupling of the
coefficient functions here is evaluated at $\mu_0$ and subsequently
evolved up to $\epsilon k_{t1}$ by the operator containing the
diagonal matrix ${\bf \Gamma}_{N_\ell}^{(C)}$ in the second line
of~\eqref{eq:partxs-mellin}. Similarly, the parton densities are
evolved from $\mu_0$ up to $\epsilon k_{t1}$. As it was shown in
ref.~\cite{Catani:2010pd}, starting at a given order in perturbation
theory one needs to include the contribution from the collinear
coefficient functions ${\bf G}$, that describe the azimuthal
correlations with the initial-state gluons. Such a contribution starts at ${\cal O}(\as^2)$
(i.e.~N$^3$LL) for gluon-fusion processes, and at yet higher orders
for quark-initiated ones. It is included in the above formulation by
simply adding to Eq.~\eqref{eq:partxs-mellin} an analogous term where
one makes the replacements
\begin{equation}
\left[{\bf C}^{c_1; T}_{N_1}(\alpha_s(\mu_0)) H(\mu_R) {\bf C}^{c_2}_{N_2}(\alpha_s(\mu_0)) \right]\to\left[{\bf G}^{c_1; T}_{N_1}(\alpha_s(\mu_0)) H(\mu_R) {\bf G}^{c_2}_{N_2}(\alpha_s(\mu_0)) \right],
\end{equation}
and
\begin{equation}
\label{eq:RGE-G}
{\bf \Gamma}_{N_\ell}^{(C)}(\alpha_s(k_{t}))\to{\bf \Gamma}_{N_\ell}^{(G)}(\alpha_s(k_{t})),
\end{equation} 
where ${\bf \Gamma}_{N_\ell}^{(G)}$ is defined analogously to
Eq.~\eqref{eq:gammaC}, and the flavour structure of ${\bf G}$ is
analogous to the one of the ${\bf C}$ matrix. In what follows this
contribution, whenever not reported, is understood. \\

Eq.~\eqref{eq:partxs-mellin} has been derived by iterating the
single-emission probability. As discussed above, higher-order
logarithmic corrections are simply included by adding higher-order
correlated blocks. Specifically, this amounts to including
higher-order logarithmic corrections to the radiator $R$ and its
derivative $R'$, as well as in the anomalous dimensions which drive
the evolution of the parton densities and coefficient functions.
 
We conclude the discussion by pointing out that even if the all-order
formulation has been conveniently obtained in Mellin space, it is
possible to evaluate Eq.~\eqref{eq:hadxs} directly in momentum space
at any given logarithmic order. We will describe how to do this in
Sec.~\ref{sec:mom-space}.  Eq.~\eqref{eq:partxs-mellin} holds for all
inclusive observables (see definition in Sec.~\ref{sec:higher-order})
that do not depend on the rapidity of the initial-state radiation.  In
the remaining part of this article we specialise to the study of the
transverse-momentum case, but analogous conclusions will apply to
other observables of the same class.

\subsection{Equivalence with impact-parameter-space formulation}
\label{sec:compare-to-bspace}

In this section we show how to relate our Eq.~\eqref{eq:hadxs} to the
impact-parameter-space formulation of~\cite{Parisi:1979se}. We show the
equivalence for the differential partonic cross
section~\eqref{eq:partxs-mellin} in the case of the transverse momentum $p_t$. An analogous proof can be
carried out in the case of the $\phi^*$.

Our starting point is the
differential partonic cross section, where we now set $\mu_0=\mu_R=M$
without loss of generality:
\begin{align}
\label{eq:master-differential}
\frac{d}{d^2 \vec{p}_t} &\hat{\bf \Sigma}_{N_1,N_2}^{c_1
  c_2}(p_t)  = {\bf
             C}^{c_1; T}_{N_1}(\alpha_s(M)) H(M)
    {\bf C}^{c_2}_{N_2}(\alpha_s(M))\,\int_0^{M}\frac{d k_{t1}}{k_{t1}} \int_0^{2\pi}
                     \frac{d\phi_1}{2\pi}\notag\\& \times e^{-{\bf R}(\epsilon k_{t1})}\exp\left\{-\sum_{\ell=1}^{2}\left(
\int_{\epsilon k_{t1}}^{M}\frac{dk_t}{k_t}\frac{\alpha_s(k_t)}{\pi}{\bf
  \Gamma}_{N_\ell}(\alpha_s(k_t)) + \int_{\epsilon
                                                   k_{t1}}^{M}\frac{dk_t}{k_t} {\bf \Gamma}_{N_\ell}^{(C)}(\alpha_s(k_{t}))\right)\right\}\notag\\
&\times\sum_{\ell_1=1}^2\left(
  {\bf R}_{\ell_1}'\left(k_{t1}\right) + \frac{\alpha_s(k_{t1})}{\pi}{\bf
  \Gamma}_{N_{\ell_1}}(\alpha_s(k_{t1})) + {\bf \Gamma}_{N_{\ell_1}}^{(C)}(\alpha_s(k_{t1}))\right) \notag\\
&\times\sum_{n=0}^{\infty}\frac{1}{n!}
                                                     \prod_{i=2}^{n+1}
                                                     \int_{\epsilon}^{1}\frac{d\zeta_i}{\zeta_i}\int_0^{2\pi}
                                                     \frac{d\phi_i}{2\pi}
  \sum_{\ell_i=1}^2\left({\bf R}_{\ell_i}'\left(k_{ti}\right) +\frac{\alpha_s(k_{ti})}{\pi}{\bf
  \Gamma}_{N_{\ell_i}}(\alpha_s(k_{ti})) + {\bf \Gamma}_{N_{\ell_i}}^{(C)}(\alpha_s(k_{ti}))\right)\notag\\
& \times \delta^{(2)}\left(\vec{p}_t-\left(\vec{k}_{t1}
                                                     + \dots + \vec{k}_{t(n+1)}\right)\right).
\end{align}
We transform the $\delta$ function into $b$-space as
\begin{equation}
\label{eq:delta-fourier}
  \delta^{(2)}\left(\vec{p}_t-\left(\vec{k}_{t1}
      + \dots +
      \vec{k}_{t(n+1)}\right)\right)
   = \int\frac{d^2 \vec
     b}{4\pi^2} e^{-i
     \vec{b}\cdot\vec{p}_t} \prod_{i=1}^{n+1}e^{i \vec b\cdot\vec {k}_{ti}},
\end{equation}
and we evaluate the azimuthal integrals, which simply amounts to replacing
each of the factors $e^{\pm i \vec b\cdot \vec k_{t}}$ with a Bessel
function $J_0(b k_t)$. It is now straightforward to see that the sum
in Eq.~\eqref{eq:master-differential} gives rise to an exponential
function, yielding
\begin{align}
\label{eq:master-differential-exp}
\frac{d}{d p_t} &\hat{\bf \Sigma}_{N_1,N_2}^{c_1
  c_2}(p_t)  = {\bf
             C}^{c_1; T}_{N_1}(\alpha_s(M)) H(M)
    {\bf C}^{c_2}_{N_2}(\alpha_s(M)) \,p_t\int\! \!b\, d b
                          J_0(p_t b)\int_0^{M}\frac{d k_{t1}}{k_{t1}} \notag\\&\times\sum_{\ell_1=1}^2\left(
  {\bf R}_{\ell_1}'\left(k_{t1}\right) +\frac{\alpha_s(k_{t1})}{\pi}{\bf
  \Gamma}_{N_{\ell_1}}(\alpha_s(k_{t1})) + {\bf
                                                                                \Gamma}_{N_{\ell_1}}^{(C)}(\alpha_s(k_{t1}))\right)
                                                                                J_0(b
                                                                                k_{t1})\notag\\
&\times\exp\left\{ -  \sum_{\ell=1}^2\int_{k_{t1}}^{M}\frac{d k_t}{k_t}
  \left({\bf R}_{\ell}'\left(k_{t}\right) +  \frac{\alpha_s(k_t)}{\pi}{\bf
  \Gamma}_{N_\ell}(\alpha_s(k_t)) +  {\bf \Gamma}_{N_\ell}^{(C)}(\alpha_s(k_{t}))\right) J_0(b k_{t})\right\}\notag\\
&\times\exp\left\{ -  \sum_{\ell=1}^2\int_{\epsilon k_{t1}}^{M}\frac{d k_t}{k_t}
  \left({\bf R}_{\ell}'\left(k_{t}\right) +  \frac{\alpha_s(k_t)}{\pi}{\bf
  \Gamma}_{N_\ell}(\alpha_s(k_t)) +  {\bf
  \Gamma}_{N_\ell}^{(C)}(\alpha_s(k_{t}))\right) (1-J_0(b
  k_{t}))\right\}.
\end{align}
We finally notice that we can set $\epsilon\to 0$ in the above formula,
given that now the cancellation of divergences is manifest. The $k_{t1}$
integrand is a total derivative and it integrates to one, leaving
\begin{align}
\label{eq:master-differential-final}
\frac{d}{d p_t} &\hat{\bf \Sigma}_{N_1,N_2}^{c_1
  c_2}(p_t) = {\bf
             C}^{c_1; T}_{N_1}(\alpha_s(M)) H(M)
    {\bf C}^{c_2}_{N_2}(\alpha_s(M)) \,p_t\int\! \!b\, d b
                          J_0(p_t b)\notag\\
&\times\exp\left\{ -  \sum_{\ell=1}^2\int_{0}^{M}\frac{d k_t}{k_t}
  \left({\bf R}_{\ell}'\left(k_{t}\right) +  \frac{\alpha_s(k_t)}{\pi}{\bf
  \Gamma}_{N_\ell}(\alpha_s(k_t)) +  {\bf
  \Gamma}_{N_\ell}^{(C)}(\alpha_s(k_{t}))\right) (1-J_0(b
  k_{t}))\right\}.
\end{align}
We now insert the resulting partonic cross section back into the
definition of the hadronic cross section~\eqref{eq:hadxs}, and use the
second and third terms in the exponent of
Eq.~\eqref{eq:master-differential-final} to evolve the parton
densities and the coefficient functions down to $b_0/b$, with
$b_0=2e^{-\gamma_E}$. After performing the inverse Mellin transform,
and neglecting N$^4$LL corrections, we obtain (hereafter we simplify the
notation for the parton densities by omitting their $x_1$ and $x_2$
dependence, which is determined by the Born kinematics $\Phi_B$)
\begin{align}
\label{eq:hadxs-bspace-final}
  \frac{d^2\Sigma(v)}{d\Phi_Bd p_t} &=\sum_{c_1, c_2}\frac{d|M_B|_{c_1 c_2}^2}{d\Phi_B}\int\! \!b\, d b
  \,p_t  J_0(p_t b)\,
  {\bf f}^{T}(b_0/b) {\bf
  C}^{c_1; T}_{N_1}(\alpha_s(b_0/b)) H(M)
  {\bf C}^{c_2}_{N_2}(\alpha_s(b_0/b)) {\bf f}(b_0/b) \notag\\
&\times\exp\left\{ -  \sum_{\ell=1}^2\int_{0}^{M}\frac{d k_t}{k_t}
  {\bf R}_{\ell}'\left(k_{t} \right)(1-J_0(b
  k_{t}))\right\}.
\end{align}
Eq.~\eqref{eq:hadxs-bspace-final} represents indeed the $b$-space
formulation of transverse-momentum resummation. Commonly, it is expressed in the equivalent
form~\cite{Collins:1984kg}\footnote{This corresponds to a change of
  scheme of the type discussed in ref.~\cite{Catani:2003zt}.}
\begin{align}
\label{eq:hadxs-bspace-theta}
  \frac{d^2\Sigma(v)}{d\Phi_Bd p_t} &=\sum_{c_1, c_2}\frac{d|M_B|_{c_1 c_2}^2}{d\Phi_B}\int\! \!b\, d b
  \,p_t  J_0(p_t b)\,
  {\bf f}^{T}(b_0/b) {\bf
  C}^{c_1;T}_{N_1}(\alpha_s(b_0/b)) H_{\css}(M)
  {\bf C}^{c_2}_{N_2}(\alpha_s(b_0/b))  {\bf f}(b_0/b) \notag\\
&\times\exp\left\{ -  \sum_{\ell=1}^2\int_{0}^{M}\frac{d k_t}{k_t}
  {\bf R}_{\css, \ell}'\left(k_{t} \right)\Theta(k_t-\frac{b_0}{b})\right\}.
\end{align}
where ${\bf R}_{\css, \ell}'$ and $H_{\css}(M)$ are the Sudakov and
hard function commonly used for a $b$-space
formulation~\cite{Collins:1984kg}. As shown in
ref.~\cite{Catani:2010pd}, and as already stressed above, both
Eqs.~\eqref{eq:hadxs-bspace-final} and~\eqref{eq:hadxs-bspace-theta}
receive an extra contribution due to the azimuthal correlations which
are parametrised by the ${\bf G}$ coefficient functions. We omit them
in this comparison for the sake of simplicity, however it is clear
that analogous considerations apply in that case.
The comparison between Eqs.~\eqref{eq:hadxs-bspace-final}
and~\eqref{eq:hadxs-bspace-theta} allows us to extract the N$^3$LL
ingredients from the latter formulation as obtained in
refs.~\cite{Catani:2011kr,Catani:2012qa,Li:2016ctv,Vladimirov:2016dll}, that will be
reported in the next section.

We start by using the relation\footnote{See appendix of
  ref.~\cite{Banfi:2012jm} for a derivation.}
\begin{align}
(1-J_0(b k_{t})) \simeq  \Theta(k_t-\frac{b_0}{b}) +
  \frac{\zeta_3}{12}\frac{\partial^3}{\partial \ln(M b/b_0)^3}
  \Theta(k_t-\frac{b_0}{b}) + \dots,
\end{align}
where we ignored N$^4$LL terms. In the above formula the derivative in
the second term of the right-hand-side is meant to act on the integral
whose bounds are set by $\Theta(k_t-\frac{b_0}{b})$.
This yields, at N$^3$LL,
\begin{align}
\label{eq:hadxs-bspace}
  \frac{d^2\Sigma(v)}{d\Phi_Bd p_t} &=\sum_{c_1, c_2}\frac{d|M_B|_{c_1 c_2}^2}{d\Phi_B}\int\! \!b\, d b
  \,p_t  J_0(p_t b)\,
  {\bf f}^{T}(b_0/b) {\bf
  C}^{c_1;T}_{N_1}(\alpha_s(b_0/b)) H(M)
  {\bf C}^{c_2}_{N_2}(\alpha_s(b_0/b)) {\bf f}(b_0/b) \notag\\
&\times\exp\left\{ -  \sum_{\ell=1}^2\left(\int_{b_0/b}^{M}\frac{d k_t}{k_t}
  {\bf R}_{\ell}'\left(k_{t}\right) + \frac{\zeta_3}{12}\frac{\partial^3}{\partial \ln(M b/b_0)^3}\int_{b_0/b}^{M}\frac{d k_t}{k_t}
  {\bf R}_{\ell}'\left(k_{t}\right) \right)\right\}.
\end{align}
The second term in the exponent of Eq.~\eqref{eq:hadxs-bspace} starts
at N$^3$LL, so up to NNLL the two definitions (the one in terms of a
$J_0$ and the one in terms of the theta function) are manifestly
equivalent. To relate the two formulations we recall the definition of
${\bf R}'$ in Eq.~\eqref{eq:rp-array} and we express the Sudakov
radiators as~\eqref{eq:rad-mat}
\begin{align}
\label{eq:R-us}
R(b) &= \sum_{\ell=1}^2 \int_{b_0/b}^{M}\frac{d k_t}{k_t}
  R_{\ell}'\left(k_{t}\right)=\sum_{\ell=1}^2\int_{b_0/b}^{M}\frac{d
                    k_{t}}{k_t}
                    \left(A_{\ell}(\as(k_t))\ln\frac{M^2}{k_t^2} +
                    B_{\ell}(\as(k_t))\right) \notag\\
R_{\css}(b) &=\sum_{\ell=1}^2 \int_{b_0/b}^{M}\frac{d k_t}{k_t}
  {R}_{\css,\ell}'\left(k_{t}\right)=\sum_{\ell=1}^2\int_{b_0/b}^{M}\frac{d
                    k_{t}}{k_t}
                    \left(A_{\css,\ell}(\as(k_t))\ln\frac{M^2}{k_t^2} +
                    B_{\css,\ell}(\as(k_t))\right).
\end{align}
The anomalous dimensions $A_\ell$ and $B_{\ell}$ relative to leg
$\ell$ and the hard function $H$ admit an expansion in the strong
coupling as
\begin{equation}
A_\ell(\as)=\sum_{n=1}^{4}\left(\frac{\as}{2\pi}\right)^nA^{(n)}_{\ell},\,\quad
B_\ell(\as)=\sum_{n=1}^{3}\left(\frac{\as}{2\pi}\right)^nB^{(n)}_{\ell},\quad
H(M)=1 + \sum_{n=1}^{2}\left(\frac{\as(M)}{2\pi}\right)^nH^{(n)}(M).
\end{equation}
The relation between the coefficients that enter at N$^3$LL can be deducted by equating
Eqs.~\eqref{eq:hadxs-bspace-final} and~\eqref{eq:hadxs-bspace-theta},
obtaining
\begin{align}
\label{eq:our-A-B-H}
A_\ell^{(4)}&=A_{\css,\ell}^{(4)} -32 A^{(1)}_{\ell} \pi^3
  \beta_0^3 \zeta_3,\notag\\
B_\ell^{(3)}&=B_{\css,\ell}^{(3)} - 16 A^{(1)}_{\ell} \pi^2
  \beta_0^2 \zeta_3,\notag\\
H^{(2)}(M)&=H_{\css}^{(2)}(M)  + \frac{8}{3} \pi
  \beta_0 \zeta_3  \left(\frac{1}{2}\sum_{\ell=1}^{2}
  A^{(1)}_{\ell}\right) .
\end{align}
The above equations constitute the ingredients for our N$^3$LL
resummation. Physically, the extra terms proportional to $\zeta_3$
arise from the fact that the ${\cal O}(\alpha_s^2)$ terms proportional
to $\delta(1-z)$ in the coefficient functions in momentum space differ
from their $b$-space counterpart. This difference precisely amounts to
the new contributions in Eqs.~\eqref{eq:our-A-B-H}. We stress that
only the combination of $A_\ell^{(4)}$, $B_\ell^{(3)}$, $H^{(2)}$ and
$C^{(2)}$ is resummation-scheme invariant, hence our choice of
absorbing the new terms into $A_\ell^{(4)}$, $B_\ell^{(3)}$, $H^{(2)}$
is indeed arbitrary. One could analogously define an alternative
scheme in which the extra terms are directly absorbed into the ${\cal
  O}(\alpha_s^2)$ coefficient functions, thus leaving the two-loop
form factor unchanged.

\section{Evaluation up to N$^3$LL}
\label{sec:n3ll}

In this section we evaluate our all-order master formulae~\eqref{eq:hadxs} and~\eqref{eq:partxs-mellin} explicitly up to
N$^3$LL accuracy. The latter equations can already be evaluated as they are by means of Monte Carlo
techniques; however, at any given logarithmic order it is possible,
and convenient, to further manipulate them in order
to evaluate them {\it directly in momentum space}, without the need of the
Mellin transform.

\subsection{Momentum-space formulation}
\label{sec:mom-space}
We firstly focus on the partonic cross
section~\eqref{eq:partxs-mellin}. There are three main ingredients:
the Sudakov radiator and its derivative, the block containing
coefficient functions $C(\alpha_s)$ and hard-virtual corrections to
the form factor $H(\mu_R)$, and the anomalous dimensions that rule the
evolution of parton densities and coefficient functions.

For colour-singlet production, the coefficients entering the Sudakov
radiator satisfy $A^{(n)}_{1}=A^{(n)}_{2}=A^{(n)}$, and
$B^{(n)}_{1}=B^{(n)}_{2}=B^{(n)}$. Coefficients
$A^{(1)}$,~$A^{(2)}$,~$A^{(3)}$,~$B^{(1)}$,~$B^{(2)}$ have been known
for several years~\cite{deFlorian:2001zd,Davies:1984hs,Becher:2010tm},
and they are collected, for instance, in the appendix of
ref.~\cite{Banfi:2012jm}. The N$^3$LL coefficient $B^{(3)}$ can be
extracted from the recent result~\cite{Li:2016ctv,Vladimirov:2016dll}. For gluon
processes it reads:
\begin{align}
B^{(3)} &=
      C_{A}^3        \left(\frac{22 \zeta _3 \zeta _2}{3}-\frac{799 \zeta _2}{81}-\frac{5 \pi ^2 \zeta _3}{9}-\frac{2533 \zeta _3}{54}-\frac{77 \zeta _4}{12}+20 \zeta _5-\frac{319 \pi ^4}{1080}+\frac{6109 \pi ^2}{1944}+\frac{34219}{1944}\right)\notag\\
   &+ C_{A}^2 n_f    \left(\frac{103 \zeta _2}{81}+\frac{202 \zeta _3}{27}-\frac{5 \zeta _4}{6}+\frac{41 \pi ^4}{540}-\frac{599 \pi ^2}{972}-\frac{10637}{1944}\right)
    + C_{A} C_F n_f  \left(2 \zeta _4-\frac{\pi ^4}{45}-\frac{\pi ^2}{12}+\frac{241}{72}\right)\notag\\
   & -\frac{1}{4}C_{F}^2 n_f  
    + C_{A} n_{f}^2  \left(-\frac{2 \zeta _3}{27}+\frac{5 \pi ^2}{162}+\frac{529}{1944}\right)
    -\frac{11}{36}C_{F} n_{f}^2  
    - 32 C_A \pi^2\beta_0^2 \zeta_3\notag\\
  &\approx -492.908 - 32 C_A \pi^2\beta_0^2 \zeta_3 ,
\end{align}
while for quark processes
\begin{align}
B^{(3)} &=
       C_{A}^2 C_{F}  \left(\frac{22 \zeta _3 \zeta _2}{3}-\frac{799 \zeta _2}{81}-\frac{11 \pi ^2 \zeta _3}{9}+\frac{2207 \zeta _3}{54}-\frac{77 \zeta _4}{12}-10 \zeta _5-\frac{83 \pi ^4}{360}-\frac{7163 \pi ^2}{1944}+\frac{151571}{3888}\right)\notag\\
    &+ C_{F}^3        \left(\frac{4 \pi ^2 \zeta _3}{3}-17 \zeta _3+60 \zeta _5-\frac{2 \pi ^4}{5}-\frac{3 \pi ^2}{4}-\frac{29}{8}\right) + C_{F}^2 n_f    \left(\frac{34 \zeta _3}{3}+2 \zeta _4-\frac{7 \pi ^4}{54}-\frac{13 \pi ^2}{36}+\frac{23}{4}\right)
     \notag\\&+ C_{A} C_{F}^2  \left(-\frac{2}{3} \pi ^2 \zeta _3-\frac{211 \zeta _3}{3}-30 \zeta _5+\frac{247 \pi ^4}{540}+\frac{205 \pi ^2}{36}-\frac{151}{16}\right)\notag\\
    &
     + C_A C_F n_f    \left(\frac{103 \zeta _2}{81}-\frac{128 \zeta _3}{27}-\frac{5 \zeta _4}{6}+\frac{11 \pi ^4}{180}+\frac{1297 \pi ^2}{972}-\frac{3331}{243}\right)+ C_{F} n_{f}^2  \left(\frac{10 \zeta _3}{27}-\frac{5 \pi ^2}{54}+\frac{1115}{972}\right)\notag\\
&     - 32 C_F \pi^2\beta_0^2 \zeta_3\approx -116.685 - 32 C_F \pi^2\beta_0^2 \zeta_3.
\end{align}
The remaining N$^3$LL anomalous dimension $A^{(4)}$ is currently
incomplete given that the four-loop cusp anomalous dimension is still
unknown. Here we compute $A^{(4)}$ according to Eq.~(71) of
ref.~\cite{Becher:2010tm} or Eq.~(4.6) of ref.~\cite{Monni:2011gb},
using the results of refs.~\cite{Li:2016ctv,Vladimirov:2016dll} for the soft anomalous
dimension, and setting the four-loop cusp anomalous dimension to
zero. For gluon-initiated processes we get
\begin{align}
A^{(4)} &=
     C_{A}^{4}               \left(\frac{121}{3} \zeta_3 \zeta_2-\frac{8789 \zeta_2}{162}-\frac{19093 \zeta_3}{54}-\frac{847 \zeta_4}{24}+132 \zeta_5+\frac{3761815}{11664}\right)\notag\\
  &+ C_{A}^{3} n_{f}         \left(-\frac{22}{3} \zeta_3 \zeta_2+\frac{2731 \zeta_2}{162}+\frac{4955 \zeta_3}{54}+\frac{11 \zeta_4}{6}-24 \zeta_5-\frac{31186}{243}\right) \notag\\
  &+ C_{A}^{2} C_F n_{f}     \left(\frac{272 \zeta_3}{9}+11 \zeta_4-\frac{7351}{144}\right)
   + C_{A}^{2} n_{f}^{2}     \left(-\frac{103 \zeta_2}{81}-\frac{47 \zeta_3}{27}+\frac{5 \zeta_4}{6}+\frac{13819}{972}\right) \notag\\
  &+ C_{A} C_F n_{f}^{2}     \left(-\frac{38 \zeta_3}{9}-2 \zeta_4+\frac{215}{24}\right)
   + C_{A} n_{f}^{3}         \left(-\frac{4 \zeta_3}{9}-\frac{232}{729}\right)
   - 64 C_A \pi^3\beta_0^3 \zeta_3\notag\\
  &\approx -2675.68 - 64 C_A \pi^3\beta_0^3 \zeta_3, 
\end{align}
while for quark-initiated ones
\begin{align}
A^{(4)} &=
  C_{A}^{3} C_F\left(\frac{121}{3} \zeta_3 \zeta_2-\frac{8789 \zeta_2}{162}-\frac{19093 \zeta_3}{54}-\frac{847 \zeta_4}{24}+132 \zeta_5+\frac{3761815}{11664}\right) \notag\\
  &+ C_{A}^{2} C_{F} n_{f} \left(-\frac{22}{3} \zeta_3 \zeta_2+\frac{2731 \zeta_2}{162}+\frac{4955 \zeta_3}{54}+\frac{11 \zeta_4}{6}-24 \zeta_5-\frac{31186}{243}\right) \notag\\
  &+ C_{A} C_{F}^{2} n_{f} \left(\frac{272 \zeta_3}{9}+11 \zeta_4-\frac{7351}{144}\right)
   + C_{A} C_{F} n_{f}^{2} \left(-\frac{103 \zeta_2}{81}-\frac{47 \zeta_3}{27}+\frac{5 \zeta_4}{6}+\frac{13819}{972}\right) \notag\\
  &+ C_{F}^{2} n_{f}^{2}   \left(-\frac{38 \zeta_3}{9}-2 \zeta_4+\frac{215}{24}\right)
   + C_{F} n_{f}^{3}       \left(-\frac{4 \zeta_3}{9}-\frac{232}{729}\right)
   - 64 C_F \pi^3 \beta_0^3 \zeta_3\notag \\
  &\approx -1189.19 - 64 C_F \pi^3 \beta_0^3 \zeta_3.
\end{align}
We have left the additional terms arising from
Eq.~\eqref{eq:our-A-B-H} unexpanded to facilitate the comparison to
the existing literature. The remaining quantities are evaluated with
$n_f=5$.
The expression of the Sudakov radiator is analogous to the $b$-space one, i.e.
\begin{align}
\label{eq:R-momentum}
R(\epsilon k_{t1}) &= \sum_{\ell=1}^2 \int_{\epsilon k_{t1}}^{M}\frac{d k_t}{k_t}
  R_{\ell}'\left(k_{t}\right)=\sum_{\ell=1}^2\int_{\epsilon k_{t1}}^{M}\frac{d
                    k_{t}}{k_t}
                    \left(A_{\ell}(\as(k_t))\ln\frac{M^2}{k_t^2} +
                    B_{\ell}(\as(k_t))\right),
\end{align}
and, as above, we define $R'$ as the logarithmic derivative of $R$
\begin{equation}
R'_{\ell}(k_{t1}) = \frac{d R_\ell(k_{t1})}{d L},
\end{equation}
where we defined 
\begin{equation}
L=\ln\frac{M}{k_{t1}}.
\end{equation}

In order to make the numerical evaluation of our master formula
Eq.~\eqref{eq:partxs-mellin} more efficient, we can make a further
approximation on the integrand without spoiling the logarithmic
accuracy of the result. Before we describe the procedure in detail, we
stress that this additional manipulation is not strictly necessary and
one could in principle implement directly Eq.~\eqref{eq:partxs-mellin}
in a Monte-Carlo program.

Since the ratios $k_{ti}/k_{t1}$ for all {\it resolved} blocks are of
order 1, we can expand $R$ and its derivative about $k_{t1}$,
retaining terms that contribute at the desired logarithmic
accuracy. At N$^3$LL one has
\begin{align}
\label{eq:R-expanded}
R(\epsilon k_{t1}) &= R(k_{t1}) + R'(k_{t1})\ln\frac{1}{\epsilon} + \frac{1}{2!}R''(k_{t1})\ln^2\frac{1}{\epsilon} + \frac{1}{3!}R'''(k_{t1})\ln^3\frac{1}{\epsilon} + \dots \notag\\
R'(k_{ti}) &= R'(k_{t1}) + R''(k_{t1})\ln\frac{1}{\zeta_i} + \frac{1}{2!}R'''(k_{t1})\ln^2\frac{1}{\zeta_i}+\dots,
\end{align}
where the dots denote N$^4$LL terms, and we have employed the usual notation
$\zeta_i=k_{ti}/k_{t1}$.

We recall that the transverse momenta of blocks in the resolved
ensemble are parametrically of the same order. This is because rIRC
safety ensures that blocks $k$ with $k_t \ll k_{t1}$ do not contribute
to the observable and are encoded in the Sudakov radiator. Therefore,
since $\ln(1/\zeta_i)$ in the above formula is the logarithm of an
${\cal O}(1)$ quantity, each term in the right-hand-side of
Eq.~\eqref{eq:R-expanded} is logarithmically subleading with respect
to the one to its left.

The logarithms $\ln(1/\epsilon)$ in the first line of
Eq.~\eqref{eq:R-expanded} are a parametrisation of the IRC divergences
arising from the combination of real-unresolved blocks and virtual
corrections, expanded at a given logarithmic order. The $\epsilon$
dependence exactly cancels against the corresponding terms in the
resolved real corrections (denoted by the same-order derivative of
$R$) upon integration over $\zeta_i$, as it will be shown below. This
is a convenient way to recast the subtraction of IRC divergences at
each logarithmic order in our formulation.

The terms proportional to $R'(k_{t1})$ are to be retained starting at
NLL, those proportional to $R''(k_{t1})$ contribute at NNLL and,
finally, the ones proportional to $R'''(k_{t1})$ are needed at
N$^3$LL. Starting from the NLL ensemble, we note that correcting a
single block with respect to its $R'(k_{t1})$ approximation
(i.e.~including for that block the subleading terms of
Eq.~\eqref{eq:R-expanded}) gives rise at most to a NNLL correction of
order ${\cal O}(\as^nL^{n-1})$ in our counting. Modifying two blocks
would lead to a relative correction of order ${\cal O}(\as^nL^{n-2})$,
i.e. N$^3$LL, and so on. Therefore, at any given logarithmic order, it
is sufficient to keep terms beyond the $R'(k_{t1})$ approximation only
for a finite number of blocks (namely a single block at NNLL, two
blocks at N$^3$LL, and so forth). Consistently, one has to expand out
the corresponding terms in the Sudakov that cancel the $\epsilon$
divergences of the modified real blocks to the given logarithmic
order. This prescription has been derived and discussed in detail at
NNLL in ref.~\cite{Banfi:2014sua}, and will be used later in this
section.

As a next step we address the evolution of the parton densities and
relative coefficient functions encoded in Eq.~\eqref{eq:partxs-mellin},
whose anomalous dimensions $\Gamma_N$ and $\Gamma_N^{(C)}$ have been
defined in Eqs.~\eqref{eq:DGLAP}, and~\eqref{eq:RGE}. Only a finite
number of terms in their perturbative series needs to be retained at a
given logarithmic accuracy: in particular, contributions from the
${\cal O}(\as^n)$ term in $\Gamma_N$ enter for a N$^{n+1}$LL
resummation (we recall that the series of $\Gamma_N$ starts at ${\cal
  O}(\as^0)$, hence these terms start contributing at NLL). On the
other hand, the contribution of the coefficient functions, and
therefore of the corresponding anomalous dimension, starts at
NNLL. Therefore the ${\cal O}(\as^m)$ term in $\Gamma_N^{(C)}$ is
necessary at N$^{m+1}$LL, since its expansion starts at ${\cal
  O}(\as)$.

We can then perform the same expansion about $k_{t1}$ for the terms in
Eq.~\eqref{eq:partxs-mellin} containing ${\bf \Gamma}$ and ${\bf
  \Gamma}^{(C)}$.  Up to N$^3$LL we expand the exponent of the
evolution operators as
\begin{align}
\int_{\epsilon k_{t1}}^{\mu_0}\frac{dk_t}{k_t}\frac{\alpha_s(k_t)}{\pi}{\bf
  \Gamma}_{N_\ell}(\alpha_s(k_t)) &= \int_{k_{t1}}^{\mu_0}\frac{dk_t}{k_t}\frac{\alpha_s(k_t)}{\pi}{\bf
  \Gamma}_{N_\ell}(\alpha_s(k_t)) + \frac{d}{d L}\int_{k_{t1}}^{\mu_0}\frac{dk_t}{k_t}\frac{\alpha_s(k_t)}{\pi}{\bf
  \Gamma}_{N_\ell}(\alpha_s(k_t))\ln\frac{1}{\epsilon} \notag\\
\label{eq:DGLAP-SUD}
&+ \frac{1}{2}\frac{d^2}{d L^2}\int_{k_{t1}}^{\mu_0}\frac{dk_t}{k_t}\frac{\alpha_s(k_t)}{\pi}{\bf
  \Gamma}_{N_\ell}(\alpha_s(k_t))\ln^2\frac{1}{\epsilon}+\dots\\
\label{eq:RGE-SUD}
\int_{\epsilon k_{t1}}^{\mu_0}\frac{dk_t}{k_t}{\bf
  \Gamma}_{N_\ell}^{(C)}(\alpha_s(k_t)) &= \int_{k_{t1}}^{\mu_0}\frac{dk_t}{k_t}{\bf
  \Gamma}_{N_\ell}^{(C)}(\alpha_s(k_t)) + \frac{d}{d L}\int_{k_{t1}}^{\mu_0}\frac{dk_t}{k_t}{\bf
  \Gamma}_{N_\ell}^{(C)}(\alpha_s(k_t))\ln\frac{1}{\epsilon} +\dots,
\end{align}
and the corresponding resolved real-emission kernels as
\begin{align}
\label{eq:DGLAP-real}
\frac{\alpha_s(k_{tj})}{\pi}{\bf
  \Gamma}_{N_\ell}(\alpha_s(k_{tj})) &= \frac{\alpha_s(k_{t1})}{\pi}{\bf
  \Gamma}_{N_\ell}(\alpha_s(k_{t1})) + \frac{d}{d L}\frac{\alpha_s(k_{t1})}{\pi}{\bf
  \Gamma}_{N_\ell}(\alpha_s(k_{t1}))\ln\frac{1}{\zeta_j} +\dots\\
\label{eq:RGE-real}
{\bf  \Gamma}_{N_\ell}^{(C)}(\alpha_s(k_{tj})) &= {\bf
  \Gamma}_{N_\ell}^{(C)}(\alpha_s(k_{t1})) +\dots,
\end{align}
where as usual $L=\ln(M/k_{t1})$. The first terms on the right-hand
side of Eqs.~\eqref{eq:DGLAP-SUD}, and~\eqref{eq:RGE-SUD} represent
the evolution operator that runs the parton densities and the
coefficient functions, respectively, from $\mu_0$ up to $k_{t1}$. The
remaining terms describe the exclusive evolution of the parton
densities and of the coefficient functions in the resolved strip. In
particular, the $\epsilon$-dependent terms completely cancel against
the corresponding terms in the real-emission kernel of
Eqs.~\eqref{eq:DGLAP-real}, and~\eqref{eq:RGE-real} upon integration
over the resolved-radiation phase space.

At NLL the coefficient functions are an identity matrix in flavour
space, and therefore their evolution operator is trivial. The
contribution of the ${\bf \Gamma}_N$ in the exponent starts at NLL, while
the exclusive evolution of the parton densities in the resolved strip
starts at NNLL since it corresponds to emissions in the hard-collinear
edge of the phase space. Therefore, at NLL one only needs to retain
the first term in the right-hand side of Eq.~\eqref{eq:DGLAP-SUD}, and
ignore everything else in
Eqs.~\eqref{eq:DGLAP-SUD},~\eqref{eq:RGE-SUD},~\eqref{eq:DGLAP-real}, and~\eqref{eq:RGE-real},
which corresponds to evaluating the parton densities at
$\mu_F=k_{t1}$. At this order, the evolution can be carried out by
means of the tree-level anomalous dimension $\gamma^{(0)}_N$.

Similarly, at NNLL one needs to take into account the second term in
the r.h.s.~of Eq.~\eqref{eq:DGLAP-SUD} and the first term in the
r.h.s.~of Eq.~\eqref{eq:DGLAP-real}, where now the anomalous dimension
${\bf \Gamma}_N$ is evaluated at one-loop accuracy (i.e.~including
$\gamma_N^{(1)}$). At this order also the coefficient functions start
contributing with their inclusive evolution, therefore one needs to
add the first term in the r.h.s.~of Eq.~\eqref{eq:RGE-SUD}. The
corresponding exclusive evolution of the coefficient functions in the
resolved strip, encoded in the r.h.s.~of Eq.~\eqref{eq:RGE-real} only
starts at N$^3$LL. At higher orders, one simply needs to add
subsequent terms from the above equations, and evaluate the anomalous
dimensions at the appropriate perturbative accuracy.

As discussed above for the Sudakov radiator, at any given logarithmic order beyond
NLL, it is sufficient to include the extra $\epsilon$-dependent terms
from Eqs.~\eqref{eq:DGLAP-SUD},~\eqref{eq:RGE-SUD} in the exponent, and
the corresponding terms in the resolved real radiation from
Eqs.~\eqref{eq:DGLAP-real},~\eqref{eq:RGE-real} only for a finite
number of emissions, namely a single emission
at NNLL, two emissions at N$^3$LL, and so forth.

Finally, we need to deal with the block
${\bf C}^{c_1;T}_{N_1}(\alpha_s(\mu_0)) H(\mu_R) {\bf
  C}^{c_2}_{N_2}(\alpha_s(\mu_0))$
in Eq.~\eqref{eq:partxs-mellin}. As discussed in the previous section,
for a generic process this block receives a contribution from the
gluon collinear correlations ${\bf G}$, as in Eq.~\eqref{eq:RGE-G}.
Since the contribution of the ${\bf G}$ functions starts at N$^3$LL,
at this order one can drop the $\epsilon$ dependence in their
evolution; namely, in the analogue of Eq.~\eqref{eq:RGE-SUD} with
${\bf \Gamma}_N^{(C)}\to{\bf \Gamma}_N^{(G)}$, only the first term on
the right-hand side needs to be retained. This amounts to evaluating
the coupling of the ${\bf G}$ coefficient functions at $k_{t1}$.

With the expansions detailed above, Eq.~\eqref{eq:partxs-mellin} becomes
\begin{align}
\label{eq:partxs-mellin-expanded}
&\hat{\bf \Sigma}_{N_1,N_2}^{c_1
  c_2}(v)   = {\bf
             C}^{c_1;T}_{N_1}(\alpha_s(\mu_0)) H(\mu_R)
    {\bf C}^{c_2}_{N_2}(\alpha_s(\mu_0)) \,\int_0^{M}\frac{d k_{t1}}{k_{t1}} \int_0^{2\pi}
                     \frac{d\phi_1}{2\pi}\notag\\&\times
                                                   e^{-{\bf R}(k_{t1})
                                                   - {\bf
                                                   R}'(k_{t1})\ln\frac{1}{\epsilon}
                                                   - \frac{1}{2!}{\bf
                                                   R}''(k_{t1})\ln^2\frac{1}{\epsilon}
                                                   -  \frac{1}{3!}{\bf
                                                   R}'''(k_{t1})\ln^3\frac{1}{\epsilon}+
                                                   \dots}\notag\\
&\times\exp\left\{-\sum_{\ell=1}^{2}\left(
\int_{k_{t1}}^{\mu_0}\frac{dk_t}{k_t}\frac{\alpha_s(k_t)}{\pi}{\bf
  \Gamma}_{N_\ell}(\alpha_s(k_t)) + \frac{d}{d L}\int_{k_{t1}}^{\mu_0}\frac{dk_t}{k_t}\frac{\alpha_s(k_t)}{\pi}{\bf
  \Gamma}_{N_\ell}(\alpha_s(k_t))\ln\frac{1}{\epsilon}\right.\right. \notag\\
& \left.\left.\hspace{10mm}+ \frac{1}{2!}\frac{d^2}{d L^2}\int_{k_{t1}}^{\mu_0}\frac{dk_t}{k_t}\frac{\alpha_s(k_t)}{\pi}{\bf \Gamma}_{N_\ell}(\alpha_s(k_t))\ln^2\frac{1}{\epsilon}+\dots\right.\right. \notag\\
& \left. \hspace{10mm}+ \int_{k_{t1}}^{\mu_0}\frac{dk_t}{k_t} {\bf \Gamma}_{N_\ell}^{(C)}(\alpha_s(k_{t})) + \frac{d}{d L}\int_{k_{t1}}^{\mu_0}\frac{dk_t}{k_t} {\bf \Gamma}_{N_\ell}^{(C)}(\alpha_s(k_{t}))\ln\frac{1}{\epsilon} + \dots\right)\Bigg\}\notag\\
                     &\times\sum_{\ell_1=1}^2\left(
  {\bf R}_{\ell_1}'\left(k_{t1}\right) + \frac{\alpha_s(k_{t1})}{\pi}{\bf
  \Gamma}_{N_{\ell_1}}(\alpha_s(k_{t1})) + {\bf \Gamma}_{N_{\ell_1}}^{(C)}(\alpha_s(k_{t1}))\right) \notag\\
&\times\sum_{n=0}^{\infty}\frac{1}{n!}
                                                     \prod_{i=2}^{n+1}
                                                     \int_{\epsilon}^{1}\frac{d\zeta_i}{\zeta_i}\int_0^{2\pi}
                                                     \frac{d\phi_i}{2\pi}
  \sum_{\ell_i=1}^2\Bigg\{ {\bf R}_{\ell_i}'\left(k_{t1}\right) + {\bf
  R}_{\ell_i}''\left(k_{t1}\right)\ln\frac{1}{\zeta_i} +
  \frac{1}{2}{\bf
  R}_{\ell_i}'''\left(k_{t1}\right)\ln^2\frac{1}{\zeta_i} + \dots
  \notag\\
&+\frac{\alpha_s(k_{t1})}{\pi}{\bf
 \Gamma}_{N_{\ell_i}}(\alpha_s(k_{t1})) + \frac{d}{d L}\left(\frac{\alpha_s(k_{t1})}{\pi}{\bf
 \Gamma}_{N_{\ell_i}}(\alpha_s(k_{t1}))\right) \ln\frac{1}{\zeta_i}+\dots \notag\\
                                                    &  + {\bf \Gamma}_{N_{\ell_i}}^{(C)}(\alpha_s(k_{t1})) + \dots\Bigg\}\Theta\left(v-V(\{\tilde{p}\},k_1,\dots, k_{n+1})\right) + \{ {\bf C} \to  {\bf G}; {\bf \Gamma}^{(C)} \to {\bf \Gamma}^{(G)}\}\,.
\end{align}
Following the procedure of ref.~\cite{Banfi:2014sua}, we can express the $\ln (1/\epsilon)$ singularities in the exponent of Eq.~\eqref{eq:partxs-mellin-expanded} as integrals over dummy real emissions as follows
\begin{align}
\ln\frac{1}{\epsilon} = \int_{\epsilon}^{1}\frac{d\zeta}{\zeta},\qquad\quad
\frac{1}{2}\ln^2\frac{1}{\epsilon} =
                                     \int_{\epsilon}^{1}\frac{d\zeta}{\zeta}\ln\frac{1}{\zeta},\qquad\quad
\frac{1}{3!}\ln^3\frac{1}{\epsilon} = \frac{1}{2}\int_{\epsilon}^{1}\frac{d\zeta}{\zeta}\ln^2\frac{1}{\zeta},
\end{align}
and subsequently expand out the divergent part of the exponent,
retaining the terms necessary at a given logarithmic order. We further
introduce the average of a function $G(\{\tilde p\},\{k_i\})$ over the
measure $d {\cal Z}$
\begin{equation}
\label{eq:dZ}
\begin{split}
\int \dZ  G(\{\tilde p\},\{k_i\})=\epsilon^{R'(k_{t1})}
   \sum_{n=0}^{\infty}\frac{1}{n!} \prod_{i=2}^{n+1}
    \int_{\epsilon}^{1} \frac{d\zeta_i}{\zeta_i}\int_0^{2\pi}
   \frac{d\phi_i}{2\pi} R'(k_{t1})G(\{\tilde p\},k_1,\dots,k_{n+1})\,,
\end{split}
\end{equation}
where we simplified the notation by using
\begin{equation}
R' (k_{t1})=\sum_{\ell=1,2} R'_{\ell}(k_{t1}).
\end{equation}
The dependence on the regulator $\epsilon$ cancels exactly in
Eq.~\eqref{eq:dZ}.  \\

We can plug Eq.~\eqref{eq:partxs-mellin-expanded} into the definition
of the hadronic cross section~\eqref{eq:hadxs}. We define the
derivatives of the parton densities by means of the DGLAP evolution
equation
\begin{equation}
\label{eq:dglap}
\frac{\partial f(\mu, x)}{\partial \ln \mu} =
\frac{\alpha_s(\mu)}{\pi} \int_x^1\frac{d z}{z}
\hat{P}(z,\alpha_s(\mu)) f(\mu,\frac{x}{z}),
\end{equation}
where $\hat{P}(z,\alpha_s(\mu)) $ is the regularised splitting
function
\begin{equation}
\hat{P}(z,\alpha_s(\mu))  = \hat{P}^{(0)}(z) + \frac{\alpha_s(\mu)}{2\pi}
\hat{P}^{(1)}(z) + \left(\frac{\alpha_s(\mu)}{2\pi}\right)^2
\hat{P}^{(2)}(z) + \dots
\end{equation}
Moreover, we introduce the following parton luminosities
\begin{align} 
\label{eq:luminosity-NLL0}
{\cal L}_{\rm NLL}(k_{t1}) = \sum_{c, c'}\frac{d|M_{B}|_{cc'}^2}{d\Phi_B} f_c\!\left(k_{t1},x_1\right)f_{c'}\!\left(k_{t1},x_2\right),
\end{align}
\begin{align}
\label{eq:luminosity-NNLL0}
&{\cal L}_{\rm NNLL}(k_{t1}) = \sum_{c, c'}\frac{d|M_{B}|_{cc'}^2}{d\Phi_B} \sum_{i, j}\int_{x_1}^{1}\frac{d z_1}{z_1}\int_{x_2}^{1}\frac{d z_2}{z_2}f_i\!\left(k_{t1},\frac{x_1}{z_1}\right)f_{j}\!\left(k_{t1},\frac{x_2}{z_2}\right)\notag\\&\Bigg(\delta_{ci}\delta_{c'j}\delta(1-z_1)\delta(1-z_2)
\left(1+\frac{\alpha_s(\mu_R)}{2\pi} H^{(1)}(\mu_R)\right) \notag\\
&+ \frac{\alpha_s(\mu_R)}{2\pi}\frac{1}{1-2\alpha_s(\mu_R)\beta_0
  L}\left(C_{c i}^{(1)}(z_1)\delta(1-z_2)\delta_{c'j}+
  \{z_1\leftrightarrow z_2; c,i \leftrightarrow c'j\}\right)\Bigg),
\end{align}
\begin{align}
\label{eq:luminosity-N3LL0}
&{\cal L}_{\rm N^3LL}(k_{t1})=\sum_{c,
  c'}\frac{d|M_{B}|_{cc'}^2}{d\Phi_B} \sum_{i, j}\int_{x_1}^{1}\frac{d z_1}{z_1}\int_{x_2}^{1}\frac{d z_2}{z_2}f_i\!\left(k_{t1},\frac{x_1}{z_1}\right)f_{j}\!\left(k_{t1},\frac{x_2}{z_2}\right)\notag\\&\Bigg\{\delta_{ci}\delta_{c'j}\delta(1-z_1)\delta(1-z_2)
\left(1+\frac{\alpha_s(\mu_R)}{2\pi} H^{(1)}(\mu_R) + \frac{\alpha^2_s(\mu_R)}{(2\pi)^2} H^{(2)}(\mu_R)\right) \notag\\
&+ \frac{\alpha_s(\mu_R)}{2\pi}\frac{1}{1-2\alpha_s(\mu_R)\beta_0 L}\left(1- \alpha_s(\mu_R)\frac{\beta_1}{\beta_0}\frac{\ln\left(1-2\alpha_s(\mu_R)\beta_0 L\right)}{1-2\alpha_s(\mu_R)\beta_0 L}\right)\notag\\
&\times\left(C_{c i}^{(1)}(z_1)\delta(1-z_2)\delta_{c'j}+ \{z_1\leftrightarrow z_2; c,i \leftrightarrow c',j\}\right)\notag\\
& +
  \frac{\alpha^2_s(\mu_R)}{(2\pi)^2}\frac{1}{(1-2\alpha_s(\mu_R)\beta_0
  L)^2}\Bigg(\left(C_{c i}^{(2)}(z_1) - 2\pi\beta_0 C_{c i}^{(1)}(z_1) \ln\frac{M^2}{\mu_R^2}\right)\delta(1-z_2)\delta_{c'j} \notag\\&+ \{z_1\leftrightarrow z_2; c,i \leftrightarrow c',j\}\Bigg) +  \frac{\alpha^2_s(\mu_R)}{(2\pi)^2}\frac{1}{(1-2\alpha_s(\mu_R)\beta_0 L)^2}\Big(C_{c i}^{(1)}(z_1)C_{c' j}^{(1)}(z_2) + G_{c i}^{(1)}(z_1)G_{c' j}^{(1)}(z_2)\Big) \notag\\
& + \frac{\alpha^2_s(\mu_R)}{(2\pi)^2} H^{(1)}(\mu_R)\frac{1}{1-2\alpha_s(\mu_R)\beta_0 L}\Big(C_{c i}^{(1)}(z_1)\delta(1-z_2)\delta_{c'j} + \{z_1\leftrightarrow z_2; c,i \leftrightarrow c',j\}\Big) \Bigg\},
\end{align}
where 
\begin{equation}
x_{1}=\frac{M}{\sqrt{s}} e^{Y},\qquad x_2=\frac{M}{\sqrt{s}} e^{-Y},
\end{equation}
and $Y$ is the rapidity of the colour singlet in the centre-of-mass
frame of the collision at the Born level. $|M_{B}|_{cc'}^2$ is the
Born squared matrix element, and $L=\ln(1/v_1)$, with $v_1=k_{t1}/M$,
$v=p_t/M$. We transform back to momentum space, thus abandoning the
matrix notation used so far, by means of the following identities,
valid up to N$^3$LL
\begin{align}
&\frac{d|M_B|_{c_1c_2}^2}{d\Phi_B}{\bf f}^{T}_{N_1}(k_{t1})\left(\sum_{\ell=1}^2\frac{\alpha_s(k_{t1})}{\pi}{\bf
  \Gamma}_{N_{\ell}}(\alpha_s(k_{t1}))\right) {\bf f}_{N_2}(k_{t1}) \notag\\
&\hspace{7cm}\to \frac{\alpha_s(k_{t1})}{\pi}\hat{P}(z,\alpha_s(k_{t1}))\otimes {\cal L}_{\rm
  NLL}(k_{t1})=-\partial_L {\cal L}_{\rm
  NLL}(k_{t1})\notag\\
&\frac{d|M_B|_{c_1c_2}^2}{d\Phi_B}{\bf f}^{T}_{N_1}(k_{t1}){\bf
             C}^{c_1;T}_{N_1}(\alpha_s(k_{t1})) H(\mu_R)
    \left(\sum_{\ell=1}^2\left(\frac{\alpha_s(k_{t1})}{\pi}{\bf
  \Gamma}_{N_{\ell}}(\alpha_s(k_{t1}))\right.\right.\notag\\
&\hspace{7cm}\left.\left.+ {\bf \Gamma}_{N_{\ell}}^{(C)}(\alpha_s(k_{t1}))\right)\right){\bf C}^{c_2}_{N_2}(\alpha_s(k_{t1})) {\bf f}_{N_2}(k_{t1})
\to-\partial_L {\cal L}(k_{t1})\notag\\
&\frac{d|M_B|_{c_1c_2}^2}{d\Phi_B}{\bf f}^{T}_{N_1}(k_{t1})\left(\sum_{\ell=1}^2\frac{d}{d L}\left(\frac{\alpha_s(k_{t1})}{\pi}{\bf
 \Gamma}_{N_{\ell}}(\alpha_s(k_{t1}))\right)\right) {\bf f}_{N_2}(k_{t1}) \to 2\frac{\beta_0}{\pi}\alpha_s^2(k_{t1}) \hat{P}^{(0)}\otimes {\cal L}_{\rm
  NLL}(k_{t1}) \notag\\
&\frac{d|M_B|_{c_1c_2}^2}{d\Phi_B}{\bf f}^{T}_{N_1}(k_{t1})\left(\sum_{\ell_i=1}^2\frac{\alpha_s(k_{t1})}{\pi}{\bf
  \Gamma}_{N_{\ell_i}}(\alpha_s(k_{t1}))\right)\left(\sum_{\ell_j=1}^2\frac{\alpha_s(k_{t1})}{\pi}{\bf
  \Gamma}_{N_{\ell_j}}(\alpha_s(k_{t1}))\right) {\bf f}_{N_2}(k_{t1}) \to\notag\\
&\hspace{1cm}\to \frac{\alpha^2_s(k_{t1})}{\pi^2}\hat{P}(z,\alpha_s(k_{t1}))\otimes \hat{P}(z,\alpha_s(k_{t1}))\otimes {\cal L}_{\rm
  NLL}(k_{t1}) \simeq  \frac{\alpha^2_s(k_{t1})}{\pi^2}\hat{P}^{(0)}\otimes \hat{P}^{(0)}\otimes {\cal L}_{\rm
  NLL}(k_{t1})
\end{align}
where we defined $\partial_L = d/d L$. Since we evaluated explicitly the sum over the emitting legs $\ell_i$,
the convolution of a regularised splitting kernel $\hat{P}^{(0)}$ with
the NLL parton luminosity is now defined as
\begin{align} 
\label{eq:Pluminosity-NLL}
\hat{P}^{(0)}\otimes{\cal L}_{\rm NLL}(k_{t1}) &\equiv \sum_{c,
  c'}\frac{d|M_{B}|_{cc'}^2}{d\Phi_B} \bigg\{\left(\hat{P}^{(0)}\otimes
  f\right)_c\left(k_{t1},x_1\right)f_{c'}\!\left(k_{t1},x_2\right)
  \notag\\
&\hspace{5cm}+
  f_c \!\left(k_{t1},x_1\right) \left(\hat{P}^{(0)}\otimes f\right)_{c'}\left(k_{t1},x_2\right) \bigg\}.
\end{align}
The term $\hat{P}^{(0)}\otimes \hat{P}^{(0)}\otimes {\cal L}_{\rm
  NLL}(k_{t1})$ is to be interpreted as
\begin{align} 
\hat{P}^{(0)}\otimes \hat{P}^{(0)}&\otimes{\cal L}_{\rm NLL}(k_{t1}) \equiv \sum_{c,
  c'}\frac{d|M_{B}|_{cc'}^2}{d\Phi_B} \bigg\{\left(\hat{P}^{(0)}\otimes \hat{P}^{(0)}\otimes
  f\right)_c\left(k_{t1},x_1\right)f_{c'}\!\left(k_{t1},x_2\right)
  \notag\\
&+
  f_c \!\left(k_{t1},x_1\right) \left(\hat{P}^{(0)}\otimes
  \hat{P}^{(0)}\otimes f\right)_{c'}\left(k_{t1},x_2\right)  + 2\left(\hat{P}^{(0)}\otimes f\right)_c \!\left(k_{t1},x_1\right) \left(\hat{P}^{(0)}\otimes f\right)_{c'}\left(k_{t1},x_2\right) \bigg\}.
\end{align}

\noindent Including terms up to N$^3$LL, we can therefore recast
Eqs.~\eqref{eq:partxs-mellin-expanded},~\eqref{eq:hadxs}
as
\begin{align}
\label{eq:master-kt-space}
&\frac{d\Sigma(v)}{d\Phi_B} = \int\frac{d k_{t1}}{k_{t1}}\frac{d
  \phi_1}{2\pi}\partial_{L}\left(-e^{-R(k_{t1})} {\cal L}_{\rm
  N^3LL}(k_{t1}) \right) \int \dZ\Theta\left(v-V(\{\tilde{p}\},k_1,\dots, k_{n+1})\right)
                             \notag\\\notag\\
& + \int\frac{d k_{t1}}{k_{t1}}\frac{d
  \phi_1}{2\pi} e^{-R(k_{t1})} \int \dZ\int_{0}^{1}\frac{d \zeta_{s}}{\zeta_{s}}\frac{d
  \phi_s}{2\pi}\Bigg\{\bigg(R' (k_{t1}) {\cal L}_{\rm
  NNLL}(k_{t1}) - \partial_L {\cal L}_{\rm
  NNLL}(k_{t1})\bigg)\notag\\
&\times\left(R'' (k_{t1})\ln\frac{1}{\zeta_s} +\frac{1}{2} R'''
  (k_{t1})\ln^2\frac{1}{\zeta_s} \right) - R' (k_{t1})\left(\partial_L {\cal L}_{\rm
  NNLL}(k_{t1}) - 2\frac{\beta_0}{\pi}\alpha_s^2(k_{t1}) \hat{P}^{(0)}\otimes {\cal L}_{\rm
  NLL}(k_{t1}) \ln\frac{1}{\zeta_s}
\right)\notag\\
&+\frac{\alpha_s^2(k_{t1}) }{\pi^2}\hat{P}^{(0)}\otimes \hat{P}^{(0)}\otimes {\cal L}_{\rm
  NLL}(k_{t1})\Bigg\} \bigg\{\Theta\left(v-V(\{\tilde{p}\},k_1,\dots,
  k_{n+1},k_s)\right) - \Theta\left(v-V(\{\tilde{p}\},k_1,\dots,
  k_{n+1})\right)\bigg\}\notag\\\notag\\
& + \frac{1}{2}\int\frac{d k_{t1}}{k_{t1}}\frac{d
  \phi_1}{2\pi} e^{-R(k_{t1})} \int \dZ\int_{0}^{1}\frac{d \zeta_{s1}}{\zeta_{s1}}\frac{d
  \phi_{s1}}{2\pi}\int_{0}^{1}\frac{d \zeta_{s2}}{\zeta_{s2}}\frac{d
  \phi_{s2}}{2\pi} R' (k_{t1})\notag\\
&\times\Bigg\{ {\cal L}_{\rm
  NLL}(k_{t1}) \left(R'' (k_{t1})\right)^2\ln\frac{1}{\zeta_{s1}} \ln\frac{1}{\zeta_{s2}} - \partial_L {\cal L}_{\rm
  NLL}(k_{t1}) R'' (k_{t1})\bigg(\ln\frac{1}{\zeta_{s1}}
  +\ln\frac{1}{\zeta_{s2}} \bigg)\notag\\
&+ \frac{\alpha_s^2(k_{t1}) }{\pi^2}\hat{P}^{(0)}\otimes \hat{P}^{(0)}\otimes {\cal L}_{\rm
  NLL}(k_{t1})\Bigg\}\notag\\
&\times \bigg\{\Theta\left(v-V(\{\tilde{p}\},k_1,\dots,
  k_{n+1},k_{s1},k_{s2})\right) - \Theta\left(v-V(\{\tilde{p}\},k_1,\dots,
  k_{n+1},k_{s1})\right) -\notag\\ &\Theta\left(v-V(\{\tilde{p}\},k_1,\dots,
  k_{n+1},k_{s2})\right) + \Theta\left(v-V(\{\tilde{p}\},k_1,\dots,
  k_{n+1})\right)\bigg\} + {\cal O}\left(\alpha_s^n \ln^{2n -
                                    6}\frac{1}{v}\right).
\end{align}
Until now we have explicitly considered the case of flavour-conserving
real emissions, for which we derived
Eq.~\eqref{eq:master-kt-space}. We now turn to the inclusion of the
flavour-changing splitting kernels, that enter purely in the
hard-collinear limit and contribute to the DGLAP evolution.

We observe that at a given logarithmic order only a finite number of
hard-collinear emissions are actually necessary. As we mentioned
several times in the above sections, at N$^3$LL one needs to account
for the effect of up to two hard-collinear resolved
partons. Therefore, the inclusion of the flavour-changing kernels can
be done directly at the level of the splitting functions and parton
luminosities in Eq.~\eqref{eq:master-kt-space}.

In the above expressions for the luminosity we have used the following
expansions in powers of the strong coupling for the functions $C$, $H$
and $G$, up to N$^3$LL:
\begin{align}
\label{eq:coeff-fun}
C_{ab}(\alpha_s(\mu)) &= \delta(1-z)\delta_{ab} + \sum_{n=1}^{2}\left(\frac{\alpha_s(\mu)}{2\pi}\right)^n\,C_{ab}^{(n)}(z),\\
H(\mu_R) &= 1+\sum_{n=1}^{2}\left(\frac{\alpha_s(\mu_R)}{2\pi}\right)^n\,H^{(n)}(\mu_R),\\
G_{ab}(\alpha_s(\mu)) &= \frac{\alpha_s(\mu)}{2\pi}\,G_{ab}^{(1)}(z),
\end{align}
where $\mu$ is the same scale at which the parton densities are
evaluated, and $\mu_R$ is the renormalisation scale.

The expressions for $C^{(1)}$ and $H^{(1)}$ have been known for a long
time, and are collected, for instance, in the appendix of
ref.~\cite{Banfi:2012jm}.  The hard-virtual coefficient $H(\mu_R)$ is
defined as the finite part of the renormalised QCD form factor in the
$\overline{\rm MS}$ renormalisation scheme, divided by the underlying
Born squared matrix element.
The hard coefficients for gluonic processes up to ${\cal O}(\as^2)$
evaluated at the invariant mass of the colour singlet $H^{(1)}(M)$ and
$H^{(2)}(M)$
read~\cite{Kramer:1996iq,Chetyrkin:1997iv,Harlander:2000mg}
\begin{align}
\label{eq:H-fun-G}
H_g^{(1)}(M) &= C_A\left(5+\frac{7}{6}\pi^2\right)-3 C_F,\notag\\
H_g^{(2)}(M) &=  \frac{5359}{54} + \frac{137}{6}\ln\frac{m_H^2}{m_t^2} 
+ \frac{1679}{24}\pi^2 + \frac{37}{8}\pi^4- \frac{499}{6}\zeta_3
               + C_A \frac{16}{3} \pi
  \beta_0 \zeta_3 \,,\qquad n_f=5,
\end{align}
where the last term in $H_g^{(2)}$ was deliberately left symbolic to
stress its origin from Eq.~\eqref{eq:our-A-B-H}. Analogously, for
quark-initiated reactions one
has~\cite{Kramer:1986sg,Matsuura:1988sm,Gehrmann:2005pd}
\begin{align}
\label{eq:H-fun-Q}
H_q^{(1)}(M) &= C_F\left( -8 + \frac{7}{6}\pi^2 \right),\notag\\
H_q^{(2)}(M) &=  -\frac{57433}{972}+\frac{281}{162}\pi^2
               +\frac{22}{27}\pi^4 +\frac{1178}{27}\zeta_3+ C_F \frac{16}{3} \pi
  \beta_0 \zeta_3 \,,\qquad n_f=5.
\end{align}
The renormalisation-scale dependence of the first two hard-function
coefficients is given by
\begin{align}
H^{(1)}(\mu_R) &= H^{(1)}(M) + 2 d_B \pi \beta_0 \ln\frac{\mu_R^2}{M^2},\\
H^{(2)}(\mu_R) &= H^{(2)}(M) + 4  d_B \left( \frac{1+d_{B} }{2} \pi^2\beta_0^2 \ln^2\frac{\mu_R^2}{M^2} + \pi^2 \beta_1
  \ln\frac{\mu_R^2}{M^2}\right)\notag\\
& + 2 \left(1+d_{B}\right) \pi\beta_0  \ln\frac{\mu_R^2}{M^2} H^{(1)}(M),
\end{align}
where $d_B$ is the strong-coupling order of the Born squared
amplitude  (e.g. $d_B=2$ for Higgs production).

The $C^{(2)}$ and $G^{(1)}$ functions for gluon-fusion processes are
obtained in refs.~\cite{Catani:2011kr,Gehrmann:2014yya}, while for
quark-induced processes they are derived in
ref.~\cite{Catani:2012qa}. In the present work we extract their
expressions using the results of
refs.~\cite{Catani:2011kr,Catani:2012qa}. For gluon-fusion processes,
the $C^{(2)}_{gq}$ and $C^{(2)}_{gg}$ coefficients normalised as in
Eq.~\eqref{eq:coeff-fun} are extracted from Eqs.~(30) and~(32) of
ref.~\cite{Catani:2011kr}, respectively, where we use the hard
coefficients of Eqs.~\eqref{eq:H-fun-G} {\it without} the new term
proportional to $\beta_0$ in the $H_g^{(2)}(M)$
coefficient.\footnote{These must be replaced by $H^{(1)}\to H^{(1)}/2$
  and $H^{(2)}\to H^{(2)}/4$ to match the convention of
  refs.~\cite{Catani:2011kr,Catani:2012qa}.} The coefficient $G^{(1)}$
is taken from Eq.~(13) of ref.~\cite{Catani:2011kr}. Similarly, for
quark-initiated processes, we extract $C^{(2)}_{qg}$ and
$C^{(2)}_{qq}$ from Eqs.~(32) and~(34) of ref.~\cite{Catani:2012qa},
respectively, where we use the hard coefficients from Eqs.~\eqref{eq:H-fun-Q} {\it without} the new term proportional to $\beta_0$ in the
$H_q^{(2)}(M)$ coefficient. The remaining quark coefficient function
$C^{(2)}_{q\bar{q}}$, $C^{(2)}_{q\bar{q}'}$ and $C^{(2)}_{qq'}$ are
extracted from Eq.~(35) of the same article.

Eq.~\eqref{eq:master-kt-space} resums all logarithmic towers of
$\ln(1/v)$ (with $v=p_t/M$) up to N$^3$LL, therefore neglecting
subleading-logarithmic terms of order $\alpha_s^n \ln^{2n-6}(1/v)$.
Constant terms of order ${\cal O}(\alpha_s^3)$ relative to the Born
will be extracted automatically from a matching to the N$^3$LO
cumulative cross section in Section~\ref{sec:results}. This will allow
us to control all terms of order $\alpha_s^n \ln^{2n-6}(1/v)$ in the
matched cross section, therefore neglecting terms
${\cal O}\left(\alpha_s^n \ln^{2n-7}(1/v)\right)$.  We have split the
result into a sum of three terms. The first term (first line of
Eq.~\eqref{eq:master-kt-space}) starts at LL and contains the full NLL
corrections. The second term of Eq.~\eqref{eq:master-kt-space} (second
to fourth lines) is necessary to achieve NNLL accuracy, while the
third term (fifth to ninth lines) is purely N$^3$LL.

Since Eq.~\eqref{eq:master-kt-space} still contains
subleading-logarithmic terms (i.e.~starting at N$^4$LL in
$\ln(M/p_t)$), one could, even if not strictly required, perform
further expansions on each of the terms of
Eq.~\eqref{eq:master-kt-space} in order to neglect at least some of
the corrections beyond the desired logarithmic order.  For instance,
for a N$^3$LL resummation, the full N$^3$LL radiator is necessary in
the first term of Eq.~\eqref{eq:master-kt-space}, while the radiator
can be evaluated at NNLL in the second term, and at NLL in third term.
Analogously, for a NNLL resummation, the NLL radiator suffices in the
second term of Eq.~\eqref{eq:master-kt-space}. Furthermore, at NNLL,
one could split $R'(k_{t1})$ into the sum of a NLL term
$\hat{R}'(k_{t1})$ and a NNLL one $\delta \hat{R}'(k_{t1})$, and
expand Eq.~\eqref{eq:master-kt-space} about the former retaining only
contributions linear in $\delta \hat{R}'(k_{t1})$.  The last two
considerations relate Eq.~\eqref{eq:master-kt-space} to Eq. (9) of
ref.~\cite{Monni:2016ktx} where this approach was first formulated at
NNLL for the Higgs-boson transverse-momentum distribution.

Eq.~\eqref{eq:master-kt-space} can be evaluated in its present form
with fast Monte Carlo techniques, as we will discuss in
Section~\ref{sec:results}.

We performed numerous tests to verify the correctness of
Eq.~\eqref{eq:master-kt-space}. Firstly, we performed the expansion of
Eq.~\eqref{eq:master-kt-space} to ${\cal O}(\alpha_s^3)$ relative to
the Born for the transverse momentum of the boson as well as for the
$\phi^*$ distribution in Drell-Yan production, and compared it to the
corresponding result from the $b$-space formulation, finding full
agreement for the N$^3$LL terms. This is a highly non-trivial test of
the logarithmic structure of Eq.~\eqref{eq:master-kt-space}. The
differential ${\cal O}(\alpha_s^2)$ expansion for both observables was
also compared to {\tt MCFM}~\cite{Campbell:2011bn} and we found that
the difference between the two predictions vanishes in the logarithmic
region. Finally, we checked numerically that the coefficient of the
scaling $\Sigma(p_t)\propto p_t^2$ in the small-$p_t$ limit of
Eq.~\eqref{eq:master-kt-space} agrees with the prediction obtained
with the $b$-space formulation. The agreement of the NNLL prediction
obtained using our formula~\eqref{eq:master-kt-space} with the
$b$-space result from the program {\tt HqT}~\cite{Bozzi:2005wk} across
the spectrum was shown in ref.~\cite{Monni:2016ktx}.

\subsection{Perturbative scaling in the $p_t\to 0$ regime}
\label{sec:small-pt-limit}
In this section we show that our formulation of the
transverse-momentum resummation of Eq.~\eqref{eq:master-kt-space}
reproduces the correct scaling in the $p_t\to 0$ limit as first
observed in~\cite{Parisi:1979se}. Moreover, we obtain a correspondence
between the logarithmic accuracy and the perturbative accuracy in this
limit.
In the following we follow the approximations made in Ref.~\cite{Parisi:1979se} to
  derive an analytic estimate for the $p_t\to 0$ scaling of the
  differential cross section. Such approximations are further
  discussed in Appendix~\ref{app:more-small-pt-limit}. 
To perform a comparison with the results of~\cite{Parisi:1979se}, we
consider NLL resummation and neglect the evolution of the parton
densities with the energy scale. However the same procedure can be
easily extended to the general case. We have
\begin{align}
\label{eq:master-kt-space-PP}
&\frac{d^2\Sigma(v)}{d^2 \vec{p}_t d\Phi_B} = \sigma^{(0)}(\Phi_B)\int\frac{d k_{t1}}{k_{t1}}\frac{d
  \phi_1}{2\pi} e^{-R(k_{t1})}  R'(k_{t1})\int \dZ \delta^{(2)}\left(\vec{p}_t-\left(\vec{k}_{t1}
      + \dots +
      \vec{k}_{t(n+1)}\right)\right),
\end{align}
where
\begin{equation}
\sigma^{(0)}(\Phi_B)\equiv\frac{d\sigma^{(0)}}{d\Phi_B},
\end{equation}
and $\dZ$ is defined in Eq.~\eqref{eq:dZ}.
In order to evaluate the integral over $\dZ$ analytically we proceed
as in Sec.~\ref{sec:compare-to-bspace}. After integrating over the
azimuthal direction of $\vec{p}_t$ we obtain
\begin{align}
\label{eq:pp-1}
\frac{d^2\Sigma(v)}{d p_t d\Phi_B}  &= \sigma^{(0)}(\Phi_B) \,p_t\int\! \!b\, d b
                          J_0(p_t b) \int\frac{d k_{t1}}{k_{t1}} e^{-R(k_{t1})}  R'(k_{t1}) J_0(b k_{t1})\notag\\
&\hspace{4cm}\times\exp\left\{-R'\left(k_{t1}\right) \int^{k_{t1}}_{0}\frac{d k_t}{k_t}
  (1-J_0(b
  k_{t}))\right\}.
\end{align}
Before proceeding to the evaluation of Eq.~\eqref{eq:pp-1}, a remark
is in order. At NLL one would be tempted to perform the replacement
(see Sec.~\ref{sec:compare-to-bspace})
\begin{align}
\label{eq:j0-to-theta}
(1-J_0(b k_{t})) \simeq  \Theta(k_t-\frac{b_0}{b}) + \dots,
\end{align}
and recast Eq.~\eqref{eq:pp-1} as
\begin{align}
\frac{d^2\Sigma(v)}{d p_t d\Phi_B}  &= \sigma^{(0)}(\Phi_B) \,p_t\int\! \!b\, d b
                          J_0(p_t b) \int\frac{d k_{t1}}{k_{t1}} e^{-R(k_{t1})}  R'(k_{t1}) J_0(b
                                    k_{t1})\left(\frac{b_0}{b k_{t1}}\right)^{R'\left(k_{t1}\right)}\notag\\
&= \sigma^{(0)}(\Phi_B) \,p_t
                                    \int\frac{d k_{t1}}{k_{t1}}
                                    e^{-R(k_{t1})}  R'(k_{t1})
                                    \left(\frac{b_0}{k_{t1}}\right)^{R'\left(k_{t1}\right)}\frac{2^{1-R'\left(k_{t1}\right)}}{\left(p_t^2+k_{t1}^2\right)^{1-R'\left(k_{t1}\right)/2}}\notag\\
&\hspace{10mm}\times\frac{\Gamma\left(1-R'\left(k_{t1}\right)/2\right)}{\Gamma(R'(k_{t1})/2)}
 ~{}_{2}\hspace{-0.1mm}F_{1}\left(\frac{2-R'\left(k_{t1}\right)}{4},1-\frac{R'\left(k_{t1}\right)}{4},1,\frac{4
  p_t^2 k_{t1}^2}{\left(p_t^2+k_{t1}^2\right)^{2}}\right).          
\end{align}
The above result is singular for $R'\left(k_{t1}\right) \geq 2$, owing
to the fact that the integrand scales as $b^{1-R'(k_{t1})}$ in the
$b\to0$ limit. This singular behaviour is however entirely due to the
approximation in Eq.~\eqref{eq:j0-to-theta}, where all
power-suppressed terms are neglected, while Eq.~\eqref{eq:pp-1} is
regular, as the integral in its exponent vanishes as ${\cal O}(b^2)$
for small $b$.
Therefore, when using
Eq.~\eqref{eq:j0-to-theta} one must regularise the $b\to 0$ limit, for
instance by means of modified logarithms as in
ref.~\cite{Bozzi:2003jy}. In our formalism, instead,
Eq.~\eqref{eq:pp-1} is evaluated numerically without further
approximations so that the $b\to 0$ region is correctly described.

It is interesting to study the scaling of Eq.~\eqref{eq:pp-1} in the
small-$p_t$ limit. In this limit, the dominant mechanism that produces
a vanishing $p_t$ involves several soft and collinear emissions with
finite transverse momentum that mutually balance in the transverse
plane. 

In this kinematic configuration one has $k_{t1}\gg p_t$, thus
expanding $k_{t1}$ about $p_t$ in Eq.~\eqref{eq:pp-1} is not allowed:
such an operation would give rise to spurious singularities at
$R'(p_t)\geq 2$, as reported several times in the
literature~\cite{Frixione:1998dw,Dasgupta:2001eq,Banfi:2001bz,Becher:2010tm,Monni:2016ktx,Ebert:2016gcn}.

We therefore evaluate the $b$ integral of
Eq.~\eqref{eq:pp-1} and observe that in the limit where
$M\gg k_{t1}\gg p_t$ it gives
\begin{align}
\label{eq:j0-integral}
\int\! \!b\, d b
                          J_0(p_t b) J_0(b k_{t1})\exp\left\{-R'\left(k_{t1}\right) \int^{k_{t1}}_{0}\frac{d k_t}{k_t}
  (1-J_0(b
  k_{t}))\right\} \simeq 4\frac{k_{t1}^{-2}}{R'\left(k_{t1}\right)},
\end{align}
namely it is constant in $p_t$ in first approximation. In this regime
Eq.~\eqref{eq:pp-1} becomes
\begin{align}
\label{eq:pp-2}
\frac{d^2\Sigma(v)}{d p_t d\Phi_B}  &= 4\, \sigma^{(0)}(\Phi_B) \,p_t\int\frac{d k_{t1}}{k^3_{t1}} e^{-R(k_{t1})} .
\end{align}
In order to directly compare with the result of
ref.~\cite{Parisi:1979se}, we specialise to the case of the Drell-Yan
process, and compute $R(k_{t1})$ at the lowest order using the
leading-order running coupling expressed in terms of the QCD scale
$\Lambda_{\rm QCD}$ (with $n_f=4$),
$$\alpha_s(k_t)=\frac{12}{25}\pi \frac{1}{\ln(k_t^2/\Lambda_{\rm QCD}^2)}.$$
We obtain ($A^{(1)}=2 C_F$ in this case)
$$R(k_{t1}) = \frac{16}{25}\ln\frac{M^2}{\Lambda_{\rm
    QCD}^2}\ln\left(\frac{\ln\frac{M^2}{\Lambda_{\rm
    QCD}^2}}{\ln\frac{k_{t1}^2}{\Lambda_{\rm
    QCD}^2}} \right) - \frac{16}{25}\ln\frac{M^2}{k_{t1}^2}.$$
We now integrate over $k_{t1}$ in Eq.~\eqref{eq:pp-2} from $\Lambda_{\rm QCD}$ up to the invariant mass of the Drell-Yan pair,
obtaining
\begin{align}
\label{eq:pp-3}
\frac{d^2\Sigma(v)}{d p_t d\Phi_B} = 4\,\sigma^{(0)}(\Phi_B) \,p_t\int_{\Lambda_{\rm QCD}}^{M}\frac{d
  k_{t1}}{k^3_{t1}} e^{-R(k_{t1})} \simeq 2 \sigma^{(0)}(\Phi_B)
  p_t\left(\frac{\Lambda_{\rm QCD}^2}{M^2}\right)^{\frac{16}{25}\ln\frac{41}{16}},
\end{align}
that reproduces the scaling of ref.~\cite{Parisi:1979se}.\footnote{In the last step we have neglected a factor of $1/\Lambda_{\rm QCD}^2\ln(M^2/\Lambda_{\rm QCD}^2)$, as done in ref.~\cite{Parisi:1979se}.}
We stress that this power-like scaling is not due, by any means, to
higher-order effects that one would be missing in performing the naive
expansion of $k_{t1}$ about $p_{t}$, but rather to a collective
kinematical effect that requires the presence of any number of
emissions. Indeed, the expansion of Eq.~\eqref{eq:pp-1} to any order
in the strong coupling only gives rise to logarithmic effects and no
terms scaling as ${\cal O}(p_t)$ arise. To reproduce the correct
scaling an all-order treatment is necessary.

In order to study how this result is modified by the inclusion of
higher-order logarithmic corrections, we evaluate Eq.~\eqref{eq:pp-2}
in the fixed-coupling-constant approximation. This is a simple toy
model for the more complicated running coupling case. At lowest order
one has
\begin{equation}
R(k_{t1}) = A^{(1)}\frac{\alpha_s}{\pi} L^2,
\end{equation}
with $A^{(1)}=2 C$ (with $C=C_A$ for gluons and $C=C_F$ for quarks),
and $L=\ln M/k_{t1}$. In the perturbative regime Eq.~\eqref{eq:pp-2}
therefore reads
\begin{align}
\label{eq:pp-fixed-coupling}
\frac{d^2\Sigma(v)}{d p_t d\Phi_B}  &\simeq 4\, \sigma^{(0)}(\Phi_B)
                                    \frac{p_t}{M^2}\frac{\pi}{2}\frac{e^{\frac{\pi}{2 C\alpha_s}}}{\sqrt{2 C\alpha_s}}\left(1+{\rm
                                    Erf}\left( \frac{\sqrt{\pi}}{\sqrt{2 C\alpha_s}}\right)\right).
\end{align}
Eq.~\eqref{eq:pp-fixed-coupling} shows that in the small-$p_t$ limit
the differential spectrum features a non-perturbative scaling in
$\alpha_s$ (see also Eq.~(2.12) of
ref.~\cite{Parisi:1979se}\footnote{Please note that only the leading
  contribution for $\alpha_s\ll 1$ is reported in the right-hand side
  of that equation.}). However, the coefficient of this scaling can be
systematically improved in perturbation theory: the inclusion of NLL
terms $\alpha_s^n L^n$ in the right-hand side of Eq.~\eqref{eq:pp-2}
contributes an ${\cal O}(1)$ correction to the right-hand side of
Eq.~\eqref{eq:pp-fixed-coupling}. Analogously, NNLL terms
$\alpha_s^n L^{n-1}$ will produce an ${\cal O}(\alpha_s)$ correction
relative to the non-perturbative factor
$e^{\pi/(2 C\alpha_s)}/\sqrt{2 C\alpha_s}$, and so on. In particular,
with our N$^3$LL calculation we have control over the terms of
relative order ${\cal O}(\alpha_s^2)$.  From this scaling we deduce
that the correspondence $L\sim 1/\alpha_s$ is still valid in the deep
infrared regime.  However, this does not mean that the above
prediction is accurate in this limit: indeed non-perturbative effects
due to soft-gluon radiation below $\Lambda_{\rm QCD}$, as well as due
to the intrinsic transverse momentum of the partons in the proton,
feature a similar scaling. This is because the colour singlet's
transverse momentum is sensitive to non-perturbative dynamics only
through kinematical recoil, that is the same mechanism that drives the
scaling~\eqref{eq:pp-3}.

\section{Numerical implementation}
\label{sec:results}
In order to have a prediction that is valid accross different
kinematic regions of the spectrum, one needs to match the resummed
calculation, valid in the small-$v$ limit, to a
fixed-order calculation that describes the hard (large-$v$)
region. In this section we discuss the matching of the result
described in the previous sections, in particular
Eq.~\eqref{eq:master-kt-space}, to a fixed-order prediction that is
NNLO accurate in the hard region of the phase space. We then describe
how to evaluate Eq.~\eqref{eq:master-kt-space} exactly using a Monte
Carlo Markov process, and discuss the implementation in a parton-level
generator that is fully differential in the Born kinematics.

\subsection{Normalisation constraint and resummation-scale dependence}
\label{sec:unitarity}
In order to match the resummed calculation to a fixed-order prediction
one has to ensure that the hard region of the phase space receives no
contamination from resummation effects. We therefore need to modify
Eq.~\eqref{eq:master-kt-space} so that at large $v$ ($v=p_t/M$ in the
transverse-momentum case) all resummation effects vanish. At N$^3$LL, it
reduces to
\begin{align}
\label{eq:master-asympt}
&\frac{d\Sigma(v)}{d\Phi_B} = {\cal L}_{\rm
  N^3LL}(\mu_F)|_{L= 0}, 
\end{align}
where ${\cal L}_{\rm N^3LL} $ is defined in
Eq.~\eqref{eq:luminosity-N3LL0}. The normalisation constraint~\eqref{eq:master-asympt} can be implemented in
several ways; in what follows we impose it by modifying the
structure of the logarithms $L$ everywhere in
Eq.~\eqref{eq:master-kt-space}, as commonly done for this observable
in the literature.

Before defining the modified logarithms, it is convenient to
have a way to estimate the resummation uncertainties due to higher-order logarithmic
corrections that are not included in the calculation. To this aim, we introduce the dimensionless
resummation scale $x_Q$ by using the identity
\begin{equation}
\label{eq:res-scale-intro}
L\equiv\ln\frac{1}{v_1} = \ln\frac{x_Q}{v_1} - \ln x_Q,
\end{equation}
and then we expand the right-hand side about $\ln(x_Q/v_1)$ to the
nominal logarithmic accuracy (in terms of $\ln(x_Q/v_1)$), neglecting
subleading corrections. In the transverse-momentum case one has
$v_1=k_{t1}/M$ and $x_Q=Q/M$, where $Q$, the resummation scale, has
dimension of a mass. A variation of $x_Q$ will therefore provide an
estimate of the size of higher-order logarithmic corrections.

The normalisation constraint can now be imposed by replacing the resummed
logarithms $\ln(x_Q/v_1)$ by
\begin{equation}
\label{eq:modified-log}
\ln\frac{x_Q}{v_1} \to \tilde{L}=
\frac{1}{p}\ln\left(\left(\frac{x_Q}{v_1}\right)^{p} + 1\right),
\end{equation}
where the positive real parameter $p$ is chosen in such a way that
resummation effects vanish rapidly enough at $v_1 \sim x_Q$.
Eq.~\eqref{eq:modified-log} amounts to imposing unitarity by
introducing in the resummed logarithms power-suppressed terms that
scale as $(x_Q/v_1)^p$, which ultimately give rise to terms of order
$v^{-p}$ in the {\it cumulative} cross section $\Sigma(v)$. Given that
the differential spectrum tends to zero with a power law ($\sim
v^{-n}$ with positive $n$) at large $v$, it follows that one should
have $p\geq n-1$ in order not to affect the correct fixed-order
scaling at large $v$. However, since we are interested in turning off
the resummation at transverse momentum values of the order of the
singlet's mass, the relevant scaling $n$ to be considered in the
choice of $p$ is the one relative to the differential distribution in
this region. We stress, finally, that the
prescription~\eqref{eq:modified-log} is only one of the possible ways
of turning off resummation effects in the hard regions of the
spectrum. For instance one could, analogously, directly constrain the
first block to have $k_{t1}\leq Q$, which would naturally suppress
radiation effects at large $v$. This solution would however lead to
more complicated integrals in the expansion of the resummation formula
used in the matching to fixed order. For this reason, we stick to
prescription~\eqref{eq:modified-log} while leaving the study of
alternative solutions for future work.

We notice that, with the prescription~\eqref{eq:modified-log}, the
single-emission event in the first line of
Eq.~\eqref{eq:master-kt-space} is not a total derivative any
longer. One can however restore this property by introducing the
jacobian factor
\begin{equation}
\label{eq:jakobian}
{\cal J}(v_1/x_Q,p) = \left(\frac{x_Q}{v_1} \right)^p \left(1+\left(\frac{x_Q}{v_1} \right)^p\right)^{-1}
\end{equation}
in all integrals over $v_1=k_{t1}/M$ in Eq.~\eqref{eq:master-kt-space}.  This
jacobian tends to one at small $v_1$ and therefore does not
modify the logarithmic structure. Moreover, in the large-$v$ region
where the single-emission event dominates, this prescription prevents
the proliferation of power-suppressed terms. The
prescription~\eqref{eq:modified-log} effectively maps the point at
which the logarithms are turned off onto infinity. This also gives us
the freedom to extend the upper bound of the integration over $k_{t1}$
from $M$ to $\infty$ in Eq.~\eqref{eq:master-kt-space} without
spoiling the logarithmic accuracy.

We therefore implement the prescription~\eqref{eq:modified-log} in the
Sudakov radiator and its derivatives. We denote all modified
quantities by a `$\sim$' superscript. The expansion about $\ln(x_Q/v)$
induces some constant terms in the Sudakov radiator that are expanded
out up to ${\cal O}(\alpha_s^2)$ and included in the hard-function
coefficients. The modified quantities in
Eq.~\eqref{eq:master-kt-space} are
\begin{align}
\label{eq:mod-radiator}
\tilde{R}(k_{t1}) &= - \tilde{L} g_1(\alpha_s(\mu_R) \tilde{L} ) -
  g_2(\alpha_s(\mu_R) \tilde{L} ) - \frac{\alpha_s(\mu_R)}{\pi}
  g_3(\alpha_s(\mu_R) \tilde{L} ) - \frac{\alpha^2_s(\mu_R)}{\pi^2}
  g_4(\alpha_s(\mu_R) \tilde{L} ),\notag\\
\tilde{H}^{(1)}(\mu_R,&x_Q) = H^{(1)}(\mu_R) +
                             \left(-\frac{1}{2}A^{(1)}\ln x_Q^2 +
                             B^{(1)}\right) \ln x_Q^2\notag\\
\tilde{H}^{(2)}(\mu_R,&x_Q) = H^{(2)}(\mu_R) +
                             \frac{(A^{(1)})^2}{8}\ln^4x_Q^2 - \left(\frac{A^{(1)}
                             B^{(1)}}{2}+\frac{A^{(1)}}{3}\pi\beta_0
                             \right)\ln^3 x_Q^2\notag\\
&+\left(\frac{-A^{(2)}+(B^{(1)})^2}{2} + \pi\beta_0
  \left(B^{(1)}+A^{(1)}\ln \frac{x_Q^2 M^2}{\mu_R^2}\right)\right)\ln^2 x_Q^2\notag\\
& - \left(-B^{(2)}+B^{(1)}2\pi\beta_0\ln \frac{x_Q^2
  M^2}{\mu_R^2}\right)\ln x_Q^2  + H^{(1)}(\mu_R)\ln x_Q^2\left( -\frac{1}{2}A^{(1)}\ln x_Q^2 +
                             B^{(1)} \right),
\end{align}
where the functions $g_i$ are given in
Appendix~\ref{app:radiator}. All derivatives of the $R$ function are
to be consistently replaced by derivatives of $\tilde{R}$ with respect
to $\tilde{L}$. Notice that no constant terms are present in the
radiator and therefore $g_i(0)=0$.

The same replacement must be consistently performed in the parton
densities. In addition, it is convenient to have the latter
evaluated at a common factorisation scale $ \mu_F$ at large $v_1$, in
order to match the fixed-order convention. Both steps can be
implemented by expressing the parton densities $f$ at the scale
$\mu_F e^{-\tilde{L}}$, and expanding out the difference between
$f(\mu_F e^{-\tilde{L}} ,x)$ and $f(k_{t1} ,x)$ neglecting regular
terms as well as logarithmic terms beyond N$^3$LL. The relevant terms
in this expansion can be absorbed into a redefinition of the
coefficient functions $C^{(i)}(z)$, thereby introducing an explicit
dependence upon $\mu_F$ and $x_Q$. We obtain
\begin{align}
\tilde{C}_{ij}^{(1)}(z,&\mu_F,x_Q) = C_{ij}^{(1)}(z) +
                                    \hat{P}_{ij}^{(0)}(z)\ln\frac{x_Q^2 M^2}{\mu_F^2},\notag\\
\tilde{C}_{ij}^{(2)}(z,&\mu_F,x_Q) = C_{ij}^{(2)}(z) +
                                    \pi\beta_0 \hat{P}_{ij}^{(0)}(z)\left(
                                    \ln^2\frac{x_Q^2 M^2}{\mu_F^2} -
                                   2 \ln\frac{x_Q^2 M^2}{\mu_F^2}
                                    \ln\frac{x_Q^2
                                    M^2}{\mu_R^2}\right) +
                                    \hat{P}_{ij}^{(1)}(z)\ln\frac{x_Q^2
                                    M^2}{\mu_F^2} \notag\\
& + \frac{1}{2}(\hat{P}^{(0)}\otimes \hat{P}^{(0)})_{ij}(z) \ln^2\frac{x_Q^2
  M^2}{\mu_F^2} + (C^{(1)}\otimes \hat{P}^{(0)})_{ij}(z) \ln\frac{x_Q^2
  M^2}{\mu_F^2} - 2\pi\beta_0 C_{ij}^{(1)}(z) \ln\frac{x_Q^2
  M^2}{\mu_R^2}.
\end{align}
Finally, we also approximate the strong coupling in the terms
proportional to $\alpha_s^2(k_{t1})$ in
Eq.~\eqref{eq:master-kt-space}, featuring the convolution of one and
two splitting functions with the NLL luminosity, by retaining only
terms relevant to N$^3$LL as
\begin{equation}
\label{eq:N3LLcoupling}
\alpha_s(k_{t1})\simeq \frac{\alpha_s(\mu_R)}{1-2\alpha_s(\mu_R)\beta_0
  \tilde{L}}.
\end{equation}
Summarising, the final formula that we employ in the matching to fixed
order will be Eq.~\eqref{eq:master-kt-space} with the following
replacements:
\begin{align}
  L &\to \tilde{L},\qquad \frac{d k_{t1}}{k_{t1}} \to {\cal 
      J}(v_1/x_Q,p)\frac{d k_{t1}}{k_{t1}}, \notag\\
  R &\to \tilde{R},\qquad
      R'\to d \tilde{R}/d\tilde{L},\qquad R''\to d
      \tilde{R'}/d\tilde{L},\qquad R'''\to d \tilde{R''}/d\tilde{L},\notag\\
  {\cal L}_{\rm NLL} &\to \tilde{\cal L}_{\rm NLL},\qquad
                       {\cal L}_{\rm NNLL} \to \tilde{\cal L}_{\rm NNLL},\qquad
                       {\cal L}_{\rm N^3LL} \to \tilde{\cal L}_{\rm N^3LL}.
\end{align}
Moreover the coupling is treated according to
Eq.~\eqref{eq:N3LLcoupling} in the terms
$\hat P^{(0)}\otimes \tilde{\cal L}_{\rm NLL}$ and
$\hat P^{(0)}\otimes \hat P^{(0)}\otimes \tilde{\cal L}_{\rm NLL}$,
and the upper bound of the $k_{t1}$ integration in
Eq.~\eqref{eq:master-kt-space} is extended to infinity. The modified
luminosity factors appearing in the previous equation are defined as
\begin{align} 
\label{eq:luminosity-NLL}
\tilde{\cal L}_{\rm NLL}(k_{t1}) = \sum_{c, c'}\frac{d|M_{B}|_{cc'}^2}{d\Phi_B} f_c\!\left(\mu_F e^{-\tilde{L}},x_1\right)f_{c'}\!\left(\mu_F e^{-\tilde{L}},x_2\right),
\end{align}
\begin{align}
\label{eq:luminosity-NNLL}
&\tilde{\cal L}_{\rm NNLL}(k_{t1}) = \sum_{c, c'}\frac{d|M_{B}|_{cc'}^2}{d\Phi_B} \sum_{i, j}\int_{x_1}^{1}\frac{d z_1}{z_1}\int_{x_2}^{1}\frac{d z_2}{z_2}f_i\!\left(\mu_F e^{-\tilde{L}},\frac{x_1}{z_1}\right)f_{j}\!\left(\mu_F e^{-\tilde{L}},\frac{x_2}{z_2}\right)\notag\\&\Bigg(\delta_{ci}\delta_{c'j}\delta(1-z_1)\delta(1-z_2)
\left(1+\frac{\alpha_s(\mu_R)}{2\pi} \tilde{H}^{(1)}(\mu_R,x_Q)\right) \notag\\
&+ \frac{\alpha_s(\mu_R)}{2\pi}\frac{1}{1-2\alpha_s(\mu_R)\beta_0
  \tilde{L}}\left(\tilde{C}_{c i}^{(1)}(z_1,\mu_F,x_Q)\delta(1-z_2)\delta_{c'j}+
  \{z_1\leftrightarrow z_2; c,i \leftrightarrow c'j\}\right)\Bigg),
\end{align}
\begin{align}
\label{eq:mod-luminosity-N3LL}
&\tilde{\cal L}_{\rm N^3LL}(k_{t1})=\sum_{c,
  c'}\frac{d|M_{B}|_{cc'}^2}{d\Phi_B} \sum_{i, j}\int_{x_1}^{1}\frac{d
  z_1}{z_1}\int_{x_2}^{1}\frac{d z_2}{z_2}f_i\!\left(\mu_F e^{-\tilde{L}},\frac{x_1}{z_1}\right)f_{j}\!\left(\mu_F e^{-\tilde{L}},\frac{x_2}{z_2}\right)\notag\\&\Bigg\{\delta_{ci}\delta_{c'j}\delta(1-z_1)\delta(1-z_2)
\left(1+\frac{\alpha_s(\mu_R)}{2\pi} \tilde{H}^{(1)}(\mu_R,x_Q) + \frac{\alpha^2_s(\mu_R)}{(2\pi)^2} \tilde{H}^{(2)}(\mu_R,x_Q)\right) \notag\\
&+ \frac{\alpha_s(\mu_R)}{2\pi}\frac{1}{1-2\alpha_s(\mu_R)\beta_0 \tilde{L}}\left(1- \alpha_s(\mu_R)\frac{\beta_1}{\beta_0}\frac{\ln\left(1-2\alpha_s(\mu_R)\beta_0 \tilde{L}\right)}{1-2\alpha_s(\mu_R)\beta_0 \tilde{L}}\right)\notag\\
&\times\left(\tilde{C}_{c i}^{(1)}(z_1,\mu_F,x_Q)\delta(1-z_2)\delta_{c'j}+ \{z_1\leftrightarrow z_2; c,i \leftrightarrow c',j\}\right)\notag\\
& +
  \frac{\alpha^2_s(\mu_R)}{(2\pi)^2}\frac{1}{(1-2\alpha_s(\mu_R)\beta_0
  \tilde{L})^2}\Bigg(\tilde{C}_{c i}^{(2)}(z_1,\mu_F,x_Q)\delta(1-z_2)\delta_{c'j} + \{z_1\leftrightarrow z_2; c,i \leftrightarrow c',j\}\Bigg) \notag\\&+  \frac{\alpha^2_s(\mu_R)}{(2\pi)^2}\frac{1}{(1-2\alpha_s(\mu_R)\beta_0 \tilde{L})^2}\Big(\tilde{C}_{c i}^{(1)}(z_1,\mu_F,x_Q)\tilde{C}_{c' j}^{(1)}(z_2,\mu_F,x_Q) + G_{c i}^{(1)}(z_1)G_{c' j}^{(1)}(z_2)\Big) \notag\\
& + \frac{\alpha^2_s(\mu_R)}{(2\pi)^2} \tilde{H}^{(1)}(\mu_R,x_Q)\frac{1}{1-2\alpha_s(\mu_R)\beta_0 \tilde{L}}\Big(\tilde{C}_{c i}^{(1)}(z_1,\mu_F,x_Q)\delta(1-z_2)\delta_{c'j} + \{z_1\leftrightarrow z_2; c,i \leftrightarrow c',j\}\Big) \Bigg\}.
\end{align}

\subsection{Matching to fixed order}
To match the above result to a fixed-order calculation we design a
scheme belonging to the class of multiplicative
matchings~\cite{Antonelli:1999kx,Dasgupta:2001eq}. This, at present,
is preferable to the more common additive $R$
scheme~\cite{Catani:1992ua}, since the ${\cal O}(\alpha_s^3)$ constant
terms of the cumulative cross section are currently unknown
analytically (except for the three-loop corrections to the form factor
that were computed in
ref.~\cite{Baikov:2009bg,Lee:2010cga,Gehrmann:2010tu}) and they can
therefore be recovered numerically from our matching procedure. This
ensures that our matched prediction controls all terms up to and
including ${\cal O}(\alpha_s^n \ln^{2n-6}(1/v))$. Moreover, the
multiplicative scheme has the feature of being less sensitive to
numerical instabilities of the fixed-order prediction close to the
infrared and collinear regions.

However, the multiplicative scheme in hadronic collisions can give
rise to higher-order terms in the high-$p_t$ tail, due to the cross
product of parton luminosities. These are effectively subleading and
therefore they never spoil the perturbative accuracy, nevertheless
they can be numerically non-negligible, especially for processes
featuring large $K$ factors like Higgs production.
In order to suppress such spurious terms, we introduce a factor $Z$
defined as 
\begin{equation}
Z = \left(1-\left(\frac{v}{v_0}\right)^u\right)^h\Theta(v_0-v),
\end{equation}
where $v_0$ is the point at which the fixed-order is recovered, while
$h$ and $u$ are positive parameters. $h$ should be larger than two in
order to avoid small kinks in the differential distribution. In our
predictions below we set $v_0=1/2$ and $h=3$, and check that
the variations $v_0=1$ and $h=1,2$ do not produce sizeable
differences. The parameter $u$ will be discussed shortly.
In what follows, with a slight abuse of notation, we denote by
$\Sigma(v,\Phi_B)$ the generic exclusive cross section
$d\Sigma(v)/d\Phi_B$. We therefore define the matched cross section as
\begin{align}
\label{eq:matching-1}
\Sigma_{\rm MAT}(v,\Phi_B) =
  \left(\Sigma_{\rm RES}(v,\Phi_B)\right)^Z\frac{\Sigma_{\rm FO}(v,\Phi_B)}{\left(\Sigma_{\rm EXP}(v,\Phi_B)\right)^{Z}},
\end{align}
where $\Sigma_{\rm FO}$ is the fixed-order cross section at order
$\alpha_s^n$ differential in the Born kinematics, and
$\Sigma_{\rm EXP}$ is the expansion of the resummed cross section
$\Sigma_{\rm RES}$ to ${\cal O}(\alpha_s^n)$. The factor $Z$ ensures
that the resummation is smoothly turned off for $v\geq v_0$. We stress
that at small $v$ the factor $Z$ leads to extra terms which are
suppressed as $(v/v_0)^u$. Therefore $u$ can be chosen in order to
make these terms arbitrarily small, although they are already very
suppressed in the small-$v$ region. In our case we simply set $u=1$.

Up to N$^3$LO we now express the fixed-order and the expanded
cross sections as
\begin{align}
\Sigma_{\rm FO}(v,\Phi_B) & = \sum_{i=0}^{3}\Sigma^{(i)}_{\rm
                            FO}(v,\Phi_B),\notag\\
\Sigma^{(i)}_{\rm
                            FO}(v,\Phi_B) &= \sigma^{(i)}(\Phi_B) -
  \int_v d v' \frac{d \Sigma^{(i)}_{\rm 
                            FO}(v',\Phi_B)}{d v'} =
  \sigma^{(i)}(\Phi_B) + \bar{\Sigma}^{(i)}_{\rm
                            FO}(v,\Phi_B),\notag\\
\Sigma_{\rm EXP}(v,\Phi_B) & = \sum_{i=0}^{3}\Sigma^{(i)}_{\rm
                            EXP}(v,\Phi_B),
\end{align}
where $\bar{\Sigma}^{(0)}_{\rm FO}(v,\Phi_B)=0$,
$\Sigma_{\rm EXP}^{(0)}(v,\Phi_B)=\sigma^{(0)}$, and we defined
$\sigma^{(i)}(\Phi_B) = d \sigma^{(i)}/d\Phi_B$ as the $i$-th order of
the total cross section differential in the Born kinematics
\begin{equation}
\sigma(\Phi_B) = \sum_{i=0}^{3}\sigma^{(i)}(\Phi_B).
\end{equation}
With this notation, Eq.~\eqref{eq:matching-1} becomes
\begin{align}
\label{eq:matching-2}
  \Sigma&_{\rm MAT}(v,\Phi_B) = \left(\frac{\Sigma_{\rm
                                   RES}(v,\Phi_B)}{\sigma^{(0)}(\Phi_B)}\right)^Z\Bigg\{\sigma^{(0)}(\Phi_B) +
                                   \sigma^{(1)}(\Phi_B) +  \bar{\Sigma}^{(1)}_{\rm
                                   FO}(v,\Phi_B) - Z\,\Sigma^{(1)}_{\rm
                                   EXP}(v,\Phi_B)\notag\\\notag\\
                                 & + \sigma^{(2)}(\Phi_B) + \bar{\Sigma}^{(2)}_{\rm
                                   FO}(v,\Phi_B) -Z\,\Sigma^{(2)}_{\rm
                                   EXP}(v,\Phi_B) + \frac{Z(1+Z)}{2}\frac{(\Sigma^{(1)}_{\rm
                                   EXP}(v,\Phi_B))^2}{\sigma^{(0)}(\Phi_B)
                                   }\notag\\
                                 & - Z\,\Sigma^{(1)}_{\rm
                                   EXP}(v,\Phi_B)\frac{\sigma^{(1)}(\Phi_B) +\bar{\Sigma}^{(1)}_{\rm
                                   FO}(v,\Phi_B) 
                                   }{\sigma^{(0)}(\Phi_B)} \notag\\\notag\\
                                 &+ \sigma^{(3)}(\Phi_B) + \bar{\Sigma}^{(3)}_{\rm
                                   FO}(v,\Phi_B) -Z\,\Sigma^{(3)}_{\rm
                                   EXP}(v,\Phi_B) -Z\frac{(1+Z)(2+Z)}{6}\frac{(\Sigma^{(1)}_{\rm
                                   EXP}(v,\Phi_B))^3}{(\sigma^{(0)}(\Phi_B))^2}\notag\\
 & + \frac{Z(1+Z)}{2} \left(\Sigma^{(1)}_{\rm
  EXP}(v,\Phi_B)\right)^2\frac{\sigma^{(1)}(\Phi_B)+\bar{\Sigma}^{(1)}_{\rm
  FO}(v,\Phi_B)}{(\sigma^{(0)}(\Phi_B))^2} - Z\,\Sigma^{(2)}_{\rm
                                   EXP}(v,\Phi_B)\frac{\sigma^{(1)}(\Phi_B)+\bar{\Sigma}^{(1)}_{\rm
                                   FO}(v,\Phi_B)}{\sigma^{(0)}(\Phi_B)} \notag\\
&                                   + Z\,\Sigma^{(1)}_{\rm
                                   EXP}(v,\Phi_B)\frac{(1+Z)\Sigma^{(2)}_{\rm
                                   EXP}(v,\Phi_B) - \sigma^{(2)}(\Phi_B) - \bar{\Sigma}^{(2)}_{\rm
                                   FO}(v,\Phi_B)}{\sigma^{(0)}(\Phi_B)}\Bigg\},
\end{align}
where terms contributing at different orders in $\alpha_s$ are
separated by an extra blank line in the above equation.

To work out the expansion, we start from the three contributions of
Eq.~\eqref{eq:master-kt-space} with the replacements discussed in
Sec.~\ref{sec:unitarity}.
The first contribution starts with a single emission, the second
features at least two emissions, and the third contributes to events
with at least three emissions. The single-emission term can be worked
out analytically, since the integrand is a total derivative, while the
remaning terms can be expanded to ${\cal O}(\alpha_s^3)$ at the
integrand level and integrated over the real-emission phase space.
When the integrand is expanded out, one can safely set $\epsilon=0$ as
the cancellation of all singularities is now manifest. The expanded
result can be expressed as a linear combination in terms of the
following three classes of integrals (we write them in terms of
$v_1 = k_{t1}/M$):
\begin{align}
I_{2}^{(n,m)}(v) &= \int_{0}^{\infty} \frac{d v_1}{v_1} \int_{0}^{2
  \pi}\frac{d \phi_1}{2\pi} \int_0^1\frac{d \zeta_2}{\zeta_2}
  \int_{0}^{2 \pi}\frac{d \phi_2}{2\pi} {\cal J}(v_1/x_Q,p)
  \tilde{L}^n \ln^m\frac{1}{\zeta_2}\notag\\
&\hspace{5cm}\times\big\{\Theta(v-V(\{\tilde{p}\},k_1,k_2)) -
                                                                \Theta(v-V(\{\tilde{p}\},k_1))\big\},\notag\\
I_{3}^{(n,m)}(v) &= \int_{0}^{\infty} \frac{d v_1}{v_1} \int_{0}^{2
  \pi}\frac{d \phi_1}{2\pi} \int_0^1\frac{d \zeta_2}{\zeta_2}
  \int_{0}^{2 \pi}\frac{d \phi_2}{2\pi} \int_0^1\frac{d \zeta_3}{\zeta_3}
  \int_{0}^{2 \pi}\frac{d \phi_3}{2\pi} {\cal J}(v_1/x_Q,p)
  \tilde{L}^n \left(\ln^m\frac{1}{\zeta_2} + \ln^m\frac{1}{\zeta_3}\right)\notag\\
&\times\big\{\Theta(v-V(\{\tilde{p}\},k_1,k_2,k_3)) - 
                                                                \Theta(v-V(\{\tilde{p}\},k_1,k_2))
  \notag\\ &\hspace{5cm}-
                                                                \Theta(v-V(\{\tilde{p}\},k_1,k_3))
                                                                +\Theta(v-V(\{\tilde{p}\},k_1))\big\},\notag\\
I^{(n)}_{3,R''}(v) &= \int_{0}^{\infty} \frac{d v_1}{v_1} \int_{0}^{2
  \pi}\frac{d \phi_1}{2\pi} \int_0^1\frac{d \zeta_2}{\zeta_2}
  \int_{0}^{2 \pi}\frac{d \phi_2}{2\pi} \int_0^1\frac{d \zeta_3}{\zeta_3}
  \int_{0}^{2 \pi}\frac{d \phi_3}{2\pi} {\cal J}(v_1/x_Q,p)
  \tilde{L}^n \ln\frac{1}{\zeta_2} \ln\frac{1}{\zeta_3}\notag\\
&\times\big\{\Theta(v-V(\{\tilde{p}\},k_1,k_2,k_3)) - 
                                                                \Theta(v-V(\{\tilde{p}\},k_1,k_2))
  \notag\\ &\hspace{5cm}-
                                                                \Theta(v-V(\{\tilde{p}\},k_1,k_3))
                                                                +\Theta(v-V(\{\tilde{p}\},k_1))\big\},
\end{align}
where $\tilde{L}$ and $\cal{J}$ are defined in
Eqs.~\eqref{eq:modified-log} and~\eqref{eq:jakobian}, respectively. We
stress that we extended the upper bound of the integration over $v_1$
to infinity, following the discussion of Sec.~\ref{sec:unitarity}. The
integral over $v_1$ can be evaluated analytically. The remaining
integrations are carried out numerically and the final results are
tabulated with fine grids as a function of $v/x_Q$.

\subsection{Event generation}
\label{sec:event-generation}
Before presenting a phenomenological application of this formalism, we
comment briefly on how Eq.~\eqref{eq:master-kt-space} is implemented
numerically using a Monte Carlo method. We follow a variant of the
procedure used in
refs.~\cite{Banfi:2001bz,Banfi:2004yd,Banfi:2014sua}. For the first
emission we generate $v_1$ uniformly according to the integration
measure $d v_1/v_1 {\cal J}(v_1/x_Q,p)$, and assign it a weight in
terms of the Sudakov radiator and parton luminosities. All the
identical emissions belonging to the ensemble $\dZ$ are generated via
a shower ordered in $v_i$. This is done by expressing the term
$\epsilon^{R'(k_{t1})}$ as
\begin{equation}
e^{-R'(k_{t1}) \ln\frac{1}{\epsilon}} = \prod_{i=2}^{n+2}e^{-R'(k_{t1}) \ln\frac{\zeta_{i-1}}{\zeta_{i}}},
\end{equation}
with $\zeta_1=1$ and $\zeta_{n+2}=\epsilon$. Each emission in $\dZ$
now has a weight
$$\frac{d\zeta_i}{\zeta_i} R'(k_{t1}) e^{-R'(k_{t1})
\ln\frac{\zeta_{i-1}}{\zeta_{i}}},$$
and therefore it can be generated by solving for $\zeta_i$ the equation
\begin{equation}
e^{-R'(k_{t1})
  \ln\frac{\zeta_{i-1}}{\zeta_{i}}} = r,
\end{equation}
with $r$ being a random number extracted uniformly in the range
$[0,1]$. The above equation has no solution for
$\zeta_i > \zeta_{i-1}$, therefore this amounts to a shower ordered in
$\zeta_i$ (or, equivalently, in $v_i$). The procedure is stopped as
soon as a $\zeta_i < \epsilon$ is generated. The azimuthal angles are
generated uniformly in the range $[0,2\pi]$ for all
emissions. Finally, the {\it special} emissions, denoted by the
subscript $s$ in Eq.~\eqref{eq:master-kt-space}, do not have an
associated Sudakov suppression since their contribution is always
finite in four dimensions. Therefore we generate them according to
their phase-space measure and weight as they appear in the master
formula.

This recipe is sufficient to evaluate Eq.~\eqref{eq:master-kt-space},
and it can be implemented in a fast numerical code. We stress that it
is an exact procedure, meaning that no truncation at any perturbative
order is involved. The algorithm leads to the generation of an
arbitrary number of emissions with $\zeta_i>\epsilon$, while all
unresolved emissions with $\zeta_i<\epsilon$ are accounted for
analytically in the Sudakov radiator. This ensures that the whole
singular part of the radiation phase space and all perturbative orders
are treated exactly. We choose conservatively $\epsilon=e^{-20}$ for
our tests, although we observe that a much larger value
(e.g. $\epsilon\sim e^{-7}$) can be chosen in practice given that
emissions below this threshold will be very soft and/or collinear,
hence improving slightly the efficiency of the event generation.

We generate Born events using the LO matrix elements and
phase-space-integrator routines of {\tt MCFM}~\cite{Campbell:2011bn},
and we use {\tt HOPPET}~\cite{Salam:2008qg} to handle the evolution of
the parton densities and the convolution with the various coefficient
functions.

For each Born event we run the above algorithm to produce
the initial-state radiation, and fill the histograms on the fly,
thereby yielding $d\Sigma_{\rm RES}(v)/d\Phi_B$. As a byproduct, this
allows us to have exclusive events with N$^3$LL accuracy for the
observables treated in this article.
For each Born event we also generate a histogram filled with the
expansion counterterm, which is computed as described in the previous
section. After the generation, the two histograms are combined with
the corresponding fixed-order cumulative distribution according to
Eq.~\eqref{eq:matching-2}. 

We point out that the Sudakov radiator has a singularity in
correspondence of the Landau pole at
$2\alpha_s(\mu_R)\beta_0\tilde{L}= 1$ (see expressions in
Appendix~\ref{app:radiator}). One could use different prescriptions to
handle this singularity, all differing by power-suppressed terms in
the perturbative expansion. We choose to set the result to zero below
the singularity which, anyway, occurs at very small $p_t$ values. We
stress that other schemes can be adopted, and that this choice has no
consequences above the scale of the singularity.

The resummation and matching as described above are implemented in the
program {\tt RadISH} that can simulate the production of any colour
singlet with arbitrary phase-space cuts on the Born kinematics. The
code will be released in due course.

\subsection{Predictions for Higgs-boson production at 13 TeV $pp$ collisions}
 \label{sec:pheno}
 We now apply the method described in the previous sections to obtain
 the inclusive transverse-momentum distribution of the Higgs boson at
 the LHC. We stress that the results shown in the following are to be
 considered as a proof of concept of our method, and a more detailed
 phenomenology discussion on the precise choice of the matching scheme
 as well as on the theory uncertainties will be the subject of a
 forthcoming publication.

We perform the calculation in the large-top-mass limit, and we match
our N$^3$LL result to the NNLO distribution that was computed in
refs.~\cite{Boughezal:2015dra,Boughezal:2015aha,Chen:2016zka}. In
particular, here we use results obtained with the code of
ref.~\cite{Caola:2015wna} with a cut on the Higgs transverse momentum
at $5$~GeV.
 The matched distribution integrates to the inclusive N$^3$LO cross
 section that is taken from ref.~\cite{Anastasiou:2015ema}.

 We consider $13$ TeV collisions, and we use parton densities from the
 {\tt PDF4LHC15\_nnlo\_mc}
 set~\cite{Butterworth:2015oua,Dulat:2015mca,Ball:2014uwa,Harland-Lang:2014zoa,Carrazza:2015hva,Watt:2012tq}.
 The value of the parameter $p$ appearing in the modified logarithms
 $\tilde L$ is chosen considering the scaling of the spectrum in the
 hard region, in order to make the matching to the fixed order smooth
 in this region. On the other hand, its value should not be too large,
 in order to prevent the peak of the distribution from being
 artificially pushed upwards due to the normalisation constraint. We
 therefore set $p=2$ as our reference value, but nevertheless checked
 that the choice $p=3$ induces negligible differences.

As central scales we employ $\mu_R=\mu_F=m_H$, and
$x_Q=Q/m_H=1/2$. The perturbative uncertainty is estimated by
performing a seven-scale variation of $\mu_R$, $\mu_F$ by a factor of
two in either direction, while keeping $1/2<\mu_R/\mu_F<2$ and
$x_Q=1/2$; moreover, for central $\mu_R$ and $\mu_F$ scales, $x_Q$ is
varied around its central value in a range that we now turn to
discuss. The total error is defined as the envelope of all above
variations.

In the case of the transverse momentum $k_{t1}$ of a colour singlet of
mass $M$, the resummation scale $Q$ is introduced by splitting the
resummed logarithms as
\begin{equation}
\ln\frac{M}{k_{t1}} = \ln\frac{Q}{k_{t1}} + \ln\frac{M}{Q}, 
\end{equation}
and subsequently assuming that 
\begin{equation}
\label{eq:logs-hier} \ln\frac{Q}{k_{t1}} \gg \ln\frac{M}{Q}.
\end{equation}
The latter condition is true at small $k_{t1}$, and it allows one to
expand $\ln(M/k_{t1})$ about $\ln(Q/k_{t1})$, retaining only terms relevant
to a given logarithmic accuracy. In this case, variations of $Q$ give
a handle to estimate the size of subleading-logarithmic terms in the
region where all-order effects are important.

However, in the matching region $k_{t1}\sim M/2$,
condition~\eqref{eq:logs-hier} is violated for $k_{t1}\gtrsim Q^2/M$. In
this regime, the variation of the resummation scale is physically
meaningless, since the logarithmic hierarchy it is based upon is not
valid at these scales.
In particular, for Higgs production, a variation of $Q$ by a factor of
two around $m_H/2$ can have a couple of drawbacks.  On the one hand,
for $Q=m_H/4$, it leads to values of $Q^2/m_H$ which are below the
peak of the distribution, implying that the corresponding
resummation-scale variation is technically reliable only to the left
of the peak.  On the other hand, for $Q=m_H$, resummation effects are
allowed to survive up to the Higgs scale, which is a fairly hard
region of the phase space, where one expects to be predictive with the
sole fixed-order calculation. In practice, however, in our matching
procedure the resummed contribution is subtracted up to the
perturbative order one is matching to, which ensures that the residual
variations of $Q$ away from the region of large logarithms induce
effects that are numerically very small.

For these two reasons, we believe that a more suitable variation range
is given by $Q\in [ m_H/3, 3 m_H/4]$, which corresponds to a variation
by a factor of $3/2$ around the central value $Q=m_H/2$. This range,
that was already adopted in ref.~\cite{Banfi:2015pju}, ensures that
the resummation-scale variation is reliable in the peak region and
that resummation effects are turned off well below the hard scale of
the reaction, hence avoiding artifacts in the matched spectrum.

\begin{figure}[ht!]
  \centering
  \includegraphics[width=0.5\textwidth]{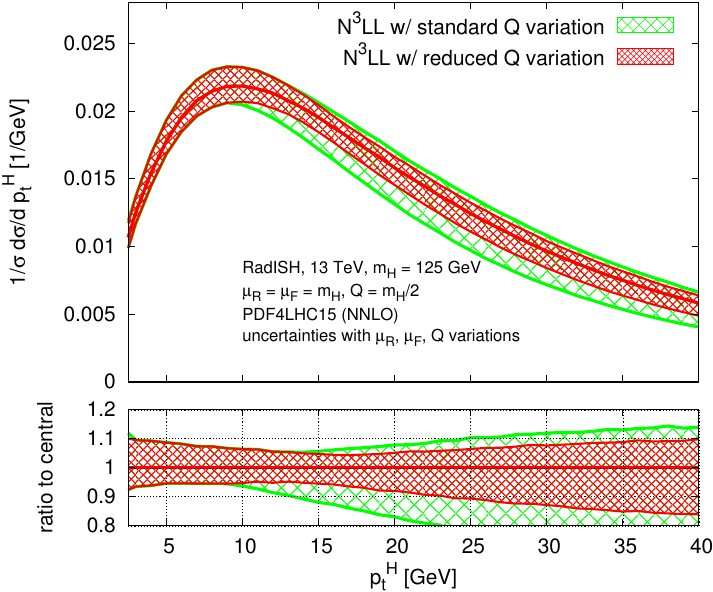}\hfill
  \includegraphics[width=0.5\textwidth]{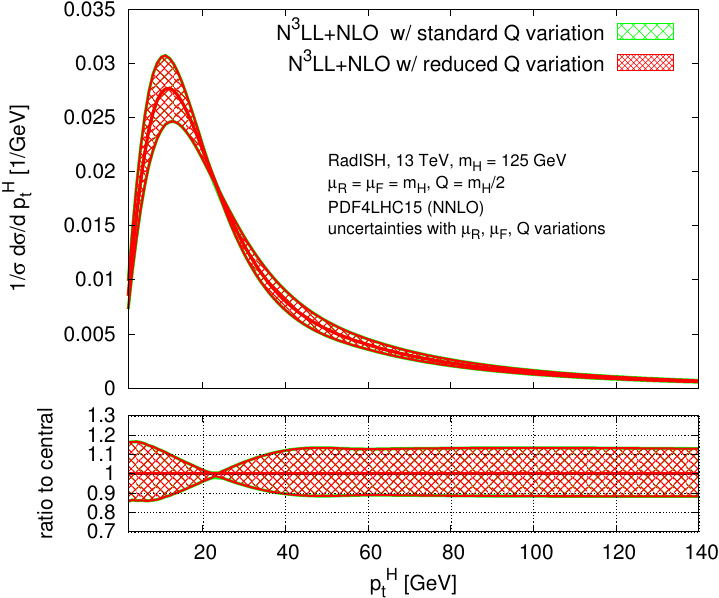}
  \caption{Comparison between two different prescriptions for the
    resummation-scale-variation range, as described in the text. The
    comparison is shown both at the resummation level (left) and with a matching to NLO (right).}
  \label{fig:Q_compare}
\end{figure}

To study the impact of this choice, in the left panel of
Figure~\ref{fig:Q_compare} we show the comparison between the pure
resummed N$^3$LL normalised spectra with two uncertainty
prescriptions: in the green coarse-textured band, $Q$ is varied by a factor of two
around $m_H/2$, while the red fine-textured band involves the aforementioned reduced
variation by a factor of $3/2$; in both cases $\mu_R$ and $\mu_F$
undergo the seven-point variation described above. As expected, the
choice $Q\in [ m_H/3, 3 m_H/4]$ reduces the impact of the
resummation-scale uncertainty in the matching region where the
logarithms are not large, while leaving the uncertainty unchanged in
the small-$p_t$ regime where the all-order treatment is necessary.

The right panel of Figure~\ref{fig:Q_compare} shows the comparison
between the two prescriptions for the matched N$^3$LL+NLO
distribution.\footnote{Preliminary results at N$^3$LL+NLO for this
  observable have been also shown at~\cite{Zhu-talk}.} In the NLO
matching, the resummed component is subtracted up to and including
${\cal O}(\alpha_s^2)$ terms relative to the Born. Therefore, in the
region where the logarithms are moderate in size, the issues due to
the large scale variation are suppressed by ${\cal O}(\alpha_s^3)$,
and we indeed observe that the two bands differ negligibly at
intermediate $p_t$ values.

We conclude that the resummation-scale variation by a factor of $3/2$
still provides a wide enough variation range to probe the size of
subleading-logarithmic corrections, while avoiding that some moderate
resummation effects persist away from the region where the logarithms
are large. We therefore adopt the modified variation in our
prescription to estimate the perturbative uncertainty.

\begin{figure}[ht!]
  \centering
  \includegraphics[width=0.5\textwidth]{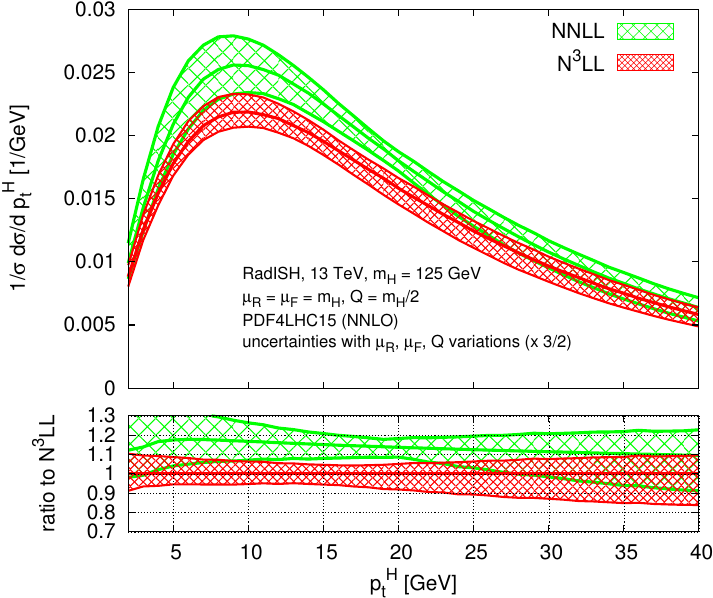}\hfill
  \includegraphics[width=0.5\textwidth]{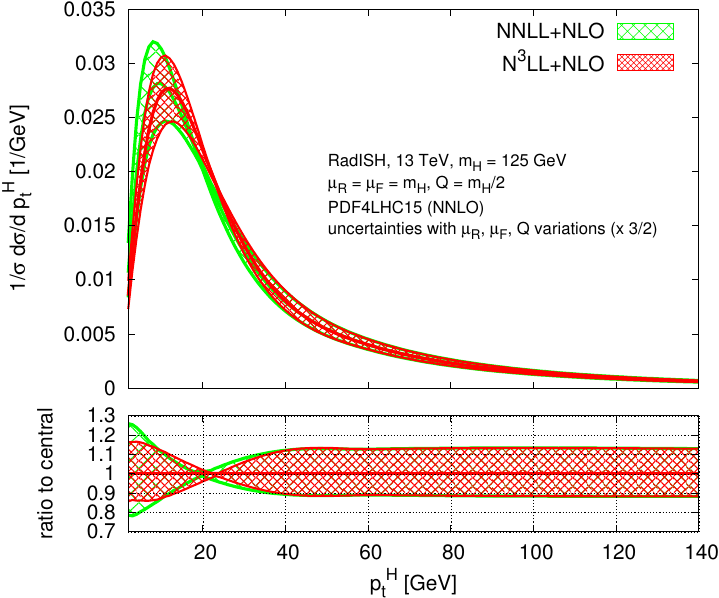}
  \caption{Left: comparison between the resummed distributions at
    N$^3$LL and NNLL; the lower panel shows the ratio of the two
    distributions. Right: comparison between the matched N$^3$LL+NLO
    and the NNLL+NLO predictions for the inclusive Higgs spectrum; the
    lower panel shows the ratio of each distribution to its central
    value.}
  \label{fig:n3ll-v-n2ll}
\end{figure}

We next turn to the comparison with NNLL. The left panel of
Figure~\ref{fig:n3ll-v-n2ll} shows a comparison between the pure
resummed predictions for the normalised spectrum at N$^3$LL and
NNLL. In this plot, the NNLL curve is normalised to the NLO total
cross section, while the N$^3$LL curve is normalised to the NNLO total
cross section. The plot shows that the inclusion of the N$^3$LL
corrections leads to a reduction in the scale uncertainty of the
resummed prediction compared to the NNLL result.\footnote{An identical
  reduction in size is observed when varying $Q$ by a factor of two
  around its central value.}

The right plot of Figure~\ref{fig:n3ll-v-n2ll} shows the matching of
the NNLL and N$^3$LL predictions to NLO. Both curves are now
normalised to the NNLO total cross section. We observe that at the
matched level, the N$^3$LL corrections amount to $\sim 10\%$ around
the peak of the spectrum, and they get slightly larger for smaller
$p_t$ values ($\lesssim 10$\,GeV). A substantial reduction of the
total scale uncertainty is observed for $p_t\lesssim 10$\,GeV.

We notice that, at the matched level, the impact of the N$^3$LL
corrections is reduced with respect to the sole resummation shown in
the left plot of Figure~\ref{fig:n3ll-v-n2ll}. This is to a good
extent due to the matching scheme that we chose here. Indeed, in a
multiplicative scheme we include the ${\cal O}(\alpha_s^2)$ constant
terms already at NNLL, although they are formally of higher-order
accuracy. While these terms enter at N$^3$LL, they are numerically
sizeable and therefore their inclusion reduces the difference between
the N$^3$LL+NLO and the NNLL+NLO predictions.

\begin{figure}[ht!]
  \centering
  \includegraphics[width=0.7\textwidth]{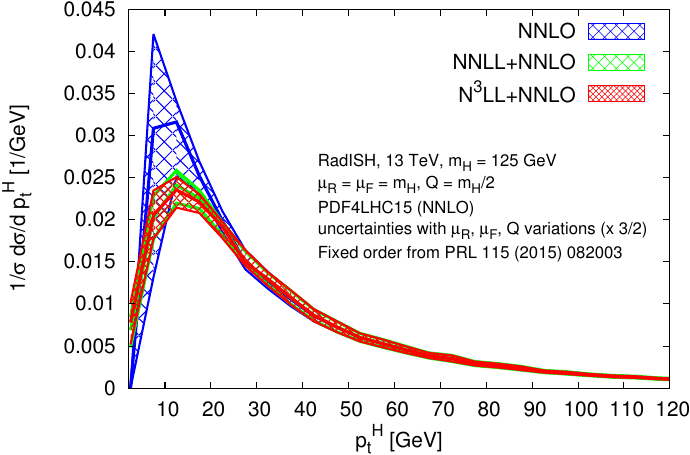}
  \caption{Comparison among the matched normalised distributions at
    N$^3$LL+NNLO, NNLL+NNLO, and NNLO. The uncertainties are obtained
    as described in the text.}
  \label{fig:n3ll+nnlo}
\end{figure}

To conclude this section, in Figure~\ref{fig:n3ll+nnlo} we report the
N$^3$LL+NNLO prediction for the normalised distribution. The latter is
compared both to NNLL+NNLO and to the pure NNLO result. All curves in
the plot are now normalised to the total N$^3$LO cross section. When
matched to NNLO, the N$^3$LL corrections give rise to a few-percent
shift of the central value with respect to the NNLL+NNLO prediction
around the peak of the distributions, while they have a somewhat
larger effect for $p_t\lesssim 10$\,GeV. We recall that some of the
N$^3$LL effects are already included in the NNLL+NNLO prediction by
means of the multiplicative matching scheme that we adopt here. As a
consequence, this reduces the difference between the N$^3$LL+NNLO and
the NNLL+NNLO curves.
We also observe that the matched N$^3$LL and NNLL predictions are only
moderately different in their theoretical-uncertainty bands. While
this is of course expected in the hard region of the spectrum, we
point out that, in the region $p_t \lesssim 30$\,GeV, the latter
feature is due (and increasingly so at smaller $p_t$) to numerical
instabilities of the fixed-order runs with one of the scales ($\mu_R$
or $\mu_F$) set to $m_H/2$. As we already observed at NLO, it is
indeed necessary to have stable fixed-order predictions for
$p_t < 10$\,GeV in order to benefit from the uncertainty reduction due
to the higher-order resummation. We leave this for future work.

\section{Conclusions}
\label{sec:conclusions}
In this article we presented a formulation of the momentum-space
resummation for global, recursive infrared and collinear safe
observables that vanish far from the Sudakov limit because of
kinematic cancellations implicit in the observable's definition. In
particular, we studied the class of inclusive observables that do not
depend on the rapidity of the QCD radiation. Members of this class
are, among others, the transverse momentum of a heavy colour singlet
and the $\phi^*$ observable in Drell-Yan pair production. We obtained
an all-order formula that is valid for all observables belonging to
this class, and we explicitly evaluated it to N$^3$LL up to effects
due to the yet unknown four-loop cusp anomalous dimension.  In the
case of the transverse momentum of a colour singlet, we proved that
our formulation is equivalent to the more common solution in
impact-parameter space at this accuracy. This evidence is also
supported by the numerous checks that we have documented. This
equivalence allowed us to extract the ingredients necessary to compute
the Sudakov radiator at N$^3$LL using the recently computed $B^{(3)}$
coefficient~\cite{Li:2016ctv,Vladimirov:2016dll}. The radiator is
universal for all observables of this class~\cite{Banfi:2014sua},
which can therefore be resummed to this accuracy with our approach.
The all-order result was shown to reproduce the correct power-like
scaling in the small-$p_t$ limit, where the perturbative component of
the coefficient of the intercept can be systematically improved by
including higher-order logarithmic corrections. We implemented our
results in the exclusive generator {\tt RadISH}, which performs the
resummation and the matching to fixed order, and allows the user to
apply arbitrary kinematic cuts on the Born phase space. Although we
explicitly treated the case of Higgs production, the code developed
here can automatically handle any colour-singlet system.

As a phenomenological application, we computed the Higgs
transverse-momentum spectrum at the LHC. In comparison to the NNLL+NLO
prediction, we find that N$^3$LL+NLO effects are moderate in size, and
lead to ${\cal O}(10\%)$ corrections near the peak of the distribution
and they are somewhat larger for $p_t \lesssim 10$\,GeV. The scale
uncertainty of the matched calculation is reduced by the inclusion of
the N$^3$LL corrections in the small transverse-momentum region. When
matched to NNLO, the effect of the N$^3$LL is pushed towards lower
$p_t$ values, leading to a few percent correction to the previously
known NNLL+NNLO prediction~\cite{Monni:2016ktx} around the peak, and
to more sizeable effects at smaller $p_t$ values. In order to further
improve the theoretical control in the small-medium transverse
momentum region, it will be necessary to consider the deviations from
the large-$m_t$ approximation. Recently, progress has been made in
this respect by computing the NLO corrections to the top-bottom
interference~\cite{Lindert:2017pky}. Higher-order effects due to the
leading tower of logarithms of $p_t/m_b$ were addressed in
ref.~\cite{Melnikov:2016emg} and were found to be moderate in
size. The procedure for the inclusion of mass effects in the context
of transverse-momentum resummation is a debated topic. While some
prescriptions are available~\cite{Grazzini:2013mca,Banfi:2013eda},
further studies are necessary to estimate these effects in the
logarithmic region at this level of accuracy.

\section{Acknowledgements}
We wish to thank A.~Banfi, C.~Bauer,
V.~Bertone, G.~Salam, G.~Zanderighi for stimulating
discussions on the topics treated here and very valuable comments on
the manuscript. We also thank F.~Caola for providing us with the
fixed-order runs for the Higgs transverse-momentum spectrum at
NNLO. The work of LR is supported by the European Research Council
Starting Grant PDF4BSM, WB is supported by the European Research
Council grant 614577 HICCUP (High Impact Cross Section Calculations
for Ultimate Precision), and the work of ER is supported by a Marie
Sk\l{}odowska-Curie Individual Fellowship of the European Commission's
Horizon 2020 Programme under contract number 659147
PrecisionTools4LHC. PM would like to thank the Erwin Schr\"odinger
Institute of Vienna for hospitality and support while part of this
work was carried out. PT would like to thank CERN's Thoretical Physics
Department for hospitality during the development of this work.

\appendix

\section{Connection with the backward-evolution algorithm at NLL}
\label{app:parton-shower}
It is interesting to relate our formulation for the
transverse-momentum resummation to a NLL-accurate backward-evolution
algorithm~\cite{Sjostrand:1985xi,Gottschalk:1986bk,Marchesini:1987cf}.
We start from Eq.~\eqref{eq:partxs-mellin}, that was deduced by
considering only flavour-conserving real splitting kernels, for the
sake of clarity. We briefly comment on the general flavour case below.

After neglecting the effect of the hard and coefficient functions,
which starts at NNLL, we recast the NLL partonic cross section as
\begin{align}
\label{eq:partxs-mellin-NLL}
&\hat{\bf \Sigma}_{N_1,N_2}^{c_1,
  c_2}(v) = \mathbb 1^{(c_1,c_2)}\int_0^{M}\frac{d k_{t1}}{k_{t1}} \int_0^{2\pi}
                     \frac{d\phi_1}{2\pi} e^{-{\bf R}(\epsilon k_{t1})}\exp\left\{-\sum_{\ell=1}^{2}
\int_{\epsilon k_{t1}}^{\mu_0}\frac{dk_t}{k_t}\frac{\alpha_s(k_t)}{\pi}{\bf
  \Gamma}_{N_\ell}(\alpha_s(k_t)) \right\}\notag\\
&\sum_{\ell_1=1}^2\left(
  {\bf R}_{\ell_1}'\left(k_{t1}\right) + \frac{\alpha_s(k_{t1})}{\pi}{\bf
  \Gamma}_{N_{\ell_1}}(\alpha_s(k_{t1})) \right) \sum_{n=0}^{\infty}\frac{1}{n!}
                                                     \prod_{i=2}^{n+1}
                                                     \int_{\epsilon}^{1}\frac{d\zeta_i}{\zeta_i}\int_0^{2\pi}
                                                     \frac{d\phi_i}{2\pi}
\notag\\
  & \times \sum_{\ell_i=1}^2\bigg({\bf R}_{\ell_i}'\left(k_{ti}\right)+\frac{\alpha_s(k_{ti})}{\pi}{\bf
  \Gamma}_{N_{\ell_i}}(\alpha_s(k_{ti}))\bigg)\Theta\left(v-V(\{\tilde{p}\},k_1,\dots, k_{n+1})\right),
\end{align}
where $\mathbb 1^{(c_1,c_2)}$ enforces the flavour of the two parton
densities to be identical to that entering the Born process,
i.e.~${\bf f}^T\mathbb 1^{(c_1,c_2)}{\bf f}=f_{c_1}f_{c_2}$. At NLL
order, the emission probabilities involve only tree-level splitting
functions, whose coupling we evaluate in the CMW scheme, as discussed
in Sec.~\ref{sec:nll}:
\begin{equation}
\frac{\alpha_s(k_{t})}{\pi}\to \frac{\alpha^{\rm CMW}_s(k_{t})}{\pi}=\frac{\alpha_s(k_{t})}{\pi}\left (1 +  \frac{\alpha_s(k_t)}{2\pi} K\right),
\end{equation}
where $K$ is defined in Eq.~\eqref{eq:K}.  In order to perform the
inverse Mellin transform of Eq.~\eqref{eq:partxs-mellin-NLL}, we
observe that, when inverted into $z$ space, each of the real-emission
probabilities acts on a generic parton distribution $f(x_{\ell_i})$ as
described in Section~\ref{sec:master-initial-state}:
\begin{align}
\label{eq:inverse-single}
\bigg(R_{\ell_i}'\left(k_{ti}\right)&+\frac{\alpha_s(k_{ti})}{\pi}
\gamma^{(0)}_{N_{\ell_i}}(\alpha_s(k_{ti})) \bigg)
f_{N_{\ell_i}}(\mu)\notag\\
&\to \frac{\alpha^{\rm CMW}_s(k_{ti})}{\pi}\Bigg(\int_{0}^{1-k_{ti}/M} d\zi P^{(0)}(\zi)
f(\mu,x_{\ell_i}) + \int_{x_{\ell_i}}^1 d \zi \frac{\hat{P}^{(0)}(\zi)}{\zi} f(\mu,\frac{x_{\ell_i}}{\zi})\Bigg),
\end{align}
where we reintroduced the regular terms in the hard-collinear
contribution to $R'_\ell$, whose $z^{(\ell)}$ upper limit was set to
$1$ in Section~\ref{sec:recoil}.

Similarly, we can now restore the remaining power-suppressed terms in
the single-emission probability that we neglected in our discussion of
Section~\ref{sec:recoil}, and recast the right-hand side of
Eq.~\eqref{eq:inverse-single} in terms of the unregularised splitting
function as\footnote{We recall that
  Eq.~\eqref{eq:inverse-single-rhs} in the case of $g\to gg$
  splitting also requires an extra symmetry factor of $2$ to account
  for the fact that the total probability to find a gluon with
  momentum fraction $z^{(\ell)}$ is the sum of the probability to find
  either of the two gluons involved in the branching, as in
  Eq.~\eqref{eq:Pgg-regularised}.}
\begin{align}
\label{eq:inverse-single-rhs}
  \frac{\alpha_s^{\rm CMW}(k_{ti})}{\pi}\int_{x_{\ell_i}}^{1-k_{ti}/M} d\zi \frac{P^{(0)}(\zi)}{\zi} f(\mu,\frac{x_{\ell_i}}{\zi}).
\end{align}
We furthermore introduce the shower Sudakov form factor $\Delta(Q_i)$,
that at NLL reads
\begin{equation}
\Delta(Q_i) = \exp\left\{ - \sum_{\ell=1}^{2}\int^{Q_i}_{\epsilon k_{t1}} \frac{d k_t}{k_t} \int_{0}^{1-k_{t}/M} d
z^{(\ell)} \frac{\alpha^{\rm CMW}_s(k_t)}{\pi}
P^{(0)}(z^{(\ell)})\right\},
\end{equation}
such that
$\Delta(M) = \exp\left\{-R_{\rm NLL}(\epsilon k_{t1})\right\}$ up to
non-logarithmic terms included in $\Delta$ but not in
$\exp\left\{-R\right\}$.

As shown in the main text, in the all-order picture, the correct
$z^{(\ell)}$ bounds for each emission depend on the radiation that was
emitted before it. Following the discussion of Section~\ref{sec:nll},
however, we recall that these effects contribute beyond NLL accuracy,
and therefore can be neglected in the present case. We then plug
Eq.~\eqref{eq:partxs-mellin-NLL} into Eq.~\eqref{eq:hadxs} and perform
the inverse Mellin transform as just described, obtaining
\begin{align}
\label{eq:backward-evo}
&\frac{d\Sigma(v)}{d\Phi_B} = \frac{d|M_B|_{c_1
                             c_2}^2}{d\Phi_B}\notag\\ &\times\int_0^{M}\frac{d k_{t1}}{k_{t1}} \int_0^{2\pi}
                     \frac{d\phi_1}{2\pi} 
                             \frac{\Delta(M)}{\Delta(k_{t1})}\sum_{\ell_1=1}^2 \int_{x_{\ell_1}}^{1-k_{t1}/M} d
  z_1^{(\ell_1)} \frac{\alpha^{\rm CMW}_s(k_{t1})}{\pi}\frac{P^{(0)}(z_1^{(\ell_1)})}{z_1^{(\ell_1)}}\sum_{n=0}^{\infty}\frac{1}{n!}
                                                     \prod_{i=2}^{n+1}
                                                     \int_{\epsilon}^{1}\frac{d\zeta_i}{\zeta_i}\int_0^{2\pi}
                                                     \frac{d\phi_i}{2\pi}\notag\\
&  \times \frac{\Delta(k_{t(i-1)})}{\Delta(k_{ti})} \sum_{\ell_i=1}^2 \int_{w_{\ell_i}}^{1-\zeta_i k_{t1}/M} d
  \zi
  \frac{\alpha^{\rm CMW}_s(k_{ti})}{\pi}\frac{P^{(0)}(\zi)}{\zi}
  f_{c_1}(\epsilon
    k_{t1},\bar{x}_1) f_{c_2}(\epsilon k_{t1},\bar{x}_2)\notag\\
&\hspace{9cm}\times\Theta\left(v-V(\{\tilde{p}\},k_1,\dots, k_{n+1})\right),
\end{align}
with $\Delta(\epsilon k_{t1}) = 1$ and 
\begin{equation}
w_{\ell_i} = x_{\ell_i}/\left(\prod_{\substack{j=1\\\ell_j = \ell_i}}^{i-1}
  z_j^{(\ell_j)}\right),\qquad \bar{x}_{1} = x_{1}/\left(\prod_{\substack{j=1\\ \ell_j = 1}}^{n+1}
  z_j^{(\ell_j)}\right),\qquad \bar{x}_{2} = x_{2}/\left(\prod_{\substack{j=1\\ \ell_j = 2}}^{n+1}
  z_j^{(\ell_j)}\right).
\end{equation}
We stress again that the $z_i^{(\ell)}$ limits in
Eq.~\eqref{eq:backward-evo} are obtained in the approximation of soft
kinematics which is valid at NLL accuracy. To implement
Eq.~\eqref{eq:backward-evo} in a Markov process we can now impose an
ordering in the transverse momentum of the emissions, which amounts to
performing the following replacement in Eq.~\eqref{eq:backward-evo}
(we remind that $\zeta_i=k_{ti}/k_{t1}$)
\begin{equation}
  \frac{1}{n!}
  \prod_{i=2}^{n+1}
  \int_{\epsilon}^{1} \frac{d\zeta_i}{\zeta_i} \to \int_{\epsilon}^{1}
  \frac{d\zeta_2}{\zeta_2} \int_{\epsilon}^{\zeta_2}
  \frac{d\zeta_3}{\zeta_3}\dots \int_{\epsilon}^{\zeta_{n}} \frac{d\zeta_{n+1}}{\zeta_{n+1}}.   
\end{equation}
With this replacement, Eq.~\eqref{eq:backward-evo} reproduces the
backward-evolution equation for a shower of primary gluons emitted off
the two initial-state legs (see e.g. Eq. (49) of
ref.~\cite{Marchesini:1987cf}), ordered in transverse momentum. The
only relevant difference with the common parton-shower formulation is
in the fact that, unlike a parton shower, Eq.~\eqref{eq:backward-evo}
does not contain a no-emission event. This term is indeed infinitely
suppressed in our case and therefore it does not contribute to the
final result. As a consequence, the cutoff (represented by $\epsilon
k_{t1}$ in our formula) is replaced by a fixed cut $Q_0$ in the
trasverse momentum of the emissions. In order for
Eq.~\eqref{eq:backward-evo} to be NLL accurate for the
transverse-momentum distribution, the recoil of all initial-state
emissions must be entirely absorbed by the colour singlet. This shows
that a branching algorithm for initial-state radiation that fulfils
the above conditions is NLL accurate for this observable (see
also~\cite{Catani:1990rr}). Analogous considerations apply to other
rIRC safe, global observables of the type~\eqref{eq:v-scaling}.  To
extend the above discussion to the generic flavour case, one is forced
to relax the assumption of $k_t$ ordering in order to implement the
above solution in a Markov-chain Monte-Carlo program.\footnote{We are
  grateful to A.~Banfi for a discussion about this aspect.} Indeed, if
some soft radiation occurs after the flavour-changing collinear
emission has taken place, then it becomes quite cumbersome to
determine the correct colour factor for the former. This is because
coherence guarantees that a soft gluon feels the effective colour
charge of the radiation at smaller angles, which now may involve
combinations of different flavours. A correct solution to this problem
requires to reformulate the evolution by ordering the radiation in
angle. This ensures that the hard-collinear emissions contributing to
the DGLAP evolution happen at last (see also the discussion in
Appendix E.2 of ref.~\cite{Banfi:2004yd}), and the colour structure of
the soft radiation is easily determined. It is possible to show that
the backward-evolution algorithm reproduces the resulting evolution
formula in that case as well, and it is therefore NLL accurate.

\section{Analytic formulae for the N$^3$LL radiator}
\label{app:radiator}
In this Appendix we report the expressions for some of the quantities
used in the article. The RGE equation for the QCD coupling reads
\begin{equation}
\frac{d\alpha_s(\mu)}{d\ln \mu^2} = \beta(\alpha_s) \equiv
-\alpha_s\left( \beta_0 \alpha_s +\beta_1\alpha_s^2 +\beta_2
  \alpha_s^3 +\beta_3 \alpha_s^4 + \dots\right).
\end{equation}
The coefficients of the $\beta$ function (with $n_f$ active flavours) are
\begin{align}
     &\beta_0 = \frac{11 C_A - 2 n_f}{12\pi}\,,\quad 
     \beta_1 = \frac{17 C_A^2 - 5 C_A n_f - 3 C_F n_f}{24\pi^2}\,,
\end{align}
\begin{eqnarray}
\beta_2 &=& \frac{2857 C_A^3+ (54 C_F^2 -615C_F C_A -1415 C_A^2)n_f
       +(66 C_F +79 C_A) n_f^2}{3456\pi^3}\,,\\
\beta_3 &=& \frac{1}{(4\pi)^4}\Bigg\{C_A C_F n_f^2 \frac14\left(\frac{17152}{243} + \frac{448}9 \zeta_3\right) + 
C_A C_F^2 n_f \frac12\left(-\frac{4204}{27} + \frac{352}{9} \zeta_3\right)\nonumber\\
&&\hspace{10mm} + \frac{53}{243} C_A n_f^3 + C_A^2 C_F n_f\frac12 \left(\frac{7073}{243} - \frac{656}9 \zeta_3\right) + 
C_A^2 n_f^2 \frac14\left(\frac{7930}{81} + \frac{224}9 \zeta_3\right)\nonumber\\
&&\hspace{10mm} + \frac{154}{243} C_F n_f^3 + 
C_A^3 n_f \frac12\left(-\frac{39143}{81} + \frac{136}3 \zeta_3\right) + C_A^4 \left(\frac{150653}{486} - \frac{44}9 \zeta_3\right)\nonumber\\
&&\hspace{10mm} + C_F^2 n_f^2 \frac14\left(\frac{1352}{27} - \frac{704}9 \zeta_3 \right) + 23 C_F^3 n_f + n_f \frac{d_F^{abcd}d_A^{abcd}}{N_A} \left(\frac{512}9 - \frac{1664}3 \zeta_3\right)\nonumber\\
&&\hspace{10mm} + n_f^2\frac{d_F^{abcd}d_F^{abcd}}{N_A} \left(-\frac{704}9 + \frac{512}3 \zeta_3\right) + \frac{d_A^{abcd}d_A^{abcd}}{N_A} \left(-\frac{80}9 + \frac{704}3 \zeta_3\right)\Bigg\}\,,
\end{eqnarray}
where
\begin{eqnarray*}
\frac{d_F^{abcd}d_F^{abcd}}{N_A} &=& \frac{N_c^4 - 6 N_c^2 + 18}{96 N_c^2},\\
\frac{d_F^{abcd}d_A^{abcd}}{N_A} &=& \frac{N_c(N_c^2 + 6)}{48},\\
\frac{d_A^{abcd}d_A^{abcd}}{N_A} &=& \frac{N_c^2 (N_c^2 + 36)}{24},
\end{eqnarray*}
and $C_A = N_c$, $C_F = \frac{N_c^2-1}{2N_c}$, and $N_c = 3$.

The lowest-order regularised Altarelli-Parisi splitting functions in
four dimensions are
\begin{eqnarray}
\label{eq:regAP}
\hat P^{(0)}_{qq}(z)&=&C_F\left[\frac{1+z^2}{(1-z)_+}+\frac32\delta(1-z)\right],\nonumber\\
\hat P^{(0)}_{qg}(z)&=&\frac12\left[z^2+(1-z)^2\right],\nonumber\\
\hat P^{(0)}_{gq}(z)&=&C_F\frac{1+(1-z)^2}{z},\nonumber\\
\hat P^{(0)}_{gg}(z)&=&2C_A\left[\frac z{(1-z)_+}+\frac{1-z}z+z(1-z)\right]+2\pi\beta_0\delta(1-z),
\end{eqnarray}
where the plus prescription is defined as
\begin{equation}
\int_0^1dz\,\frac{f(z)}{(1-z)_+}=\int_0^1dz\,\frac{f(z)-f(1)}{1-z}.
\end{equation}
The corresponding unregularised Altarelli-Parisi splitting functions
in four dimensions are
\begin{eqnarray}
\label{eq:regAP}
 P^{(0)}_{qq}(z)&=&C_F\frac{1+z^2}{1-z},\nonumber\\
 P^{(0)}_{qg}(z)&=&\frac12\left[z^2+(1-z)^2\right],\nonumber\\
 P^{(0)}_{gq}(z)&=&C_F\frac{1+(1-z)^2}{z},\nonumber\\
 P^{(0)}_{gg}(z)&=&C_A\left[\frac z{1-z}+\frac{1-z}z+z(1-z)\right] \rightarrow C_A\left[2\frac{z}{1-z}+z(1-z)\right],
\end{eqnarray}
where in the last step we exploited the symmetry of the $P^{(0)}_{gg}(z)$ splitting function 
in $z \rightarrow 1-z$.

Next we report the functions that enter the definition of the Sudakov radiator
(Eq.~\eqref{eq:mod-radiator}) up to NNLL. To simplify the notation we
set $\lambda=\alpha_s(\mu_R)\beta_0 L$. They read
\begin{align}
  g_{1}(\alpha_s L) =& \frac{A^{(1)}}{\pi\beta_{0}}\frac{2 \lambda +\ln (1-2 \lambda )}{2  \lambda }, \\
  g_{2}(\alpha_sL) =& \frac{1}{2\pi \beta_{0}}\ln (1-2 \lambda )
  \left(A^{(1)} \ln \frac{1}{x_Q^2}+B^{(1)}\right)
  -\frac{A^{(2)}}{4 \pi ^2 \beta_{0}^2}\frac{2 \lambda +(1-2
    \lambda ) \ln (1-2 \lambda )}{1-2
    \lambda} \notag\\
  &+A^{(1)} \bigg(-\frac{\beta_{1}}{4 \pi \beta_{0}^3}\frac{\ln
    (1-2 \lambda ) ((2 \lambda -1) \ln (1-2 \lambda )-2)-4
    \lambda}{1-2 \lambda}\notag\\
&\hspace{10mm}-\frac{1}{2 \pi \beta_{0}}\frac{(2 \lambda(1
    -\ln (1-2 \lambda ))+\ln (1-2 \lambda ))}{1-2\lambda} \ln
    \frac{\mu_R^2}{x_Q^2 M^2}\bigg)\,,\\
 g_{3}(\alpha_sL) =& \left(A^{(1)} \ln\frac{1}{x_Q^2}+B^{(1)}\right)
    \bigg(-\frac{\lambda }{1-2 \lambda} \ln
   \frac{\mu _{R}^2}{x_Q^2M^2}+\frac{\beta_{1}}{2 \beta_{0}^2}\frac{2 \lambda
   +\ln (1-2 \lambda )}{1-2 \lambda}\bigg)\notag\\
&   -\frac{1}{2 \pi\beta_{0}}\frac{\lambda}{1-2\lambda}\left(A^{(2)}
       \ln\frac{1}{x_Q^2}+B^{(2)}\right)-\frac{A^{(3)}}{4 \pi ^2 \beta_{0}^2}\frac{\lambda ^2}{(1-2\lambda )^2} \notag\\
&   +A^{(2)} \bigg(\frac{\beta_{1}}{4 \pi  \beta_{0}^3 }\frac{2 \lambda  (3
   \lambda -1)+(4 \lambda -1) \ln (1-2 \lambda )}{(1-2 \lambda
   )^2}-\frac{1}{\pi \beta_{0}}\frac{\lambda ^2 }{(1-2 \lambda )^2}\ln\frac{\mu_R^2}{x_Q^2 M^2}\bigg) \notag\\
   & +A^{(1)} \bigg(\frac{\lambda  \left(\beta_{0} \beta_{2} (1-3 \lambda
   )+\beta_{1}^2 \lambda \right)}{\beta_{0}^4 (1-2 \lambda)^2}
   +\frac{(1-2 \lambda) \ln (1-2 \lambda ) \left(\beta_{0} \beta_{2} 
   (1-2 \lambda )+2 \beta_{1}^2 \lambda \right)}{2\beta_{0}^4 (1-2 \lambda)^2} 
   \notag\\&\hspace{10mm}+\frac{\beta_{1}^2}{4 \beta_{0}^4}
   \frac{(1-4 \lambda ) \ln ^2(1-2 \lambda )}{(1-2 \lambda)^2}-\frac{\lambda ^2 }{(1-2 \lambda
   )^2} \ln ^2\frac{\mu_R^2}{x_Q^2 M^2}\notag\\
   &
\hspace{10mm}   -\frac{\beta_{1}}{2 \beta_{0}^{2}}\frac{(2 \lambda  (1-2 \lambda)+(1-4 \lambda) \ln (1-2 \lambda ))
   }{(1-2\lambda )^2}\ln\frac{\mu_R^2}{x_Q^2 M^2}\bigg).
\end{align}
The new N$^3$LL $g_4$ coefficient reads
\begin{eqnarray}
g_4(\alpha_s L) & = & \frac{A^{(4)} (3-2 \lambda ) \lambda ^2}{24 \pi ^2 \beta_0^2 (2 \lambda -1)^3}\nonumber\\
&& + \frac{A^{(3)}}{48 \pi 
\beta_0^3 (2 \lambda -1)^3}\Bigg\{3 \beta_1 (1-6 \lambda ) \ln (1-2 \lambda )+2 \lambda  \Bigg(\beta_1 (5 \lambda  (2 \lambda -3)+3)\nonumber\\
&&\hspace{10mm} +6 \beta_0^2 (3-2 \lambda ) \lambda  \ln
   \frac{\mu_R^2}{x_Q^2M^2}\Bigg)+12 \beta_0^2 (\lambda -1) \lambda
    (2 \lambda -1) \ln \frac{1}{x_Q^2}\Bigg\} \nonumber\\
   && + \frac{A^{(2)}}{24
   \beta_0^4 (2 \lambda -1)^3} \Bigg\{32 \beta_0 \beta_2 \lambda ^3-2 \beta_1^2 \lambda 
   (\lambda  (22 \lambda -9)+3)\nonumber\\
   &&\hspace{10mm}+12 \beta_0^4 (3-2 \lambda ) \lambda ^2 \ln
   ^2\frac{\mu_R^2}{x_Q^2M^2}+6 \beta_0^2 \ln
   \frac{\mu_R^2}{x_Q^2M^2}\times\nonumber\\
   &&\hspace{10mm}\left(\beta_1 (1-6 \lambda ) \ln (1-2
   \lambda )+2 (\lambda -1) \lambda  (2 \lambda -1) \left(\beta_1+2 \beta_0^2 \ln\frac{1}{x_Q^2}\right)\right)\nonumber\\
   &&\hspace{10mm}+3 \beta_1 \Bigg(\beta_1 \ln (1-2
   \lambda ) (2 \lambda +(6 \lambda -1) \ln (1-2 \lambda )-1)\nonumber\\
   &&\hspace{10mm}-2 \beta_0^2 (2 \lambda -1) (2
   (\lambda -1) \lambda -\ln (1-2 \lambda )) \ln \frac{1}{x_Q^2}\Bigg)\Bigg\}\nonumber\nonumber\\
   && + \frac{\pi  A^{(1)}}{12 \beta_0^5 (2 \lambda -1)^3} \Bigg\{\beta_1^3 (1-6 \lambda ) \ln ^3(1-2 \lambda )+3 \ln (1-2 \lambda )
   \Bigg(\beta_0^2 \beta_3 (2 \lambda -1)^3\nonumber\\
   &&\hspace{10mm}+\beta_0 \beta_1
   \beta_2 \left(1-2 \lambda  \left(8 \lambda ^2-4 \lambda +3\right)\right)+4 \beta_1^3
   \lambda ^2 (2 \lambda +1)\nonumber\\
   &&\hspace{10mm}+\beta_0^2 \beta_1 \ln \frac{\mu_R^2}{
   x_Q^2M^2} \left(\beta_0^2 (1-6 \lambda ) \ln \frac{\mu_R^2}{
   x_Q^2M^2}-4 \beta_1 \lambda \right)\Bigg)\nonumber\\
   &&\hspace{10mm}+3 \beta_1^2 \ln ^2(1-2 \lambda
   ) \left(2 \beta_1 \lambda +\beta_0^2 (6 \lambda -1) \ln \frac{\mu_R^2}{x_Q^2M^2}\right)\nonumber\\
   &&\hspace{10mm}+3 \beta_0^2 (2 \lambda -1) \ln
\frac{1}{x_Q^2} \Bigg(-\beta_1^2 \ln ^2(1-2 \lambda ) +2 \beta_0^2
   \beta_1 \ln (1-2 \lambda ) \ln \frac{\mu_R^2}{
   x_Q^2M^2}\nonumber\\
   &&\hspace{10mm}+4 \lambda 
   \left(\lambda  \left(\beta_1^2-\beta_0 \beta_2\right)+\beta_0^4
   (\lambda -1) \ln ^2\frac{\mu_R^2}{x_Q^2M^2}\right)\Bigg)\nonumber\\
   &&\hspace{10mm}+2 \lambda  \Bigg(\beta_0^2 \beta_3 ((15-14 \lambda )
   \lambda -3)+\beta_0 \beta_1 \beta_2 (5 \lambda  (2 \lambda -3)+3)\nonumber\\
   &&\hspace{10mm}+4
   \beta_1^3 \lambda ^2+2 \beta_0^6 (3-2 \lambda ) \lambda  \ln
   ^3\frac{\mu_R^2}{x_Q^2M^2}+3 \beta_0^4 \beta_1 \ln
   ^2\frac{\mu_R^2}{x_Q^2M^2}\nonumber\\
   &&\hspace{10mm}+6 \beta_0^2 \lambda  (2 \lambda +1)
   \left(\beta_0 \beta_2-\beta_1^2\right) \ln \frac{\mu_R^2}{x_Q^2M^2}-8 \beta_0^6 \left(4 \lambda ^2-6 \lambda +3\right) \zeta_3\Bigg)\Bigg\}\nonumber\\
   && + \frac{B^{(3)} (\lambda -1) \lambda }{4 \pi  \beta_0 (1-2 \lambda )^2}+ \frac{B^{(2)} \left(\beta_1 \ln (1-2 \lambda )-2 (\lambda -1) \lambda  \left(\beta_1-2 \beta_0^2 \ln \frac{\mu_R^2}{x_Q^2M^2}\right)\right)}{4\beta_0^2
   (1-2\lambda )^2}\nonumber\\
   && + \frac{\pi  B^{(1)}}{4 \beta_0^3 (1-2 \lambda )^2} \Bigg\{4 \lambda  \left(\lambda 
   \left(\beta_1^2-\beta_0 \beta_2\right)+\beta_0^4 (\lambda -1) \ln
   ^2\frac{\mu_R^2}{x_Q^2M^2}\right)\nonumber\\
   &&\hspace{10mm}-\beta_1^2 \ln ^2(1-2 \lambda )+2 \beta_0^2 \beta_1
   \ln (1-2 \lambda ) \ln \frac{\mu_R^2}{x_Q^2M^2}\Bigg\}.
\end{eqnarray}

\section{Additional considerations on the small-$p_t$ scaling}
\label{app:more-small-pt-limit}
In this Appendix we discuss further the analysis of the $p_t\to 0$ limit of the
differential cross section carried out in Sec.~\ref{sec:small-pt-limit}.
Specifically, following Ref.~\cite{Parisi:1979se}, we have made a
number of approximations to derive the scaling of the
integral~\eqref{eq:pp-1} with $\Lambda^2_{\rm QCD}/M^2$. Such
approximations, however, are too rough to capture the correct
${\cal O}(1)$ normalisation of the scaling, and in this appendix we
quantify the difference from the correct result obtained by directly
integrating Eq.~\eqref{eq:pp-1}.

For the sake of convenience, we define the quantity
\begin{equation}
\Delta \equiv \lim_{p_t \rightarrow 0} \frac{1}{\sigma^{(0)}(\Phi_B)	}  \frac{d^2\Sigma(v)}{p_t d p_t d\Phi_B} \equiv \lim_{p_t \rightarrow 0} \frac{d \bar \sigma}{p_t d p_t}\,,
\end{equation}
where $\frac{d^2\Sigma(v)}{p_t d p_t d\Phi_B} $ was obtained in
Eq.~\eqref{eq:pp-1}.  The $p_t\to 0$ limit simply corresponds to
setting $J_0(p_t b) = 1 + \mathcal O(p_t^2)$, leading to
\begin{align}\label{eq:app-ds}
\Delta & = \int\! \!b\, d b
                           \int\frac{d k_{t1}}{k_{t1}} e^{-R(k_{t1})}  R'(k_{t1}) J_0(b k_{t1})\notag\\
&\hspace{4cm}\times\exp\left\{-R'\left(k_{t1}\right) \int^{k_{t1}}_{0}\frac{d k_t}{k_t}
  (1-J_0(b
  k_{t}))\right\}.
\end{align}
The result given in Eq.~\eqref{eq:pp-2} and~\eqref{eq:pp-3} has been
obtained by further approximating the integral in
Eq.~\eqref{eq:j0-integral} with its asymptotic behaviour. While this
approximation is sufficient to capture the correct scaling of the
cross section at $p_t=0$, it leads to an inaccurate estimate of the
${\cal O}(1)$ normalisation. For comparison, we then recall
Eq.~\eqref{eq:pp-2}:
\begin{equation}\label{eq:app-ds-approx}
\Delta_{\textrm{approx.}} = 4 \int \frac{d k_{t1} }{k_{t1}^3} e^{-R(k_{t1})}\,.
\end{equation}

\begin{figure}[ht!]
  \centering
  \includegraphics[width=0.7\textwidth]{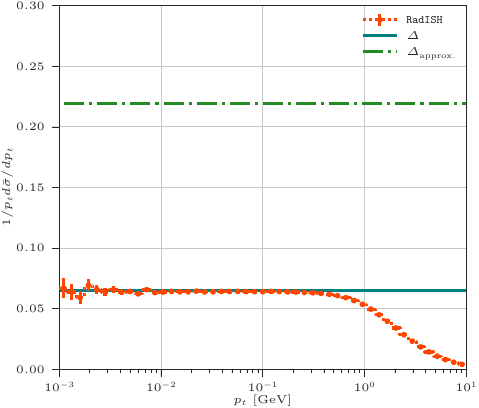}
  \caption{Comparison of different approximations for the small-$p_t$ limit.}
  \label{fig:ppapprox}
\end{figure}

To numerically quantify the difference between
Eqs.~\eqref{eq:app-ds-approx} and~\eqref{eq:app-ds}, we consider the
case of $Z$ production with $m_Z = 91.1876$~GeV.
We evaluate numerically Eq.~\eqref{eq:app-ds} and
Eq.~\eqref{eq:app-ds-approx} and we compare them with the prediction
obtained with the \texttt{RadISH} code.
For the sake of simplicity, we use everywhere the LL expressions for $R$
and its derivative $R'$.
To regulate the $b\rightarrow 0$ ($+\infty$) limits in
Eq.~\eqref{eq:app-ds} we set a lower (upper) limit of $b_0/m_Z$
($b_0/Q_0$) in the $b$ integral, where $Q_0$ is the Landau singularity
of the integrand which reads
\begin{equation}
Q_0=m_Z\,e^{-\frac{1}{2\as \beta_0}}\,.
\end{equation}
Correspondingly, we integrate over $k_{t1} $ in both
Eqs.~\eqref{eq:app-ds} and~\eqref{eq:app-ds-approx} between $Q_0$ and
$m_Z$ and we use the same limits in the numerical results obtained
with \texttt{RadISH}.
The results are displayed in Fig.~\ref{fig:ppapprox}, where we observe
that the numerical result obtained with \texttt{RadISH}
converges in the $p_t \rightarrow 0$ limit to the result given in
Eq.~\eqref{eq:app-ds}.
Both results differ significantly from the approximated result
Eq.~\eqref{eq:app-ds-approx}, which overestimates the normalisation by
more than a factor of three.
We observe that such a difference can be also obtained by
computing the ratio of Eq.~\eqref{eq:app-ds-approx} to the $p_t\to 0$
limit of the original $b$-space formulation.\footnote{We thank G. Salam for useful discussions on this topic.}

\bibliographystyle{JHEP}
\bibliography{directspace}

\end{document}